\documentclass{aa}
\usepackage{graphics}
\usepackage{times}
\newcommand{\eV}{{\rm e\hspace{-1pt}V}}
\newcommand{\Teff}{T_{\rm eff}}
\newcommand{\taur}{\overline{\tau}} 
\newcommand{\Vmac}{\Xi_{\rm rt}}
\newcommand{\vmic}{\xi_{\rm t}}
\newcommand{\SH}{S_{\rm H}}
\newcommand{\Emin}{E_{\rm min}}
\newcommand{\eps}[1]{\log\varepsilon_{\rm #1}}
\newcommand{\kms}{km\,s$^{-1}$}
\newcommand{\FeI}{\ion{Fe}{i}}
\newcommand{\FeII}{\ion{Fe}{ii}}

\newcommand{\term}[3]{$#1^{#2}{\rm #3}$}

\begin{document}

\title{Kinetic equilibrium of iron in the atmospheres of cool dwarf stars}
\subtitle{II. Weak \FeI\ lines in the solar spectrum}
\titlerunning{Kinetic equilibrium of iron in the solar spectrum}
\author{Thomas Gehren\inst{1} \and Andreas J. Korn\inst{1}
\and Jianrong Shi\inst{1,2}}
\offprints{T. Gehren,\\ \email{gehren@usm.uni-muenchen.de}}
\institute{Institut f\"ur Astronomie und Astrophysik der Universit\"at 
M\"unchen\\ Universit\"ats-Sternwarte M\"unchen (USM), Scheinerstr. 1, D-81679 
M\"unchen, Germany \and National Astronomical Observatories, Chinese Academy of 
Sciences, Beijing 100012, China} 
\date{Received / Accepted}
\abstract{%
NLTE line formation calculations of \FeI\ in the solar atmosphere are extended 
to include weak lines in the visual spectrum of the Sun. Previously established 
atomic models are used to discriminate between different ways of treating 
collisional interaction processes. As indicated by the analysis of strong \FeI\ 
lines, the influence of deviations from LTE in the solar atmosphere on the Fe 
abundance is small for all lines. To derive a common {\em solar} \FeI\ abundance 
from both strong {\em and} weak lines fine-tuning of the microturbulence 
velocity parameter and the van der Waals damping constants is required. The 
solar \FeI\ abundances based on all available $f$-values are dominated by the 
large scatter already found for the stronger lines. In particular the bulk of 
the data from the work of May et al. and O'Brian et al. is not adequate for 
accurate abundance work. Based on $f$-values measured by the Hannover and Oxford 
groups alone, the \FeI\ LTE abundances are $\eps{\FeI,\odot} = 7.57$ for the 
empirical and $\eps{\FeI,\odot} = 7.48 \ldots 7.51$ for the line-blanketed solar 
model. The solar Fe ionization equilibrium obtained for different atomic and 
atmospheric models rules out NLTE atomic models with a low efficiency of 
hydrogen collisions. At variance with Paper I, it is now in better agreement 
with \emph{laboratory} \FeII\ $f$-values for all types of line-blanketed models. 
Our final model assumptions consistent with a {\em single} unique solar Fe 
abundance $\eps{Fe,\odot} \sim 7.48 \ldots 7.51$ calculated from NLTE line 
formation are (a) a line-blanketed solar model atmosphere, (b) an iron model 
atom with hydrogen collision rates $0.5 < \SH < 5$ times the standard value to 
compensate for the large photoionization cross-sections, (c) a microturbulence 
velocity $\vmic = 1.0$ \kms, (d) van der Waals damping parameters decreased by 
$\Delta\log C_6 = -0.10 \ldots -0.15$ as compared to Anstee \& O'Mara's 
calculations, depending on $\SH$, (e) \FeII\ $f$-values as published by Schnabel 
et al., and (f) \FeI\ $f$-values published by the Hannover and Oxford groups. 
} 
\maketitle 

\section{Introduction}

Our previous attempt to understand the formation of the iron spectrum in cool
dwarf stars (Gehren et al. \cite{GBMRS01}, Paper I) was successful in isolating
some of the important interaction processes encountered in stellar atmospheres
of spectral types F and G. The compensating influence of (a) strong collisional
coupling of the highly excited ($> 7.3$ \eV) \FeI\ terms to the \term{a}{6}{D}
ground state of \FeII, (b) hydrogen collision cross sections, and (c)
photoionization from the low-excitation terms was shown to dominate the
synthesis of line profiles and the abundances of solar lines. 

The lines used for the analysis were selected for strength because it is planned
to extend the investigation to extremely metal-poor stars where the NLTE effects
are predicted to be much more important. In such stars only lines are detected
that are strong in the Solar spectrum. The comparison of observed solar flux
spectra with synthesized line profiles is thus hampered by all the problems 
usually occurring whenever line-broadening starts to play a role. 

The treatment of van der Waals damping had been based on relatively simple
approximations for a long time (Uns\"old \cite{U68}, Kurucz \cite{K92}), often 
resulting in significant underestimates of the damping constant. For a treatment 
of NLTE effects this was completely inacceptable, thus in Paper I we applied the 
quantum mechanical calculations of Anstee \& O'Mara (\cite{AO91}, \cite{AO95}) 
without any corrections. Although the results show substantial improvements 
there were still multiplets for which corrections would seem adequate from 
profile fitting. This is not easily explained although the calculations refer to 
simple LS coupling schemes whereas some of the upper \FeI\ terms involved are 
affected by mixing from different configurations. It appears that the Anstee \& 
O'Mara damping constants in some multiplets lead to line abundances that are 
slightly {\em smaller} than those obtained from weaker lines. 

Granular hydrodynamics are a second item that affects our results (Asplund et
al. \cite{ANTS00}). Relying on horizontally homogeneous, plane-parallel
atmospheric stratifications implies that dynamic movements are replaced by
approximate velocity fields, usually termed micro- and macroturbulence. For
obvious reasons such an artificial replacement could depend on atmospheric depth
as found in the empirical solar model of Holweger \& M\"uller (\cite{HM74}).
Whereas such a stratification $\xi(\taur)$ can in principle also be constructed
for other {\em solar} models, this is not always possible for other stars.
Therefore, our fit to the solar \FeI\ line spectrum was based on a single
microturbulence velocity $\overline{\xi}$. The values assumed for the strong
lines of Paper I ($\vmic = 1.00$ \kms\ for the empirical and $\vmic = 0.85$ 
\kms\ for the line-blanketed atmospheric model) were smaller than usually 
adopted for both types of model atmospheres. Thus, based on turbulence lines 
alone (lines whose equivalent widths are dominated by broadening due to 
microturbulence velocities), the abundances derived for both \FeII\ and \FeI\ 
would be slightly too high. 

After having examined more than 100 strong \FeI\ lines arising from excitation
energies between 0 and 5 \eV\ including some of the stronger turbulence lines we 
have found that combinations of certain atomic model properties lead to 
acceptable solar flux profile fits if varying {\em macroturbulence} velocities 
$\Vmac$ (Gray \cite{G77}) are applied. Due to the fact that a plane-parallel 
atmospheric model can not represent granular hydrodynamics with infinite 
accuracy, we have not tried to improve our NLTE profile fits beyond certain 
limits that are characterized by $\sim 1$\% {\em rms} deviation from the 
observed fluxes. Yet it became clear that atomic models with different strengths 
of collisional interaction led essentially to similarly good fits. This could be 
explained as a consequence of different \FeI\ abundances or uncertain $f$-values 
and van der Waals damping parameters. Unfortunately, the solar \FeII\ abundances 
are at least as uncertain due to significantly different sets of $f$-values. 
Thus the solar ionization equilibrium of iron could not be established because 
the absolute abundances were uncertain from both ends. 

As explained above part of the uncertainty remaining after modelling the strong 
lines is due to line-broadening by microturbulence and damping. Our 
understanding of the kinetic equilibrium of \FeI\ could therefore be 
considerably improved by extending the NLTE line formation analysis to lines 
that are substantially weaker than those of Paper I. Such lines would not be 
detected in metal-poor stars, but they would help to select the atomic model 
producing the best fit to the solar spectrum. Our present investigation is thus 
extended to a large number of lines with equivalent widths smaller than $\sim 
100$m\AA. This includes lines of all degrees of excitation, although recently 
identified Rydberg transitions in the infrared with excitation energies well 
above 7 \eV\ (Johansson et al. \cite{JNGSGSCF94}, Schoenfeld et al. 
\cite{SCGJNSG95}) were excluded because no $f$-values are available. The 
following section gives a short representation of the assumptions concerning 
both atomic and atmospheric models. Section 3 introduces the sample of \FeI\ 
lines with results of NLTE line formation and profile synthesis. The last 
section presents our conclusions and a comparison with those of Paper I. We note 
in advance that the present analysis is still not able to produce a unique 
atomic model that can be applied to all kinds of stars. Such an investigation is 
left to a forthcoming paper, in which we will extend the analysis to a number of 
(mostly metal-poor) reference stars. 

\section{Model assumptions}

\subsection{Atomic models}

Basic atomic models are the same as those of Paper I. Because they are described 
there at considerable length we will not repeat the details here. The main 
differences between them are characterized by 
\begin{itemize}
\item
the strength of the {\em neutral hydrogen collisions}, represented by a
collision enhancement factor, $\SH$, which is 0 in the case of no hydrogen
collisions. All other cases describe the factor with which the collision formula
proposed by Drawin (\cite{D68}, \cite{D69}; see also Steenbock \& Holweger
\cite{SH84}) is multiplied. We note that $\SH \rightarrow \infty$ leads to LTE.
Our final choice resulted in $\SH = 5$, a value that is significantly greater
than found previously for other atoms such as Al (Baum\"uller \& Gehren
\cite{BG96}, \cite{BG97}) or Mg (Zhao et al. \cite{ZBG98}). Note that the role
of hydrogen collisions is more important for \FeI\ than it is for \ion{Al}{i} or
\ion{Mg}{i}, because photoionization of \FeI\ levels is substantially stronger 
than that of the other atoms for levels of {\em all} excitation energies; the 
large value of $\SH$ is therefore to be considered as a compensation for the 
large photoionization cross-sections calculated by Bautista (\cite{B97}). 
\item
the treatment of the highly excited levels of \FeI. Due to the strong
photoionization from virtually all \FeI\ levels the collisional coupling between
levels above a certain limit $\Emin$ of excitation energy and between these
levels and the \FeII\ parent terms is of critical quality. Electron collisions
are treated by the van Regemorter (\cite{V62}) approximation in case of {\em
allowed bb} collisions, by that of Allen (\cite{A73}) for {\em forbidden bb}
collisions, and according to Seaton's (\cite{S62}) recipe for {\em bf}
collisions. As is obvious already from the year of appearance of these 
references collisions are the ''weak point'' of our considerations. At optical 
depths of the solar atmosphere from where most of the \FeI\ lines emerge, the 
resulting interaction by electron collisions is too weak to produce a tight 
coupling of the higher terms to the continuum. As a consequence, hydrogen 
collisions tend to result in a relative thermalization of only the lower \FeI\ 
terms (see Paper I, Figs. 6b and 6h). Thus even with strong hydrogen collisions 
($\SH > 1$) only the {\em source functions} are thermalized but not the level 
populations or line opacities. Such a situation always leads to uncomfortably 
{\em strong} NLTE effects in the solar spectrum. We have therefore forced 
thermalization with respect to \FeII\ of all terms above $\Emin$, where 
different models specified  $\Emin = 6.7$, $7.0$, and $7.3$ \eV, respectively. 
In Paper I we decided to use $\Emin = 7.3$ \eV\ for the final model, because 
that choice guaranteed that none of the lines and levels investigated in the 
solar spectrum was directly affected. 
\end{itemize}

\begin{figure}
\resizebox{\hsize}{!}{\includegraphics{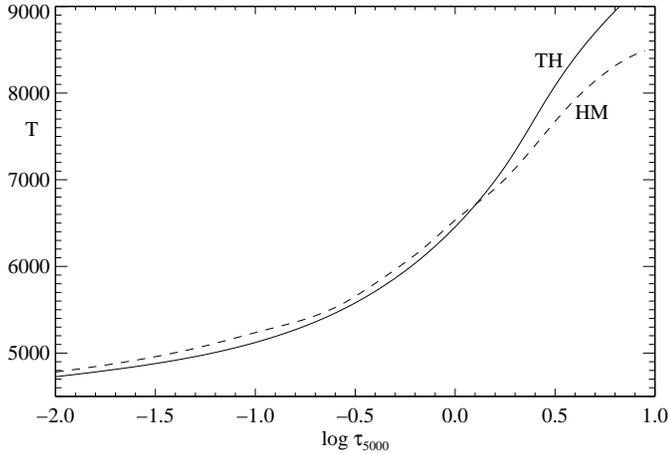}} \caption[]{Photospheric 
solar temperature distributions of the HM empirical model (dashes) and the TH 
line-blanketed model (continuous curve)} \label{soltemp} 
\end{figure}
\subsection{Atmospheric models}

The two plane-parallel horizontally homogeneous atmospheric models used in our 
analysis are the semi-empirical solar model of Holweger \& M\"uller (HM, 
\cite{HM74}) and our line-blanketed solar model (TH, see Paper I). Their most 
important difference with respect to line formation is the temperature 
stratification, with $T_{\rm HM}(\taur) - T_{\rm TH}(\taur) \sim 150$ K at 
optical depths between 0.1 and 1.0. The two stratifications are displayed in 
Fig. \ref{soltemp}, and the most important result of the temperature difference 
is that typically the \emph{stronger} lines are calculated with \emph{weaker} 
line wings in the empirical solar model. Therefore a proper fit of \ion{Fe}{I} 
line profiles using the HM empirical model always requires slightly higher 
damping parameters than for the TH model. 

Other important parameters of the models are those determining non-thermal 
spectral line core broadening. In Paper I we have chosen $\xi = 1.00$ (HM) and 
$0.85$ \kms\ (TH), $\Xi_{\rm rt} = 2.5$ (HM) and $3.2$ \kms\ (TH), respectively. 
There is clear evidence that both micro- and macroturbulence vary with depth of 
line formation, however, only $\Xi_{\rm rt}$ was allowed to vary between $\sim 
2.0$ \kms\ for some of the most saturated Doppler profiles and $\sim 4.0$ \kms\ 
for very weak lines. For a more realistic analysis of both weak and strong lines 
in this paper we have added a second value of $\xi = 1.00$ \kms\ for the TH 
model and recalculated the non-LTE populations and line profiles. No such 
alternative was examined for the HM model although this would probably reduce 
the solar \ion{Fe}{I} abundances by similar amounts as for the TH model.   

The empirical HM model is used here only as a comparison for abundance 
discussions. It had been established as a reference for LTE conditions in the 
solar photosphere, and therefore we have \emph{not} attempted to calculate 
non-LTE populations for its temperature distribution. All the other level  
populations in this paper thus refer to the TH model for which we distinguish 
between the (sets of) model assumptions given in Table \ref{models}.
\begin{table}
\begin{center}
\caption{TH models used in the present calculations}\label{models}
\begin{tabular}{ccccccc}
    & Type & $\overline{\xi}$&$\SH$&$E_{\rm min}$&$\Delta\log C_6$& Name\\
\noalign{\smallskip}\hline\noalign{\smallskip}   
  0 &  LTE & 0.85 &     &     &      & LTE(0.85)\\
  1 & NLTE & 0.85 & 0.0 & 7.3 &      & 0+(0.85) \\
  2 & NLTE & 0.85 & 5.0 & 7.3 &      & 5+(0.85) \\
  3 & NLTE & 0.85 & 5.0 &     &      & 5-(0.85) \\
  5 &  LTE & 1.00 &     &     &      & LTE(1.00)\\
  6 & NLTE & 1.00 & 5.0 & 7.3 &      & 5+(1.00) \\
  7 & NLTE & 1.00 & 1.0 & 7.3 &      & 1+(1.00) \\
  8 & NLTE & 1.00 & 1.0 & 7.3 & $-0.4$ & 1+(1.00) \\
  9 & NLTE & 1.00 & 0.5 & 7.3 & $-0.4$ & 0.5+(1.00)\\ 
\noalign{\smallskip}\hline 
\end{tabular}
\end{center}
\end{table}
Here, $\SH$ and $E_{\rm min}$ refer to the model atom interaction described in 
section 2.1, whereas $\Delta\log C_6$ in the last two entries specifies a 
\emph{decrease} of the damping constants with respect to the Anstee \& O'Mara 
standard. This type of model leads to a substantially improved fit of turbulence 
lines \emph{and} those broadened by van der Waals damping (see below). Note that 
for each model both NLTE populations and line profiles have been recalculated. 
When deriving solar \FeI\ abundances in section 3, some of these models are used 
to interpolate between different damping parameters (model 7 and 8).  

\section{The solar weak line spectrum}

\begin{table*}
\caption[]{\ion{Fe}{I} lines in the solar flux spectrum including lines of Paper I, which have been recalculated with current model settings and $f$-values eliminating some errors in the previous data set. 
Sources of $f$-values and remarks are noted at the end of the table. NLTE models are  described in the text. Equivalent widths are in m\AA. {\bf The complete table is only reproduced in the A\&A version}}
\label{lintab}
{\tiny
\begin{tabular}{rr@{ -- }lrrrlr|rrrr|rrrrrr|p{16pt}r}
\hline\noalign{\smallskip}
Mult & \multicolumn{2}{c}{Transition} & $\lambda$ [\AA]&$E$ [eV]& $\log gf$ & & $\log C_6$ 
& \multicolumn{4}{l}{$\log \varepsilon(\ion{Fe}{I})_\odot$ (0.85)} & 
\multicolumn{6}{l}{$\log \varepsilon(\ion{Fe}{I})_\odot$ (1.00)}& Rem & 
$W_\lambda$\\
\multicolumn{8}{c}{ } & LTE & 0+ & 5+ & 5- & HM & LTE & 5+ & 1+ & 1+ & 0.5+\\ 
\multicolumn{16}{c}{ } & -0.4 & -0.4 \\
\noalign{\smallskip}\hline\noalign{\smallskip}
   1 & $a^5{\rm D  _0}$ & $z^7{\rm D^o_1}$ & $5250.216$ & $0.121$ & $-4.94$ &e& $-32.051$ & 7.57 & 7.71 & 7.61 & 7.74 & 7.66 & 7.49 & 7.53 & 7.55 & 7.56 & 7.59 &  acf  & $  71.2$ \\
   1 & $a^5{\rm D  _1}$ & $z^7{\rm D^o_1}$ & $5225.533$ & $0.110$ & $-4.79$ &e& $-32.052$ & 7.60 & 7.73 & 7.64 & 7.77 & 7.69 & 7.51 & 7.55 & 7.57 & 7.59 & 7.60 &  af   & $  76.1$ \\
   1 & $a^5{\rm D  _2}$ & $z^7{\rm D^o_3}$ & $5247.057$ & $0.087$ & $-4.95$ &e& $-32.057$ & 7.56 & 7.72 & 7.59 & 7.75 & 7.65 & 7.48 & 7.53 & 7.55 & 7.57 & 7.58 &  af   & $  68.7$ \\
   1 & $a^5{\rm D  _4}$ & $z^7{\rm D^o_5}$ & $5166.282$ & $0.000$ & $-4.20$ &e& $-32.070$ & 7.51 & 7.66 & 7.64 & 7.71 & 7.69 & 7.42 & 7.51 & 7.51 & 7.52 & 7.53 &  cdfh & $ 109.0$ \\
   2 & $a^5{\rm D  _2}$ & $z^7{\rm F^o_2}$ & $4445.480$ & $0.087$ & $-5.44$ &e& $-31.996$ & 7.54 & 7.65 & 7.54 & 7.66 & 7.65 & 7.50 & 7.52 & 7.53 & 7.54 & 7.55 &  a    & $  43.3$ \\
   2 & $a^5{\rm D  _3}$ & $z^7{\rm F^o_4}$ & $4427.309$ & $0.052$ & $-2.92$ &a& $-32.000$ & 7.58 & 7.66 & 7.58 & 7.65 & 7.73 & 7.54 & 7.56 & 7.56 & 7.69 & 7.70 &  dei  & $ 199.1$ \\
   2 & $a^5{\rm D  _4}$ & $z^7{\rm F^o_4}$ & $4347.237$ & $0.000$ & $-5.50$ &e& $-32.006$ & 7.54 & 7.65 & 7.55 & 7.66 & 7.66 & 7.49 & 7.51 & 7.53 & 7.53 & 7.54 &  ac   & $  43.6$ \\
   3 & $a^5{\rm D  _1}$ & $z^7{\rm P^o_2}$ & $4232.720$ & $0.110$ & $-4.93$ &e& $-31.968$ & 7.47 & 7.57 & 7.49 & 7.59 & 7.59 & 7.40 & 7.44 & 7.45 & 7.46 & 7.47 &  ac   & $  60.2$ \\
  13 & $a^5{\rm F  _1}$ & $z^7{\rm F^o_1}$ & $6625.026$ & $1.011$ & $-5.35$ &m& $-31.927$ & 7.52 & 7.64 & 7.53 & 7.65 & 7.63 & 7.51 & 7.53 & 7.54 & 7.54 & 7.55 &  ag   & $  16.4$ \\
  13 & $a^5{\rm F  _2}$ & $z^7{\rm F^o_2}$ & $6574.233$ & $0.990$ & $-5.02$ &a& $-32.400$ & 7.56 & 7.69 & 7.57 & 7.69 & 7.68 & 7.54 & 7.56 & 7.58 & 7.58 & 7.60 &  a    & $  29.7$ \\
  13 & $a^5{\rm F  _3}$ & $z^7{\rm F^o_3}$ & $6498.945$ & $0.958$ & $-4.70$ &f& $-31.934$ & 7.58 & 7.70 & 7.59 & 7.73 & 7.67 & 7.54 & 7.56 & 7.58 & 7.59 & 7.61 &  ag   & $  48.1$ \\
  13 & $a^5{\rm F  _4}$ & $z^7{\rm F^o_4}$ & $6400.323$ & $0.915$ & $-4.32$ &a& $-31.939$ & 7.49 & 7.60 & 7.51 & 7.63 & 7.56 & 7.44 & 7.45 & 7.48 & 7.49 & 7.51 &  bf   & $  65.5$ \\
  13 & $a^5{\rm F  _5}$ & $z^7{\rm F^o_5}$ & $6280.620$ & $0.859$ & $-4.39$ &f& $-31.947$ & 7.55 & 7.71 & 7.60 & 7.75 & 7.66 & 7.52 & 7.52 & 7.52 & 7.53 & 7.53 &  dc   & $  68.4$ \\
  14 & $a^5{\rm F  _4}$ & $z^7{\rm P^o_4}$ & $6120.250$ & $0.915$ & $-5.95$ &m& $-31.926$ & 7.53 & 7.64 & 7.54 & 7.65 & 7.69 & 7.53 & 7.53 & 7.53 & 7.53 & 7.53 &  gc   & $   5.6$ \\
  15 & $a^5{\rm F  _2}$ & $z^5{\rm D^o_1}$ & $5405.775$ & $0.990$ & $-1.88$ &p& $-31.870$ & 7.50 & 7.52 & 7.51 & 7.60 & 7.64 & 7.47 & 7.49 & 7.50 & 7.64 & 7.64 &  e    & $ 271.5$ \\
  15 & $a^5{\rm F  _3}$ & $z^5{\rm D^o_2}$ & $5371.489$ & $0.958$ & $-1.65$ &n& $-31.870$ & 7.41 & 7.44 & 7.43 & 7.52 & 7.57 & 7.40 & 7.40 & 7.40 & 7.53 & 7.53 &  e    & $ 306.3$ \\
  15 & $a^5{\rm F  _4}$ & $z^5{\rm D^o_3}$ & $5328.038$ & $0.915$ & $-1.47$ &n& $-31.880$ & 7.46 & 7.52 & 7.47 & 7.56 & 7.62 & 7.44 & 7.46 & 7.47 & 7.61 & 7.61 &  e    & $ 397.9$ \\
  15 & $a^5{\rm F  _4}$ & $z^5{\rm D^o_4}$ & $5397.128$ & $0.915$ & $-1.99$ &n& $-31.880$ & 7.47 & 7.51 & 7.48 & 7.56 & 7.64 & 7.46 & 7.46 & 7.47 & 7.63 & 7.63 &  i    & $ 241.7$ \\
  15 & $a^5{\rm F  _5}$ & $z^5{\rm D^o_4}$ & $5269.537$ & $0.859$ & $-1.32$ &n& $-31.890$ & 7.44 & 7.51 & 7.44 & 7.54 & 7.61 & 7.43 & 7.45 & 7.45 & 7.61 & 7.61 &  e    & $ 501.5$ \\
  34 & $a^3{\rm F  _2}$ & $z^5{\rm F^o_2}$ & $6851.640$ & $1.608$ & $-5.32$ &b& $-31.786$ & 7.46 & 7.58 & 7.47 & 7.58 & 7.60 & 7.46 & 7.47 & 7.49 & 7.49 & 7.50 &  acf  & $   3.9$ \\
  34 & $a^3{\rm F  _3}$ & $z^5{\rm F^o_3}$ & $6739.540$ & $1.557$ & $-4.79$ &p& $-31.795$ & 7.37 & 7.49 & 7.38 & 7.52 & 7.48 & 7.37 & 7.38 & 7.40 & 7.40 & 7.41 &  acf  & $  12.3$ \\
  34 & $a^3{\rm F  _4}$ & $z^5{\rm F^o_4}$ & $6581.220$ & $1.485$ & $-4.68$ &p& $-31.806$ & 7.42 & 7.54 & 7.43 & 7.55 & 7.54 & 7.41 & 7.42 & 7.44 & 7.44 & 7.45 &  ac   & $  20.5$ \\
  34 & $a^3{\rm F  _4}$ & $z^5{\rm F^o_5}$ & $6710.310$ & $1.485$ & $-4.88$ &m& $-31.813$ & 7.54 & 7.65 & 7.55 & 7.66 & 7.64 & 7.53 & 7.54 & 7.55 & 7.55 & 7.56 &  acf  & $  16.2$ \\
  36 & $a^3{\rm F  _2}$ & $z^3{\rm F^o_2}$ & $5216.274$ & $1.608$ & $-2.15$ &o& $-31.670$ & 7.45 & 7.45 & 7.48 & 7.59 & 7.53 & 7.35 & 7.39 & 7.41 & 7.51 & 7.51 &  defh & $ 130.3$ \\
  36 & $a^3{\rm F  _3}$ & $z^3{\rm F^o_3}$ & $5194.941$ & $1.557$ & $-2.09$ &o& $-31.680$ & 7.45 & 7.46 & 7.45 & 7.55 & 7.55 & 7.36 & 7.38 & 7.39 & 7.50 & 7.50 &  eh   & $ 129.6$ \\
  38 & $a^3{\rm F  _2}$ & $y^5{\rm D^o_2}$ & $4798.734$ & $1.608$ & $-4.25$ &b& $-31.612$ & 7.59 & 7.70 & 7.60 & 7.71 & 7.71 & 7.58 & 7.59 & 7.60 & 7.60 & 7.61 &  gh   & $  34.3$ \\
  38 & $a^3{\rm F  _3}$ & $y^5{\rm D^o_3}$ & $4772.820$ & $1.557$ & $-2.90$ &a& $-31.627$ & 7.68 & 7.81 & 7.72 & 7.86 & 7.81 & 7.61 & 7.65 & 7.66 & 7.72 & 7.73 &  dg   & $  93.7$ \\
  41 & $a^3{\rm F  _3}$ & $z^5{\rm G^o_4}$ & $4404.750$ & $1.557$ & $-0.10$ &p& $-31.560$ & 7.40 & 7.48 & 7.41 & 7.50 & 7.57 & 7.39 & 7.41 & 7.41 & 7.54 & 7.55 &  h    & $ 786.0$ \\
  41 & $a^3{\rm F  _4}$ & $z^5{\rm G^o_5}$ & $4383.545$ & $1.485$ & $ 0.20$ &o& $-31.580$ & 7.39 & 7.49 & 7.42 & 7.48 & 7.58 & 7.39 & 7.41 & 7.41 & 7.55 & 7.56 &  h    & $1345.6$ \\
  42 & $a^3{\rm F  _4}$ & $z^3{\rm G^o_3}$ & $4147.669$ & $1.485$ & $-2.10$ &o& $-31.520$ & 7.47 & 7.51 & 7.48 & 7.58 & 7.58 & 7.41 & 7.41 & 7.41 & 7.51 & 7.51 &  ch   & $ 131.2$ \\
  42 & $a^3{\rm F  _4}$ & $z^3{\rm G^o_5}$ & $4271.760$ & $1.485$ & $-0.16$ &o& $-31.550$ & 7.37 & 7.44 & 7.38 & 7.46 & 7.55 & 7.36 & 7.36 & 7.36 & 7.51 & 7.51 &  h    & $ 846.7$ \\
  43 & $a^3{\rm F  _2}$ & $y^3{\rm F^o_2}$ & $4071.738$ & $1.608$ & $-0.02$ &o& $-31.440$ & 7.33 & 7.41 & 7.33 & 7.41 & 7.51 & 7.31 & 7.31 & 7.31 & 7.47 & 7.47 &  h    & $ 860.9$ \\
  43 & $a^3{\rm F  _3}$ & $y^3{\rm F^o_3}$ & $4063.594$ & $1.557$ & $ 0.06$ &a& $-31.470$ & 7.31 & 7.45 & 7.35 & 7.46 & 7.55 & 7.35 & 7.35 & 7.36 & 7.50 & 7.50 &  hj   & $ 900.8$ \\
  43 & $a^3{\rm F  _4}$ & $y^3{\rm F^o_4}$ & $4045.812$ & $1.485$ & $ 0.28$ &o& $-31.490$ & 7.36 & 7.43 & 7.38 & 7.46 & 7.55 & 7.36 & 7.36 & 7.36 & 7.51 & 7.51 &  hj   & $1250.5$ \\
  62 & $a^5{\rm P  _1}$ & $y^5{\rm D^o_2}$ & $6297.800$ & $2.223$ & $-2.73$ &q& $-31.565$ & 7.57 & 7.67 & 7.58 & 7.71 & 7.64 & 7.48 & 7.50 & 7.52 & 7.53 & 7.55 &  h    & $  75.3$ \\
  62 & $a^5{\rm P  _3}$ & $y^5{\rm D^o_2}$ & $6151.620$ & $2.176$ & $-3.27$ &q& $-31.569$ & 7.51 & 7.62 & 7.51 & 7.61 & 7.55 & 7.45 & 7.46 & 7.47 & 7.48 & 7.50 &  h    & $  51.3$ \\
  63 & $a^5{\rm P  _1}$ & $y^5{\rm F^o_2}$ & $6015.250$ & $2.223$ & $-4.68$ &m& $-31.539$ & 7.52 & 7.62 & 7.53 & 7.63 & 7.65 & 7.52 & 7.53 & 7.55 & 7.54 & 7.55 &  ac   & $   4.5$ \\
  64 & $a^5{\rm P  _1}$ & $z^3{\rm P^o_1}$ & $6082.720$ & $2.223$ & $-3.59$ &p& $-31.545$ & 7.53 & 7.63 & 7.54 & 7.65 & 7.61 & 7.51 & 7.52 & 7.53 & 7.54 & 7.55 &  g    & $  35.8$ \\
  64 & $a^5{\rm P  _1}$ & $z^3{\rm P^o_2}$ & $6240.660$ & $2.223$ & $-3.23$ &p& $-31.560$ & 7.48 & 7.57 & 7.49 & 7.61 & 7.55 & 7.43 & 7.44 & 7.45 & 7.47 & 7.48 &  g    & $  50.3$ \\
  66 & $a^5{\rm P  _1}$ & $y^5{\rm P^o_2}$ & $5198.711$ & $2.223$ & $-2.14$ &g& $-31.440$ & 7.58 & 7.65 & 7.60 & 7.70 & 7.63 & 7.51 & 7.53 & 7.54 & 7.56 & 7.57 &  dfk  & $ 103.9$ \\
  66 & $a^5{\rm P  _2}$ & $y^5{\rm P^o_1}$ & $5079.223$ & $2.198$ & $-2.07$ &g& $-31.430$ & 7.55 & 7.64 & 7.59 & 7.68 & 7.67 & 7.50 & 7.52 & 7.53 & 7.57 & 7.59 &  bfk  & $ 107.5$ \\
  66 & $a^5{\rm P  _2}$ & $y^5{\rm P^o_2}$ & $5145.099$ & $2.198$ & $-2.88$ &a& $-31.439$ & 7.26 & 7.36 & 7.27 & 7.38 & 7.34 & 7.22 & 7.24 & 7.25 & 7.27 & 7.28 &  d    & $  54.3$ \\
  66 & $a^5{\rm P  _2}$ & $y^5{\rm P^o_3}$ & $5250.646$ & $2.198$ & $-2.18$ &a& $-31.460$ & 7.69 & 7.75 & 7.72 & 7.82 & 7.76 & 7.59 & 7.62 & 7.65 & 7.72 & 7.73 &  eh   & $ 108.1$ \\
  68 & $a^5{\rm P  _1}$ & $x^5{\rm D^o_1}$ & $4447.717$ & $2.176$ & $-1.34$ &g& $-31.270$ & 7.61 & 7.70 & 7.64 & 7.73 & 7.79 & 7.58 & 7.61 & 7.63 & 7.74 & 7.76 &  d    & $ 185.4$ \\
  68 & $a^5{\rm P  _2}$ & $x^5{\rm D^o_3}$ & $4494.563$ & $2.198$ & $-1.14$ &g& $-31.300$ & 7.48 & 7.53 & 7.50 & 7.59 & 7.61 & 7.45 & 7.47 & 7.48 & 7.60 & 7.60 &  df   & $ 206.8$ \\
  69 & $a^5{\rm P  _2}$ & $y^7{\rm P^o_3}$ & $4447.130$ & $2.198$ & $-2.73$ &a& $-31.280$ & 7.63 & 7.72 & 7.64 & 7.74 & 7.68 & 7.55 & 7.57 & 7.59 & 7.61 & 7.63 &  d    & $  66.4$ \\
  69 & $a^5{\rm P  _3}$ & $y^7{\rm P^o_2}$ & $4442.840$ & $2.176$ & $-2.79$ &g& $-31.290$ & 7.55 & 7.64 & 7.55 & 7.64 & 7.64 & 7.51 & 7.51 & 7.51 & 7.53 & 7.54 &  bl   & $  64.5$ \\
  71 & $a^5{\rm P  _3}$ & $z^5{\rm S^o_2}$ & $4282.402$ & $2.176$ & $-0.78$ &a& $-31.240$ & 7.17 & 7.23 & 7.18 & 7.26 & 7.33 & 7.15 & 7.15 & 7.15 & 7.28 & 7.29 &  bl   & $ 193.8$ \\
 109 & $a^3{\rm P  _2}$ & $y^5{\rm D^o_1}$ & $6392.543$ & $2.279$ & $-4.03$ &m& $-31.553$ & 7.57 & 7.68 & 7.58 & 7.69 & 7.66 & 7.56 & 7.57 & 7.58 & 7.58 & 7.59 &  g    & $  19.0$ \\
 109 & $a^3{\rm P  _2}$ & $y^5{\rm D^o_3}$ & $6608.030$ & $2.279$ & $-4.03$ &b& $-31.570$ & 7.56 & 7.66 & 7.56 & 7.67 & 7.65 & 7.55 & 7.56 & 7.57 & 7.57 & 7.58 &  g    & $  18.0$ \\
 111 & $a^3{\rm P  _0}$ & $z^3{\rm P^o_1}$ & $6978.850$ & $2.484$ & $-2.48$ &p& $-31.523$ & 7.59 & 7.64 & 7.63 & 7.73 & 7.62 & 7.53 & 7.54 & 7.57 & 7.61 & 7.62 &  af   & $  78.1$ \\
 111 & $a^3{\rm P  _1}$ & $z^3{\rm P^o_0}$ & $6663.450$ & $2.424$ & $-2.45$ &p& $-31.521$ & 7.54 & 7.60 & 7.58 & 7.70 & 7.58 & 7.48 & 7.50 & 7.52 & 7.56 & 7.57 &  afh  & $  80.8$ \\
 111 & $a^3{\rm P  _1}$ & $z^3{\rm P^o_1}$ & $6750.150$ & $2.424$ & $-2.61$ &p& $-31.528$ & 7.60 & 7.67 & 7.63 & 7.73 & 7.63 & 7.53 & 7.54 & 7.56 & 7.60 & 7.61 &  af   & $  76.7$ \\
 111 & $a^3{\rm P  _2}$ & $z^3{\rm P^o_2}$ & $6421.350$ & $2.279$ & $-1.95$ &p& $-31.560$ & 7.45 & 7.48 & 7.49 & 7.60 & 7.52 & 7.38 & 7.43 & 7.44 & 7.52 & 7.52 &  bfh  & $ 110.0$ \\
 113 & $a^3{\rm P  _1}$ & $y^5{\rm P^o_2}$ & $5678.600$ & $2.424$ & $-4.67$ &m& $-31.426$ & 7.51 & 7.61 & 7.51 & 7.61 & 7.65 & 7.51 & 7.52 & 7.53 & 7.53 & 7.54 &  cd   & $   3.2$ \\
 113 & $a^3{\rm P  _2}$ & $y^5{\rm P^o_3}$ & $5436.590$ & $2.279$ & $-2.96$ &a& $-31.451$ & 7.19 & 7.29 & 7.20 & 7.31 & 7.27 & 7.16 & 7.17 & 7.18 & 7.19 & 7.20 &  b    & $  45.6$ \\
 114 & $a^3{\rm P  _1}$ & $y^3{\rm D^o_1}$ & $5141.739$ & $2.424$ & $-1.96$ &p& $-31.350$ & 7.37 & 7.42 & 7.38 & 7.49 & 7.41 & 7.36 & 7.29 & 7.30 & 7.36 & 7.37 &  dh   & $  88.1$ \\
 114 & $a^3{\rm P  _2}$ & $y^3{\rm D^o_2}$ & $4924.769$ & $2.279$ & $-2.24$ &q& $-31.370$ & 7.71 & 7.74 & 7.73 & 7.83 & 7.78 & 7.64 & 7.64 & 7.65 & 7.70 & 7.72 &  df   & $  97.6$ \\
 114 & $a^3{\rm P  _2}$ & $y^3{\rm D^o_3}$ & $5049.819$ & $2.279$ & $-1.33$ &q& $-31.390$ & 7.51 & 7.53 & 7.52 & 7.62 & 7.65 & 7.46 & 7.48 & 7.49 & 7.60 & 7.62 &  df   & $ 164.9$ \\
 115 & $a^3{\rm P  _2}$ & $x^5{\rm D^o_2}$ & $4574.720$ & $2.279$ & $-2.97$ &b& $-31.278$ & 7.65 & 7.77 & 7.68 & 7.78 & 7.73 & 7.60 & 7.61 & 7.62 & 7.65 & 7.67 &  a    & $  59.8$ \\
 116 & $a^3{\rm P  _2}$ & $z^5{\rm S^o_2}$ & $4439.880$ & $2.279$ & $-3.00$ &g& $-31.237$ & 7.52 & 7.62 & 7.53 & 7.63 & 7.61 & 7.48 & 7.49 & 7.50 & 7.51 & 7.52 &  a    & $  52.9$ \\
 152 & $z^7{\rm D^o_1}$ & $e^7{\rm D  _2}$ & $4233.602$ & $2.482$ & $-0.60$ &g& $-30.640$ & 7.41 & 7.46 & 7.39 & 7.46 & 7.55 & 7.37 & 7.37 & 7.37 & 7.51 & 7.51 &  ef   & $ 278.4$ \\
 152 & $z^7{\rm D^o_2}$ & $e^7{\rm D  _3}$ & $4250.119$ & $2.469$ & $-0.41$ &g& $-30.660$ & 7.45 & 7.53 & 7.46 & 7.54 & 7.63 & 7.44 & 7.46 & 7.46 & 7.61 & 7.62 &  ade  & $ 355.4$ \\
 152 & $z^7{\rm D^o_3}$ & $e^7{\rm D  _2}$ & $4187.039$ & $2.449$ & $-0.55$ &g& $-30.640$ & 7.39 & 7.47 & 7.40 & 7.49 & 7.55 & 7.38 & 7.39 & 7.41 & 7.54 & 7.54 &  ade  & $ 297.2$ \\
 152 & $z^7{\rm D^o_3}$ & $e^7{\rm D  _3}$ & $4222.213$ & $2.449$ & $-0.97$ &g& $-30.650$ & 7.41 & 7.51 & 7.44 & 7.53 & 7.56 & 7.41 & 7.42 & 7.42 & 7.56 & 7.56 &  ade  & $ 198.5$ \\
 ... & ... & ... & ... & ... &  ... & & ... & ... & ... & ... & ... & ... & ... & ... & ...& ... & ...& ... & ... \\
\noalign{\smallskip}\hline
\end{tabular}}
%
\begin{footnotesize}
{\bf Sources of $f$-values:} (a) O'Brian et al. (\cite{OWLWB91}), (b) May et al. 
(\cite{MRW74}), (c) Meylan et al. (\cite{MFWK93}), (e) Blackwell et al. 
(\cite{BIPS79}), (f) Blackwell et al. (\cite{BIPW76}), (g) Blackwell et al. 
(\cite{BPSS82a}), (h) Blackwell et al. (\cite{BPSS82b}), (i) Bridges \& 
Kornblith (\cite{BK74}), (j) Garz \& Kock (\cite{GK69}), (k) Wolnik et al. 
(\cite{WBW70}), (l) Richter \& Wulff (\cite{RW70}), (m) Gurtovenko \& Kostik 
(\cite{GK81}), (n) Blackwell et al. (\cite{BPS79}), (o) Blackwell et al. 
(\cite{BPSS80}), (p) Bard et al. (\cite{BKK91}), (q) Bard \& Kock 
(\cite{BK94})\\
{\bf Line synthesis remarks:} (a) no blend, no asymmetry, (b) resolved blend(s), (c) 
continuum adjusted, (d) unresolved blend(s), (e) core asymmetry, (f) blue and red wing 
deficit, (g) only red wing deficit, (h) core too wide, (i) core too narrow, (j) 
core too deep, (k) core too shallow, (l) all faint lines included
\end{footnotesize}
\end{table*}

Iron is the element with probably the greatest number of lines visible in the 
solar spectrum. This is the combined result of a relatively high element 
abundance and of a very complex atomic configuration. In particular for 
\ion{Fe}{I} nearly 10\,000 lines have been identified in the laboratory (Nave et 
al. \cite{NJLTB94}), and possibly hundreds of thousands more are too weak to be 
detected. However, for only a small subset of these lines accurate $f$-values 
are known; most of them are laboratory data while only a subset has been derived 
from the solar spectrum itself. Our ability to identify the lines with 
laboratory $f$-values in the solar spectrum and calculate their solar 
\ion{Fe}{I} abundances is therefore strongly influenced by the accuracy of the 
data, and it is this dependence that makes an analysis of the complete solar 
iron spectrum next to impossible as we will demonstrate below. 

The term ''weak line'' refers to all line strengths that had not been considered 
in Paper I, and it does not necessarily indicate a particularly small line 
strength. Thus, all lines in the list of Nave et al. have been examined if an 
$f$-value was available. Among them were only $\sim 500$ lines with equivalent 
widths below 100 m\AA\ that were not too strongly blended by other lines. Some 
of the lines retained in our sample are still blended but are either 
well-resolved or at least permit the analysis of one line wing. From this list 
we had to exclude lines in spectral regions that in the solar spectrum were 
overly affected by weak line haze and continuum uncertainties. These lie in the 
blue-green ($4400 \ldots 4800$ \AA) and in the yellow ($5500 \ldots 5900$ \AA). 
The source of these spectral impurities is unknown although part of the blue 
could well be contaminated by a complicated pattern of \ion{Fe}{I} 
autoionization transitions. Bautista's (\cite{B97}) calculations show that they 
are there, but the accuracy of their wavelength positions is probably not very 
high. The total number of \ion{Fe}{I} lines including weak and strong lines was 
therefore reduced to 410, and during subsequent NLTE analyses their number once 
again shrank to the final value of 391 lines. 

One of the more surprising results of this evaluation of the solar \ion{Fe}{I} 
spectrum is that the number of truly \emph{weak} lines with both an acceptable 
spectral environment and laboratory $f$-value is so small. This is the case for 
lines in a range of solar equivalent widths from 3 to 30 m\AA. This has also 
been noticed among others by Rutten \& van der Zalm (\cite{RZ84}). If laboratory 
analyses were extended into the near infrared the line list could be greatly 
extended because of decreasing blend problems. The blue and near-ultraviolet  
spectral regions have been ignored here because of the problems localizing the  
continuum below 4200 \AA. 

\subsection{Spectrum synthesis}

The final set of lines is reproduced in Table \ref{lintab} together with all 
relevant data. The sources of the $f$-values as well as the remarks in the 
second last column are noted at the end of the table. The damping constants are 
calculated according to Anstee \& O'Mara's (\cite{AO91},\cite{AO95}) theory as 
in Paper I, and they are given here in terms of van der Waals damping constants. 
The equivalent widths in the last column are integrated on the basis of the best 
synthetic fit of the solar \emph{flux} profile. We emphasize that they are not 
used for the line analysis which is solely based on profile fits. Rather, they 
are derived from the theoretical profile \emph{after} the final profile fitting 
procedure. Their accuracy is low, which is uncritical since they are used for 
graphical purposes only. 

\subsubsection{Oscillator strengths}

In order to determine abundance \emph{ratios} in spectral lines of stars other 
than the Sun it is often sufficient to know the product $gf\varepsilon_\odot$, 
which can be obtained in the solar flux spectrum with no particular knowledge of 
the $f$-value. Were it not for consistency and identification checks and for the 
determination of the solar iron abundance itself, no oscillator strengths would 
be needed. Such consistency checks include the specification of broadening 
parameters such as microturbulence and damping constants, because both can to a 
certain degree replace abundances or oscillator strengths. Therefore a critical 
analysis of the $f$-values is necessary. As mentioned above, oscillator 
strengths available for \ion{Fe}{I} lines come from essentially three different 
methods: 
\begin{itemize}
\item Theory has made important progress in the last 20 years. This is not 
only seen in the \emph{bf} cross-sections we used in our kinetic equilibrium 
calculations but also in a virtually complete set of calculated $f$-values made 
available by Kurucz (\cite{K92}). The main obstacle in using these data lies in 
the problem of estimating their accuracy. Therefore we have used Kurucz' data 
for their original \emph{statistical} purpose computing particle interaction via 
the statistical equilibrium equations whenever laboratory $f$-values were 
missing. But we have not applied his $f$-values during subsequent spectrum 
synthesis. 
\item Laboratory methods have made some progress, too, and the number of 
laboratory $f$-values is steadily increasing. It is this origin we have put most 
confidence in, although the reliability of the various sources as judged from 
their ability to fit the solar flux spectrum is surprisingly different as we 
will show below. 
\item The inversion method, i.e. measuring \emph{solar} $f$-values by synthesis 
of solar equivalent widths, has become a popular method to fill the missing data 
gap in the \ion{Fe}{I} line list. Whether based on equivalent widths or line 
profiles, this method always reproduces an assumed abundance scale. This is -- 
in most cases -- the meteoritic \ion{Fe}{I} abundance, sometimes it is tied to 
some otherwise established solar iron abundance such as that of the Oxford group 
(cf. Gurtovenko \& Kostik \cite{GK81}). It \emph{never} carries information 
about the oscillator strength itself. 
\end{itemize}
Of these three methods we have applied only the results of the last two methods 
to spectrum synthesis, and in solar abundance determinations we confine our 
sample to those lines for which laboratory $f$-values are available. 

\subsubsection{Line broadening} 
 
During the analysis of the strong \ion{Fe}{I} lines we have discussed collision 
broadening at some length in Paper I, where it was documented that the results 
of the broadening theory of Anstee \& O'Mara (\cite{AO91},\cite{AO95}) provided 
the necessary adjustment between weak and strong line abundances at least in a 
qualitative way. We have followed this approach in the present investigation, 
replaced the old van der Waals damping constants by the new collision 
parameters, however, staying aware of the dependence of abundance analyses upon 
atmospheric models. Thus, the empirical model of Holweger \& M\"uller 
(\cite{HM74}) requires significantly \emph{higher} damping constants than our 
theoretical model in order to fit solar strong line profiles with the same 
abundance as the weak lines. In fact, we have added to our sample of NLTE models 
two more items with \emph{reduced} damping constants in order to explore their 
influence on the mean \ion{Fe}{I} abundance. We come back to this point in 
subsection 3.2.2. 

The introduction of weak lines, among them many lines broadened by 
microturbulence, has considerably enhanced our possibility to judge the solar 
line spectrum and the necessary atomic data. So the present analysis required an 
extension of the parameter space covered by non-thermal motions to put both weak 
and turbulence lines on a common abundance level. In fact, irrespective of the 
source of $f$-values, lines between 50 and 120 m\AA\ tend to require 
systematically higher abundances than weak or very strong lines if the value of 
Paper I, $\xi = 0.85$ \kms\ was used. We introduced a second mean value of $\xi 
= 1.00$ \kms\ which seems more appropriate for our present investigation. Note 
that this value has only limited influence on the strong lines, so our former 
results stay essentially unchanged. 

\begin{figure}
\resizebox{\hsize}{!}{\includegraphics{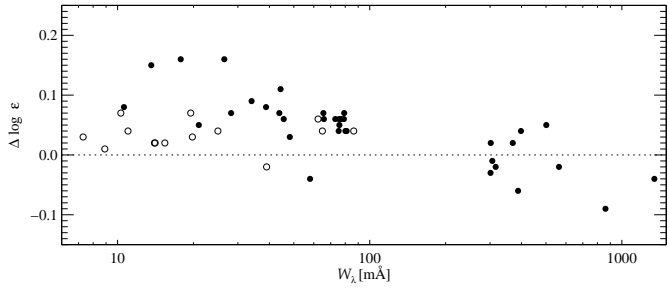}} \caption[]{Abundance 
differences between lines synthesized in our plane-parallel LTE (TH) model and 
those obtained from a hydrodynamical solar model of Asplund et al. 
(\cite{ANTS00}). Lines that were synthesized in our plane-parallel model with 
continuum adjustment are drawn as open circles} \label{asplund} 
\end{figure}

As will be shown in subsection \ref{profiles}, the details of turbulent line 
broadening are still unsatisfactory for a number of medium-strong lines. Whereas 
all weak lines with equivalent widths below $W_\lambda \sim 70$ m\AA\ and most 
of the very strong lines are well represented by the synthetic line profiles, 
some lines around $W_\lambda \sim 70 \ldots 120$ m\AA\ are not reproduced by any 
choice of model parameters. This was noticed already in Paper I when trying to 
fit \ion{Fe}{II} multiplet 42 or \ion{Fe}{I} multiplets 1 or 36. The present 
selection of \ion{Fe}{I} lines includes quite a lot of such lines that seem to 
document the ultimate difference between plane-parallel and hydrodynamical 
models. Following this difference it is interesting to compare the results of 
the two completely different model realizations of non-thermal motions. 
Therefore the results of Asplund et al. (\cite{ANTS00}) have been confronted 
with our data in Fig. \ref{asplund}.  

It is true that the mean abundance of the 49 lines in common is different by 
$\Delta\log\varepsilon \simeq 0.05$ (or even slightly more for turbulence 
lines), and this could be interpreted as the difference between plane-parallel 
and hydrodynamical models. But a closer view reveals that most of the weaker 
lines belong to a category that requires some continuum adjustment with respect 
to the solar flux atlas of Kurucz et al. (\cite{KFBT84}). There are some 
spectral regions that suffer from unknown continuum depressions, and whenever 
such an adjustment was used in \emph{our} calculations, the abundance 
differences between our respective models shrank to a mean 
$\Delta\log\varepsilon \simeq 0.03$, more probably near the true difference 
between the models. It is interesting in this respect that the bulk of 
turbulence line abundances between 60 and 90 m\AA\ is \emph{systematically} 
higher than those calculated from the hydrodynamical model. This is also found 
in our own data when strong lines and turbulence lines are compared, and it 
would mean that exactly this type of lines is not particularly well synthesized 
by plane-parallel models. 

We emphasize, however, that a single value for the \emph{microturbulence} 
velocity cannot be assumed to reproduce all types of core saturation found in 
turbulence lines. Our simple approximation is inconsistent in that it ignores 
the corresponding variations found and accepted for the \emph{macroturbulence} 
velocity, and a free fit of the $\xi$ parameter for each line profile would have 
produced slightly improved results. Comparison with Asplund et al. 
(\cite{ANTS00}) finally shows that both weak and strong lines are not strongly 
affected by dynamic processes, which means that the conventional replacement of 
laminar flow patterns by a micro-/macroturbulence approach is still surprisingly 
valid. 

\subsubsection{Line profiles and equivalent widths}
\label{profiles} 

\begin{figure}
\resizebox{\columnwidth}{!}{\includegraphics{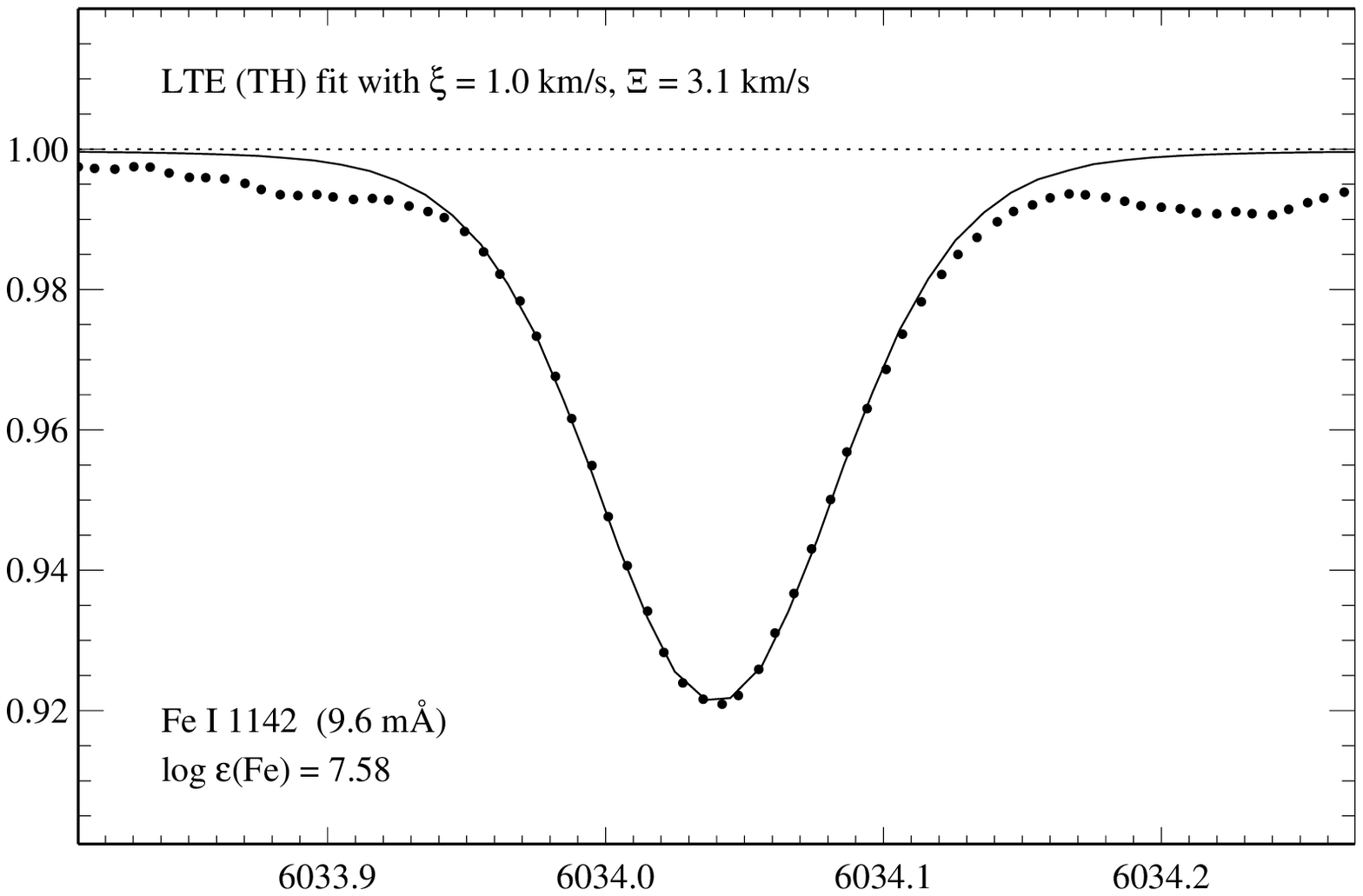}} \vspace{-3mm} 
\resizebox{\columnwidth}{!}{\includegraphics{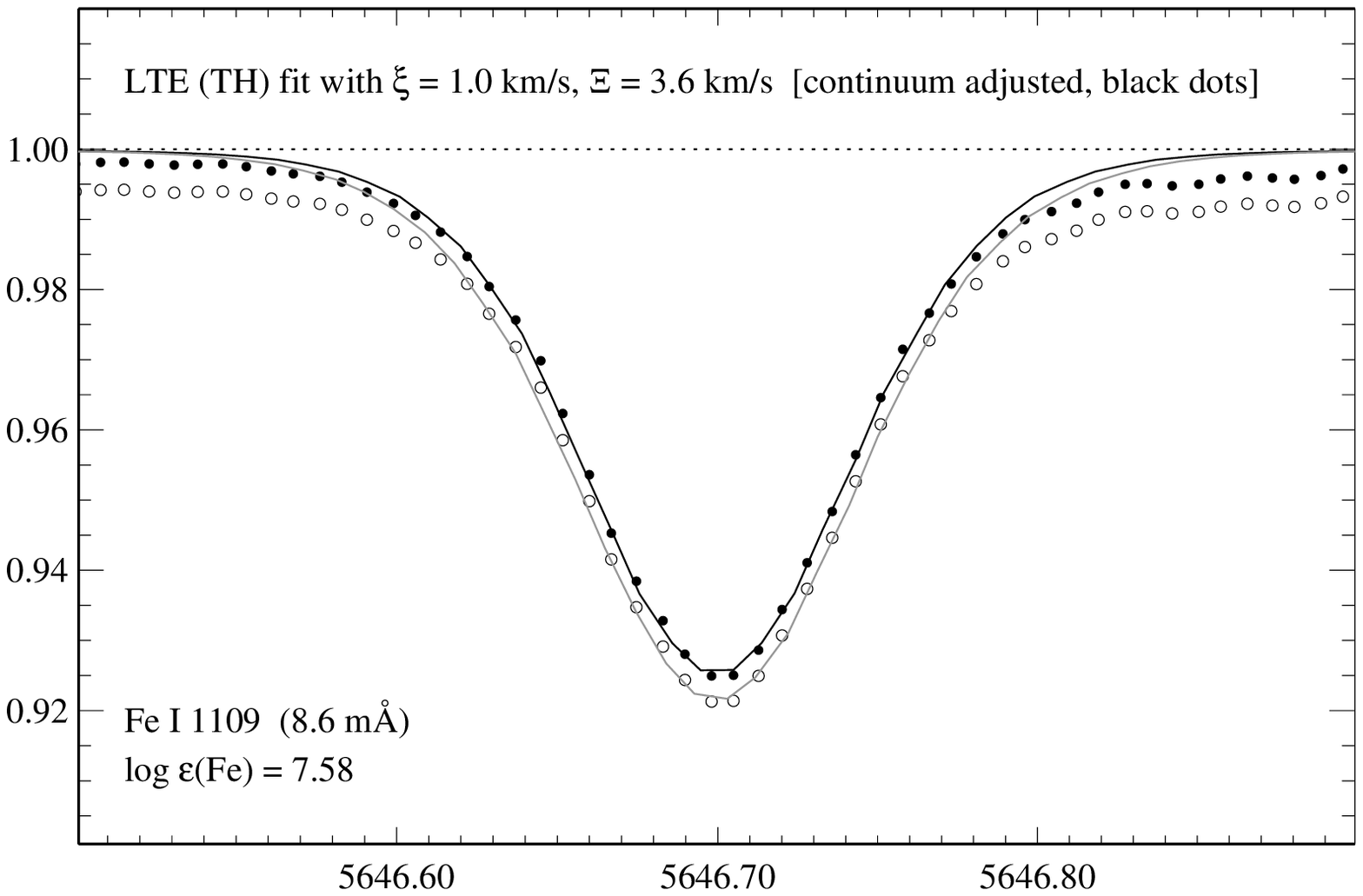}} \vspace{0mm} 
\caption[]{Typical problems with the adjustment of the local solar flux atlas 
continuum. \emph{Top:} Weak line in Mult 1142 with no changes of the local 
continuum necessary. \emph{Bottom:} Mult 1109 shown with and without continuum 
adjustment. The original atlas spectrum is reproduced with open circles and 
fitted by the grey curve with $\eps{Fe} = 7.63$ and $\Xi = 4.0$ \kms} 
\label{adjcont} 
\end{figure}

The overwhelming majority of publications is devoted to the investigation of 
equivalent widths which is mostly due to the easy access to such data in the 
literature. The critical examination of line \emph{profiles} instead makes 
available an increased amount of information about line formation and stellar 
atmospheric conditions. Our present work on NLTE effects in \ion{Fe}{I} lines is 
based on roughly 4000 line profiles, and their evaluation is coded in a very 
coarse set of remarks in Table \ref{lintab}. Such remarks combine the average 
profile properties of all models for a particular line, and the following 
description will show only typical properties. \\[1mm] 
\noindent 
{\bf Very weak lines} ($W_\lambda < 10$ m\AA):\\[1mm] 
Only 10\% of the total sample consist of very weak lines. Most of them could be  
selected to be free from known blends, but only 10 of them were unaffected by 
problems with continuum adjustment. It is this latter quality that makes the 
analysis of very weak lines so ambiguous. This can be seen in Fig. \ref{adjcont} 
where the LTE profile fits for two lines are shown. Continuum adjustment is by 
far not always as small as 0.5\% as it is for the line in Mult 1109, and 
ignoring it may lead to abundances higher by up to 0.15 dex in single cases. 

It is no straightforward procedure to decide which lines to submit to continuum 
adjustment, because this requires a look at the whole spectral region. 
Consequently, we have adjusted the atlas continuum only if there is a continuum 
depression over at least 10 \AA. In some cases we tried to synthesize faint 
background lines in order to estimate their influence on the continuum position. 
While weak lines should be least affected by broadening and therefore yield most 
reliable abundances, the continuum placement destroys a substantial part of this 
argumentation.\\[1mm] 
\begin{figure*}
\begin{center}
\hbox{\resizebox{\columnwidth}{!}{\includegraphics{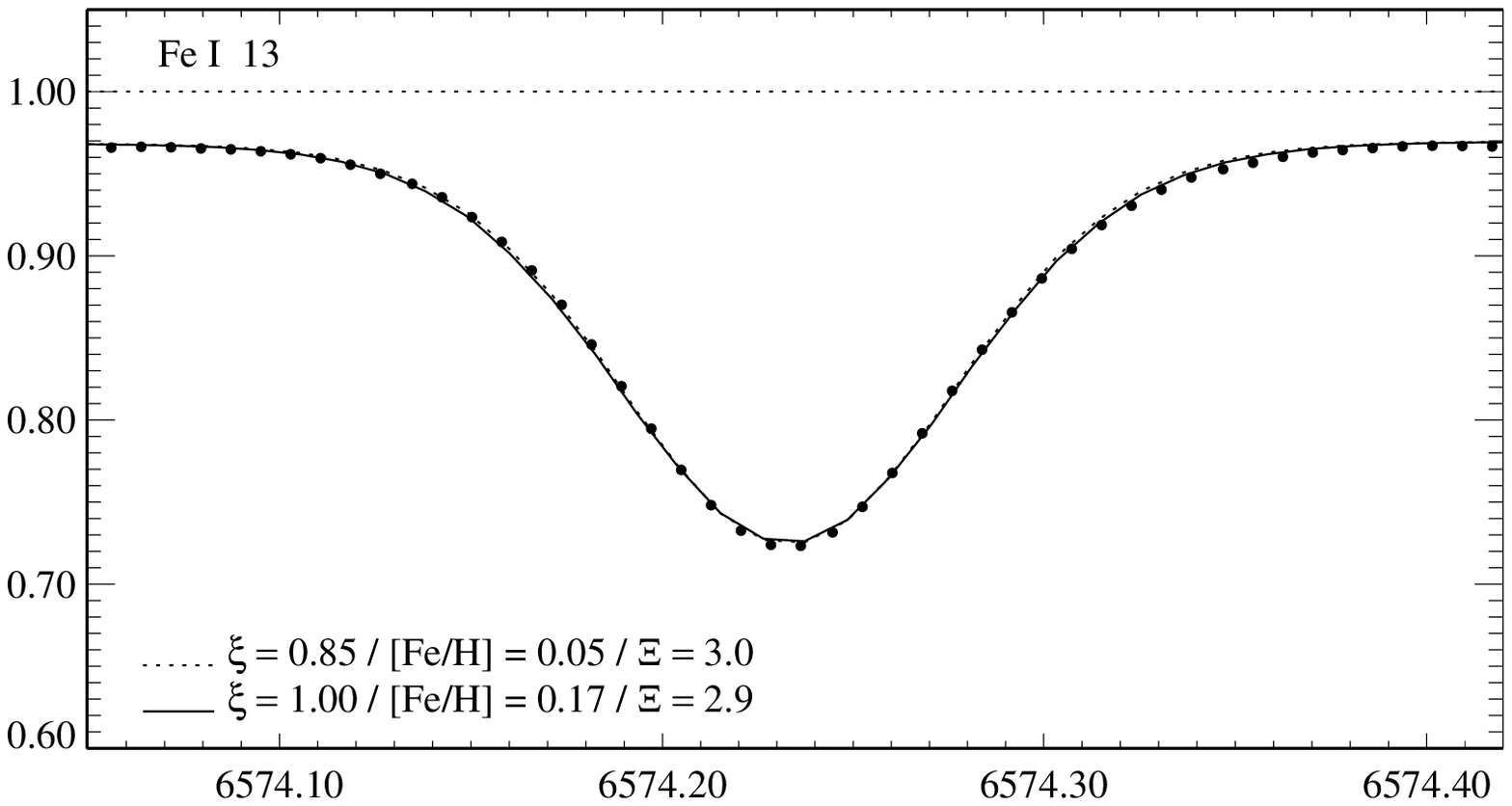}}\hfill 
  \resizebox{\columnwidth}{!}{\includegraphics{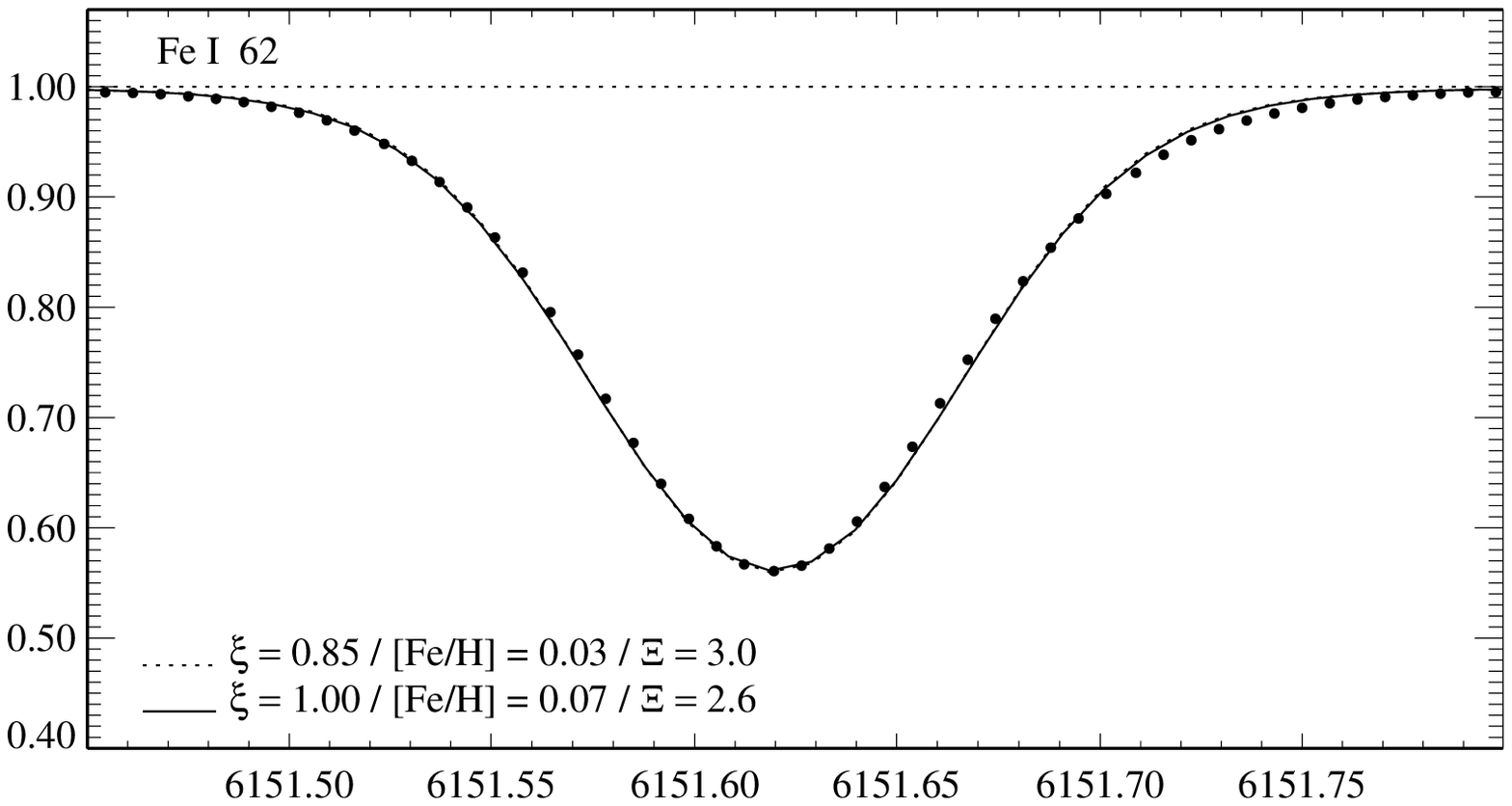}}}
\vspace{0mm} 
\hbox{\resizebox{\columnwidth}{!}{\includegraphics{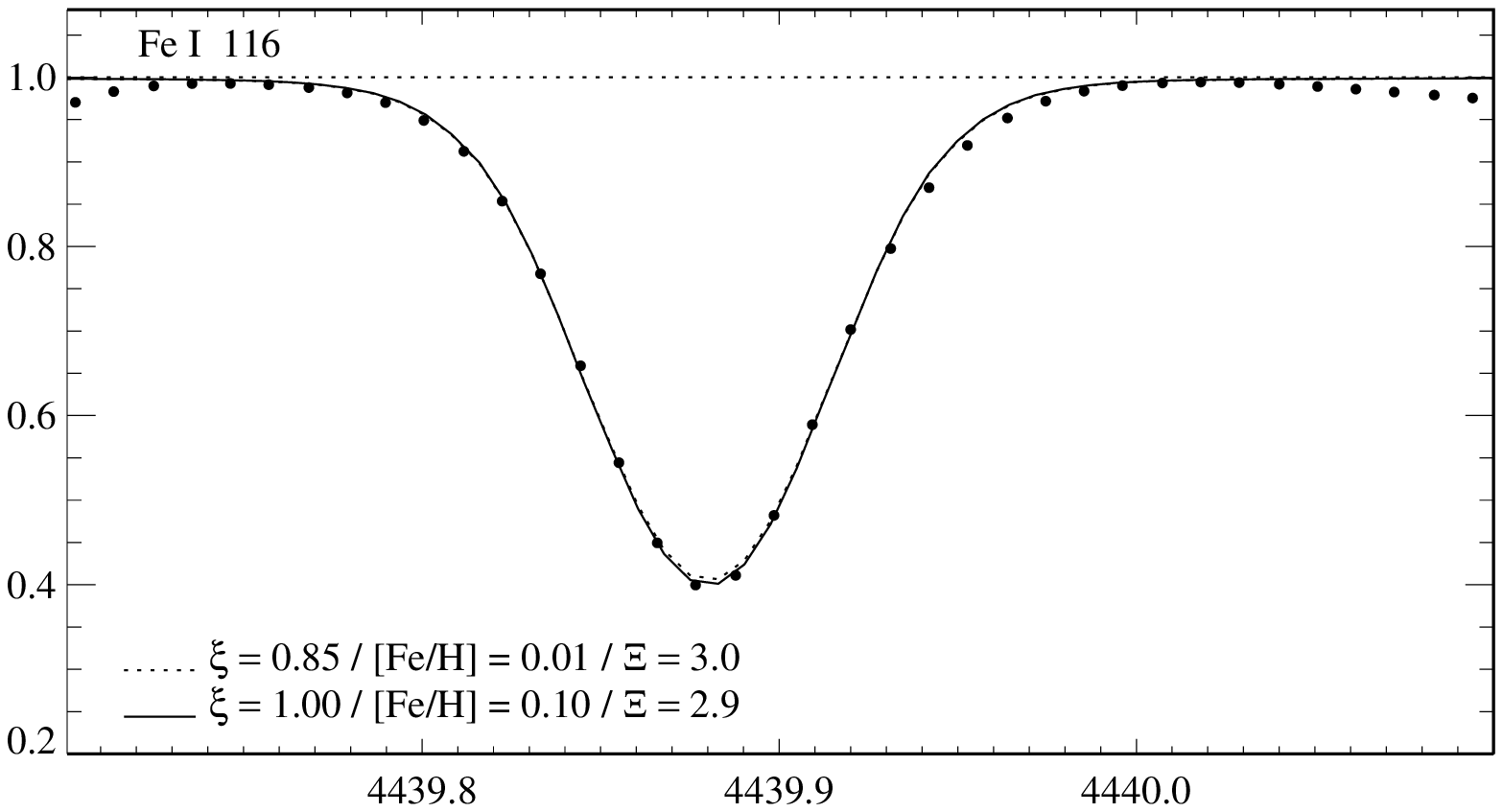}}\hfill 
  \resizebox{\columnwidth}{!}{\includegraphics{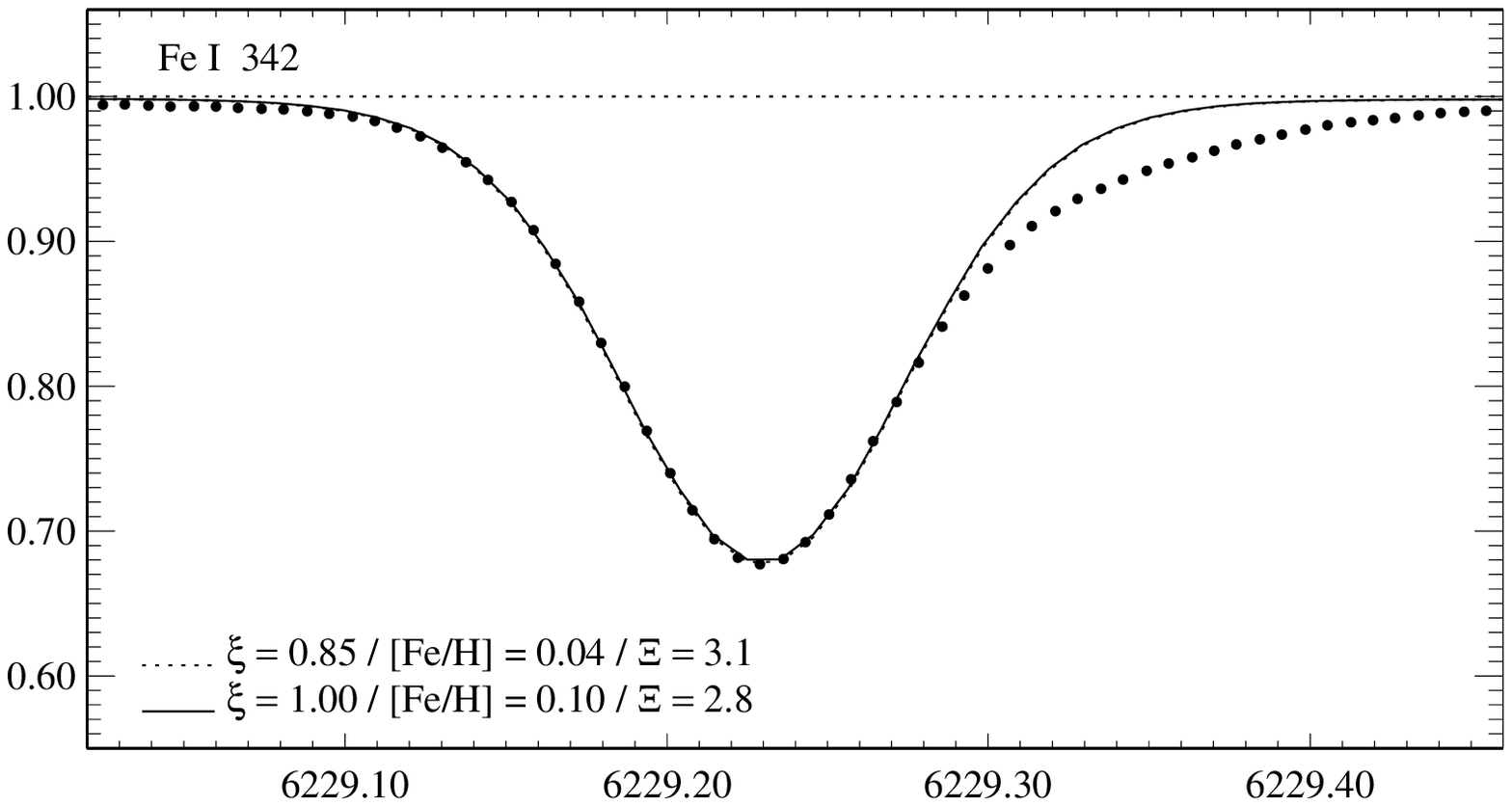}}}
\vspace{0mm} 
\hbox{\resizebox{\columnwidth}{!}{\includegraphics{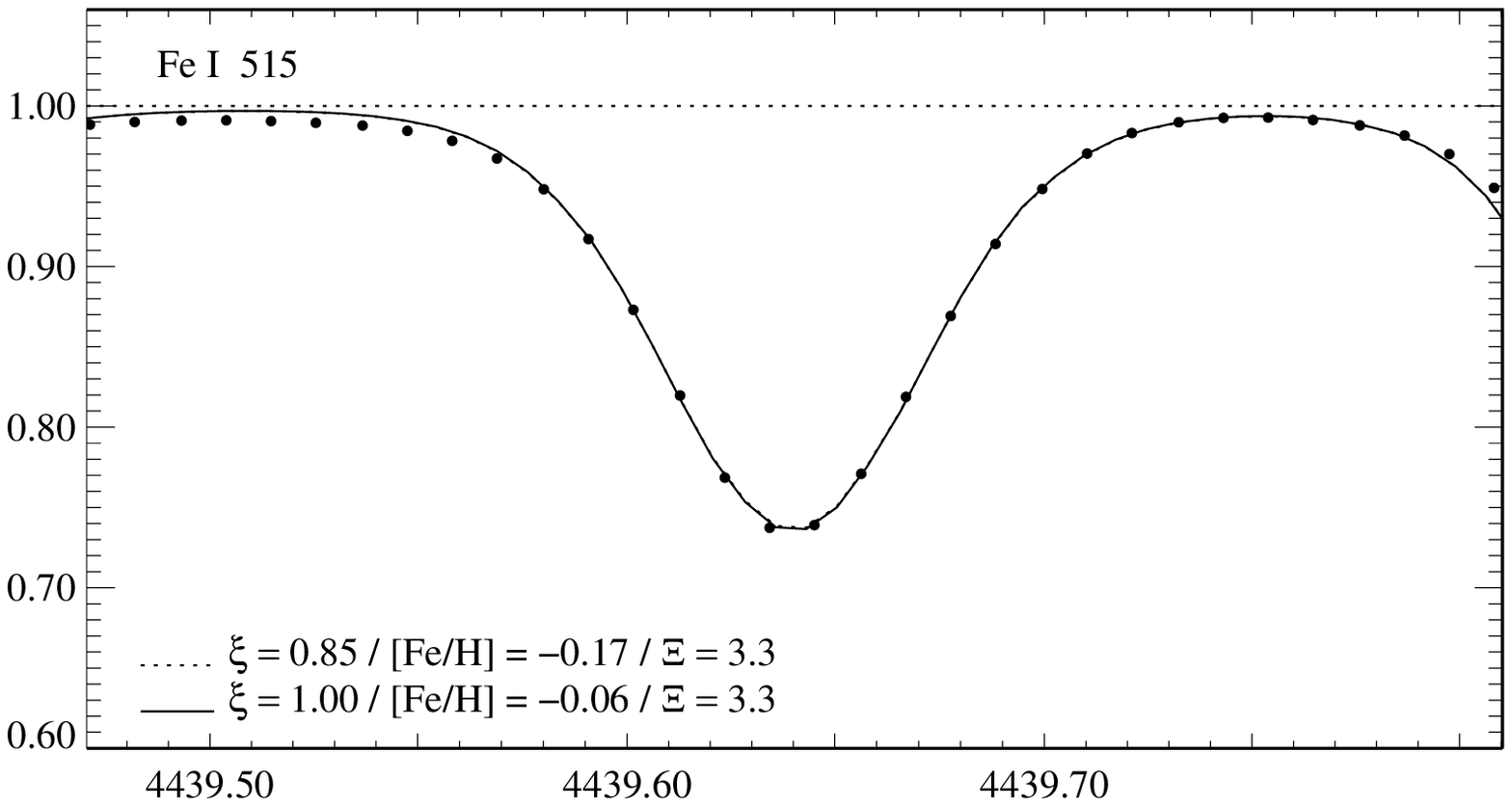}}\hfill 
  \resizebox{\columnwidth}{!}{\includegraphics{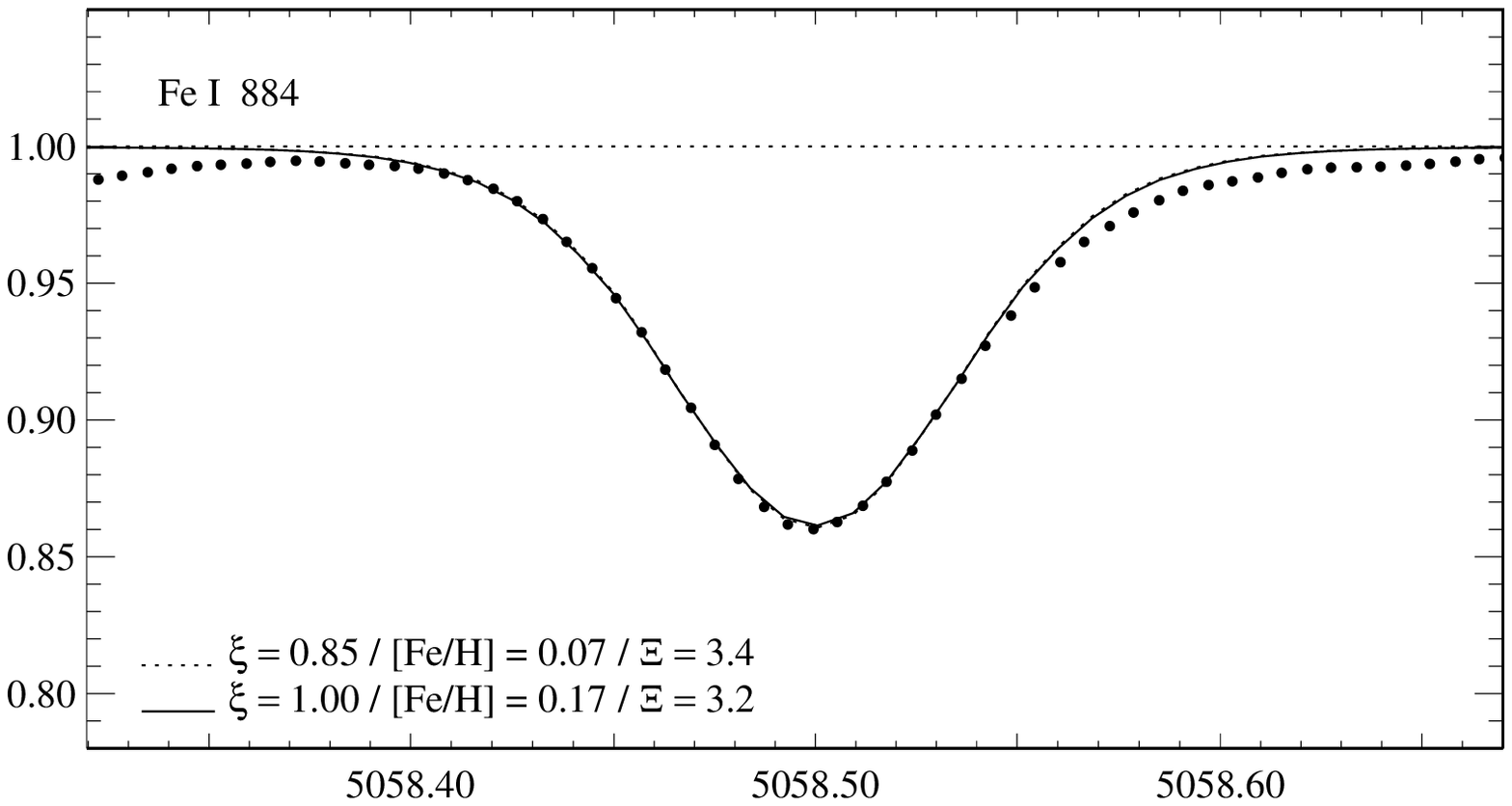}}}
\vspace{0mm} 
\hbox{\resizebox{\columnwidth}{!}{\includegraphics{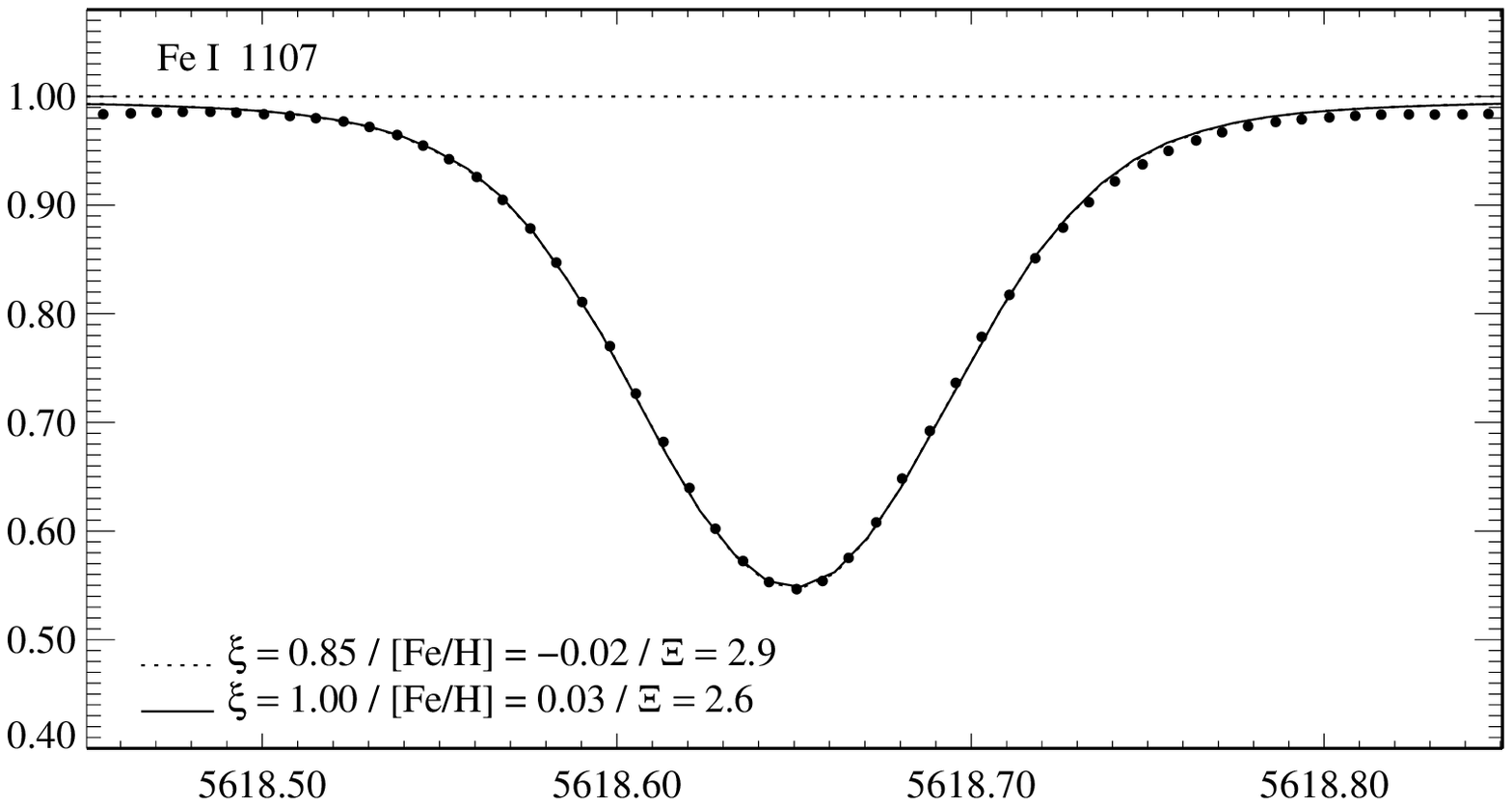}}\hfill 
  \resizebox{\columnwidth}{!}{\includegraphics{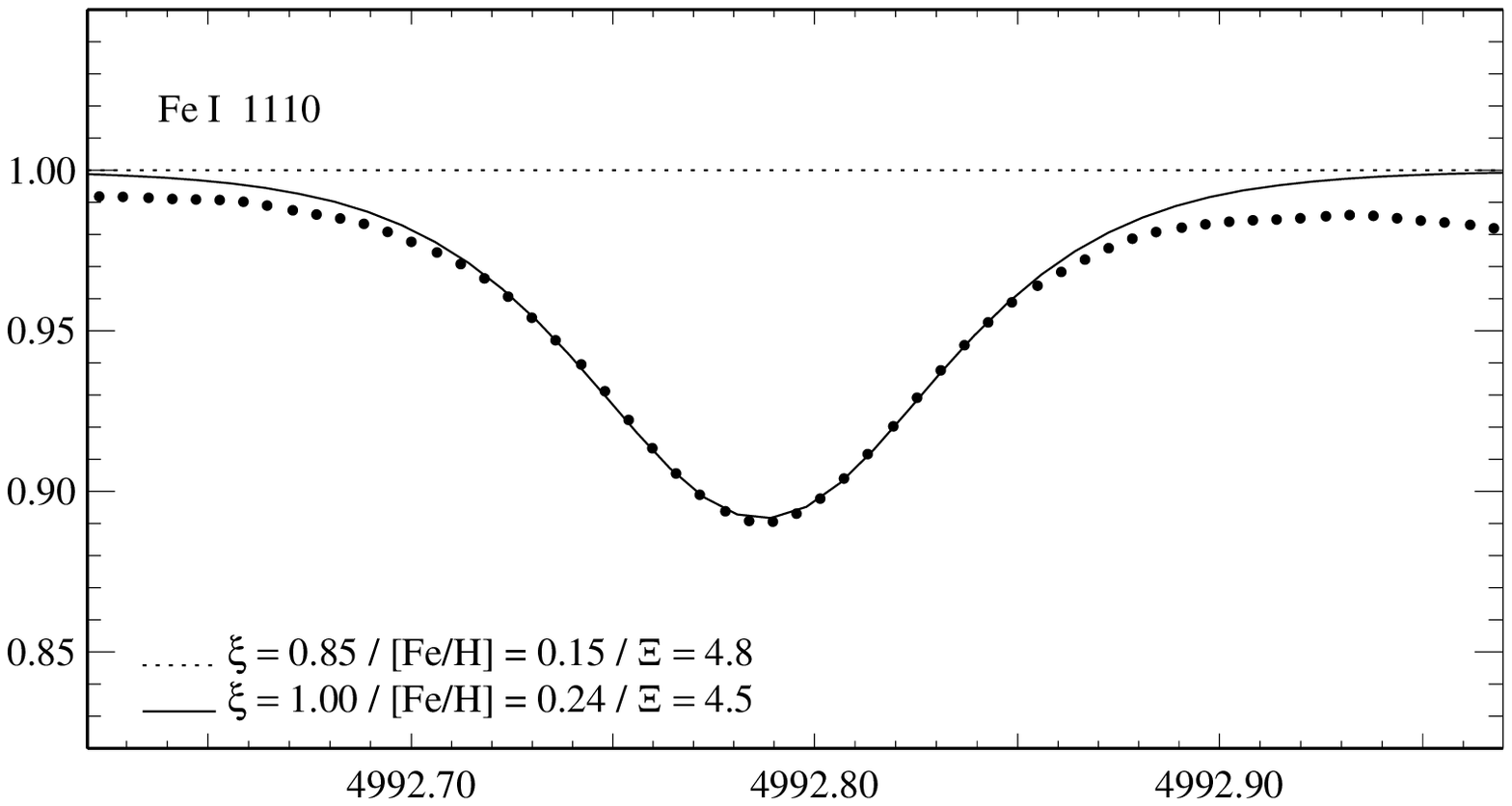}}}
\vspace{0mm} 
\hbox{\resizebox{\columnwidth}{!}{\includegraphics{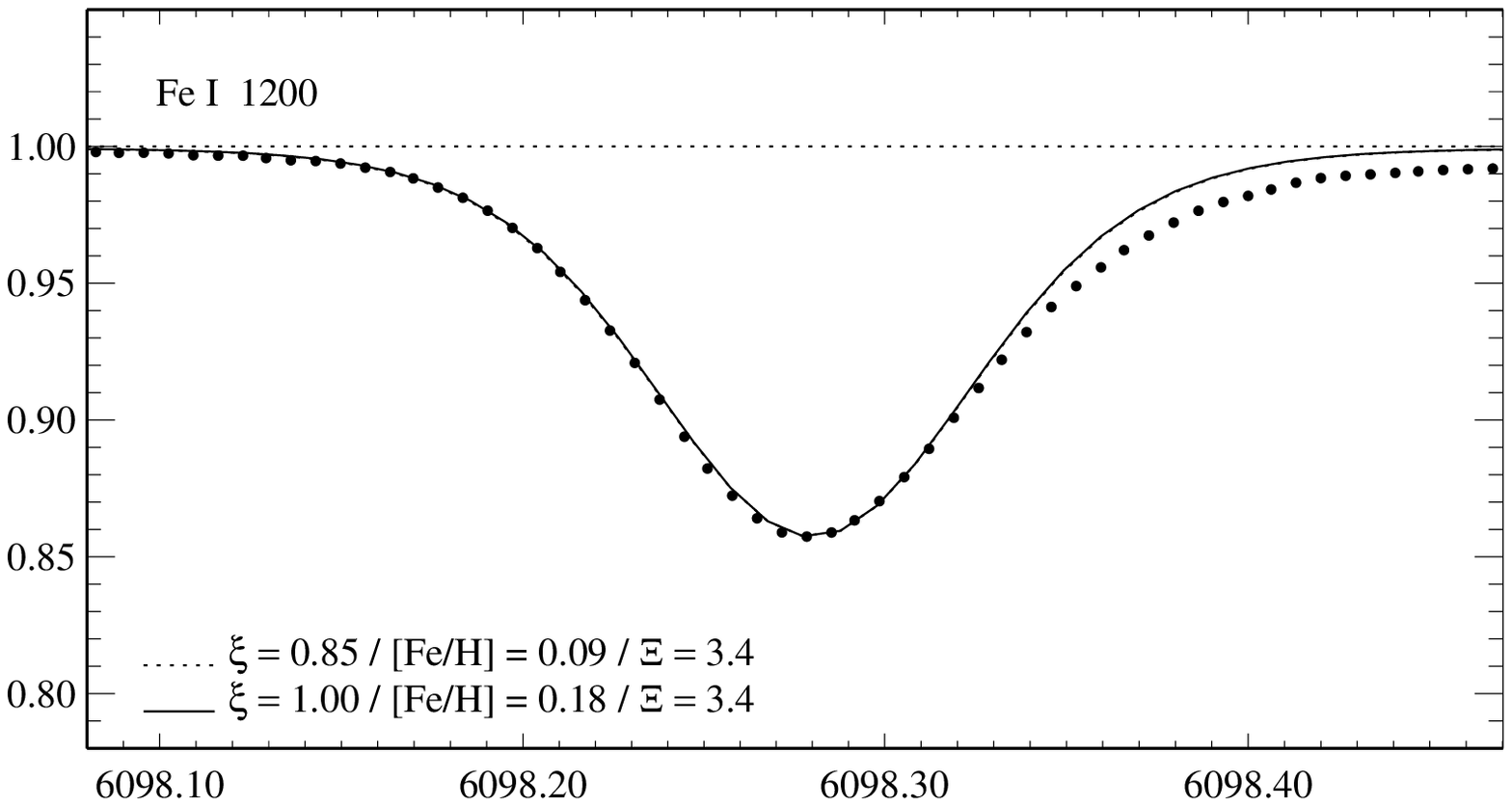}}\hfill 
  \resizebox{\columnwidth}{!}{\includegraphics{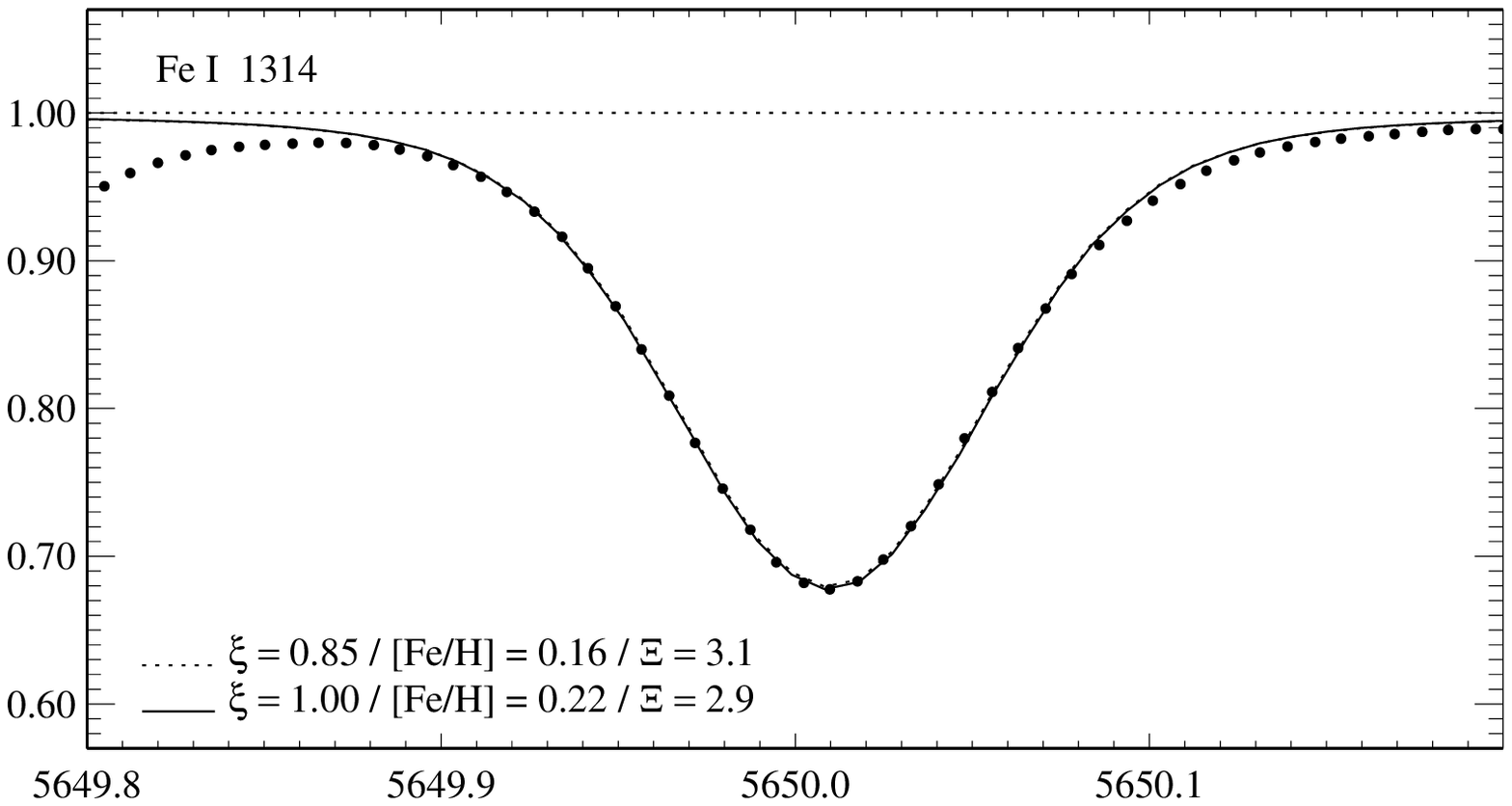}}}
\vspace{0mm} \caption[]{Profiles of weak lines ($10 < W_\lambda < 60$ m\AA) of 
\FeI\ in the solar flux spectrum (filled circles). Synthetic profile fits are for 
LTE and HM (~\makebox[0.7cm]{\hrulefill}~) or TH (~\makebox[0.7cm]{\dotfill}~) 
atmospheres. Fit parameters are indicated} \label{weaklines} 
\end{center}
\end{figure*}
\noindent {\bf Weak lines} ($10 < W_\lambda < 60$ m\AA):\\[1mm] These lines 
constitute the majority of the sample with more than half in this range of 
equivalent widths. Up to 30 m\AA\ the lines do not depend significantly upon 
microturbulence, but their abundance change increases to $-0.03$ per 0.1 \kms\ 
at 60 m\AA. A number of weak lines that are fairly representative of our sample 
is reproduced in Fig. \ref{weaklines}, together with LTE profile fits for both 
the HM and TH models. They are shown in particular to demonstrate the abundance 
\emph{differences} between the two models. It should be mentioned here that this 
subsample of \FeI\ lines produces by far the best profile fits, followed by the 
strong lines, the very weak lines, and the turbulence lines, in order of 
decreasing fit quality. The profiles of the weak lines are not dictated by core 
saturation or line wing broadening but, nearly exclusively, by external line 
broadening due to solar rotation and macroturbulence. As is the case for some of 
the very weak lines, some weaker lines in Fig. \ref{weaklines} require a high 
macroturbulence of $\Xi 
> 4$ \kms\ in order to adjust the wings. 

We note that the \emph{quality} of the profile fit is the same for both 
atmospheric models, irrespective of the abundance differences. Thus most of the 
very weak and weak lines show a systematic abundance difference of 
$\Delta(\eps{Fe,\odot})_{\rm HM-TH} = 0.06 \ldots 0.12$ (see below). As with the 
very weak lines, there is also no problem when fitting the profiles of the weak 
lines with different NLTE models (not shown in Fig. \ref{weaklines}). However, 
the kinematic properties of all lines with equivalent widths below 100 m\AA\ are 
reproduced in a number of profiles that show systematic bisector curvature and a 
\emph{red line wing deficit}. An even more critical inspection of some of the 
profiles reveals synthetic line cores that tend to be too broad even for $\xi = 
0.85$ (TH) or $1.00$ (HM) \kms, respectively. This is evident in particular for 
lines that are formed further up in the atmosphere, and -- together with the red 
wing asymmetries -- it clearly documents the pitfalls of static atmospheric 
models. Some of the weak lines are also affected by a bad definition of the 
local continuum, which either lead to a removal of a significant number of lines 
originally selected or ended in a multi-line synthesis with a number of faint 
background lines included. Such results are not given too much weight in the 
abundance analysis.\\[1mm] 
\begin{figure}
\resizebox{\hsize}{!}{\includegraphics{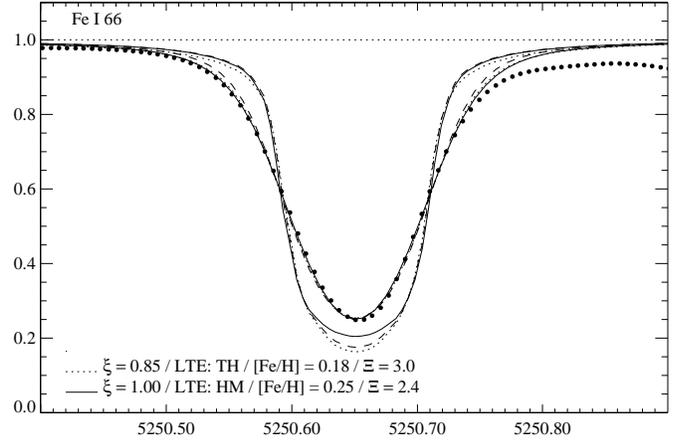}} \caption[]{LTE profiles of 
\FeI\ 66, 5250.646 \AA. Models are as in Figs. \ref{weaklines} and 
\ref{turblines}. Additionally, a TH LTE model with $\xi = 1.0$ \kms\ is plotted 
for comparison (dashes). The deep profiles are uncorrected for rotation and 
macroturbulence, the ''v''-shaped profiles include external broadening} 
\label{macro} 
\end{figure}
\begin{figure*}
\begin{center}
\hbox{\resizebox{\columnwidth}{!}{\includegraphics{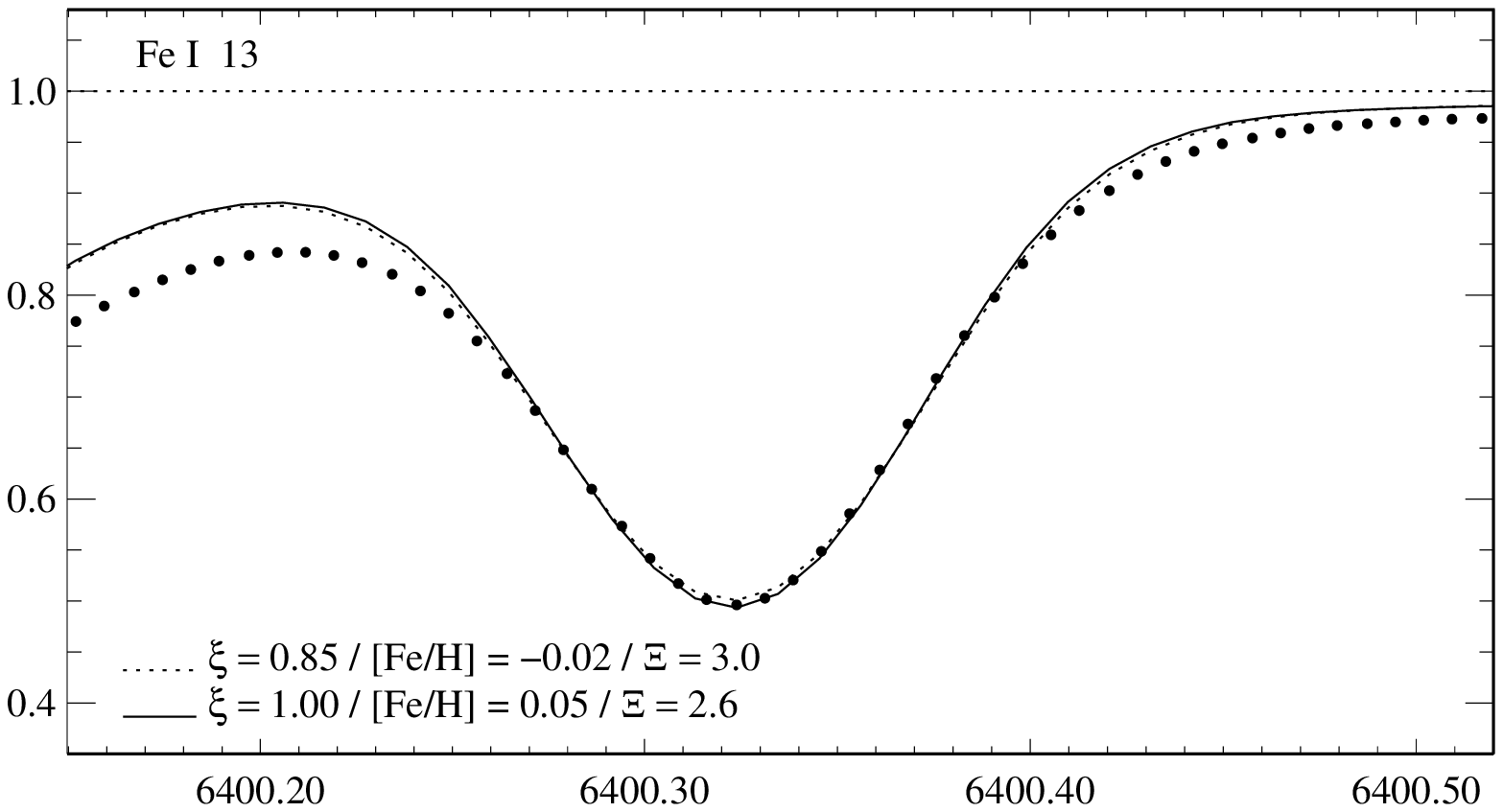}}\hfill 
  \resizebox{\columnwidth}{!}{\includegraphics{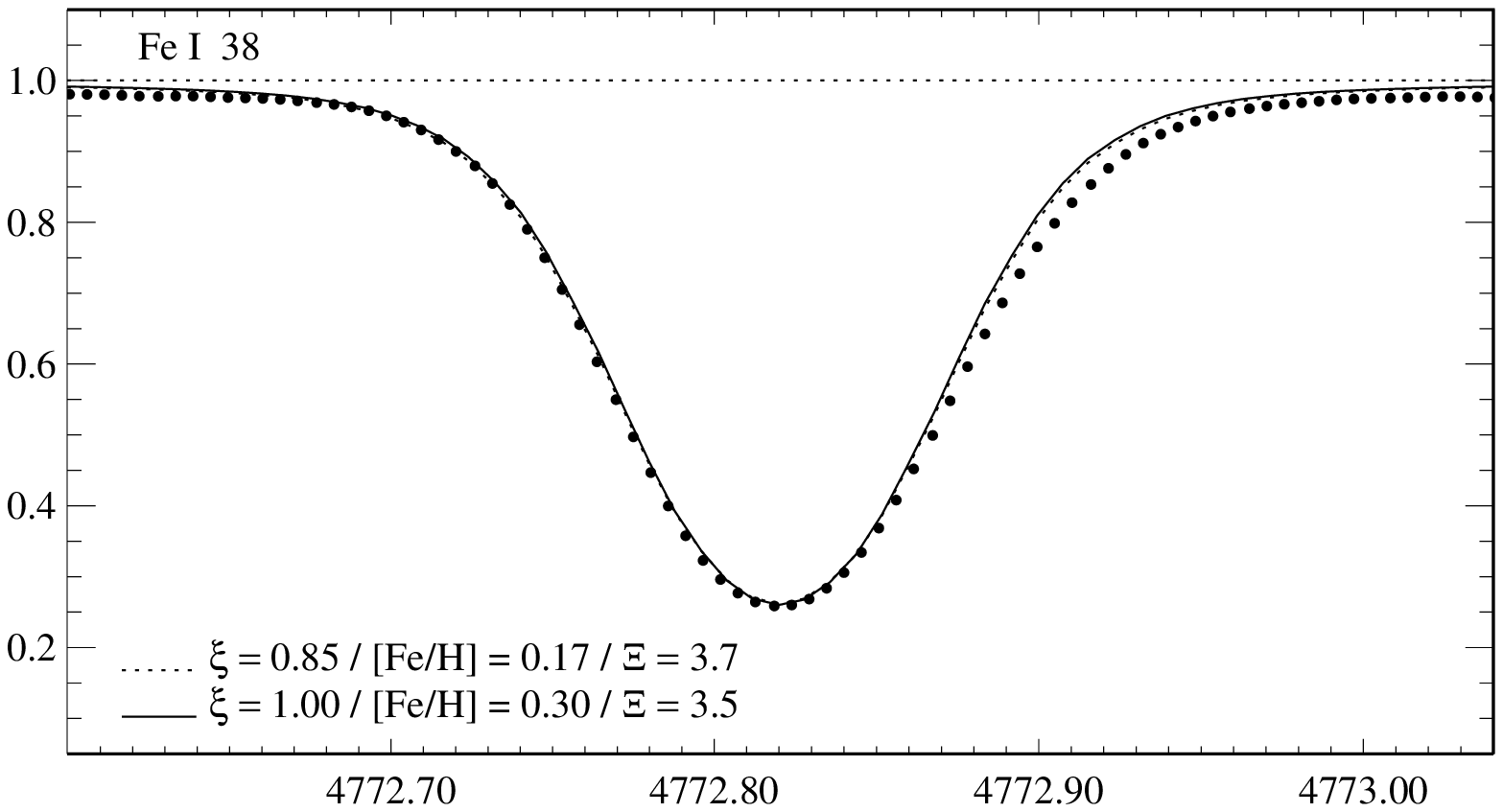}}}
\vspace{0mm} 
\hbox{\resizebox{\columnwidth}{!}{\includegraphics{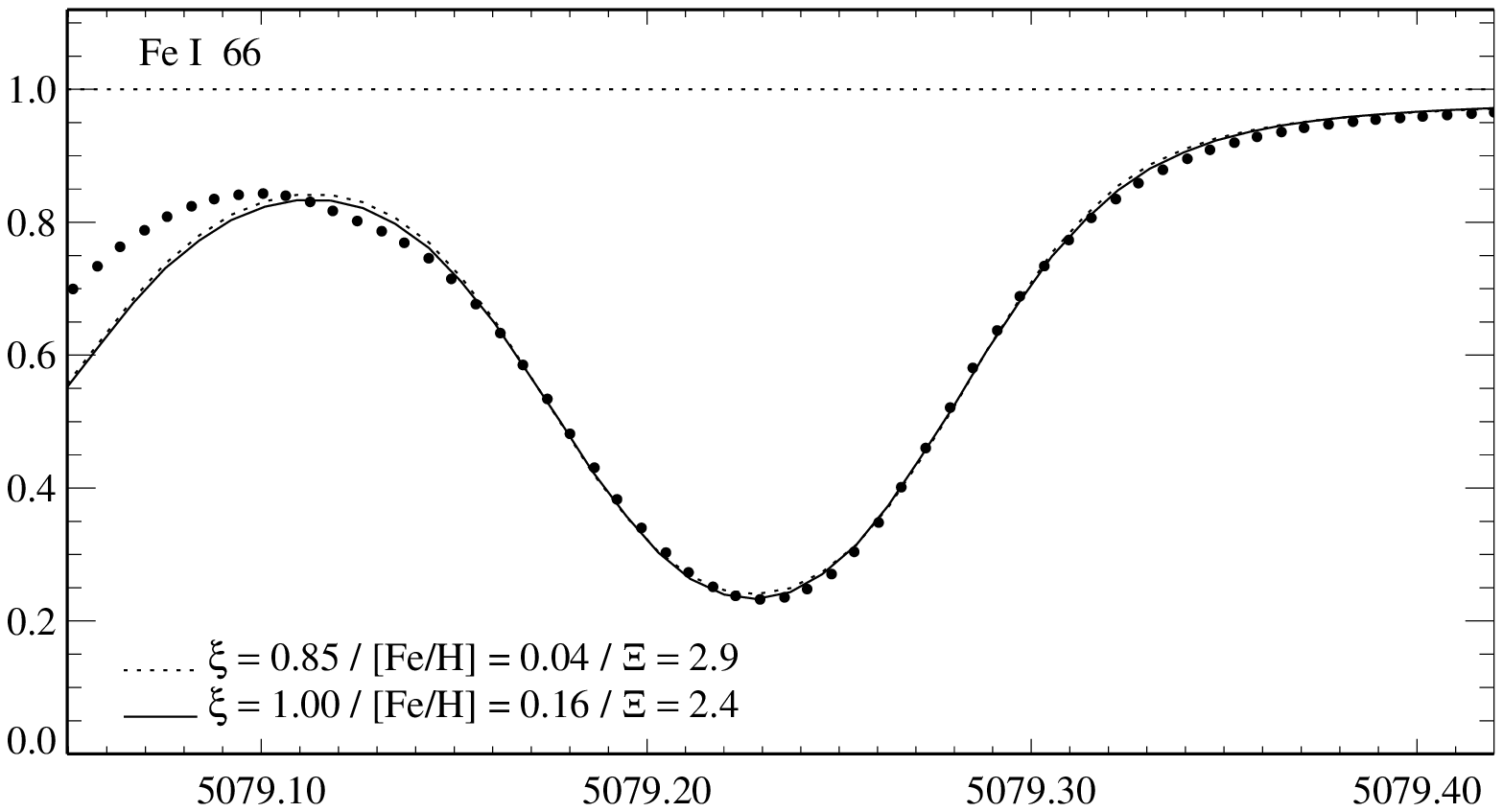}}\hfill 
  \resizebox{\columnwidth}{!}{\includegraphics{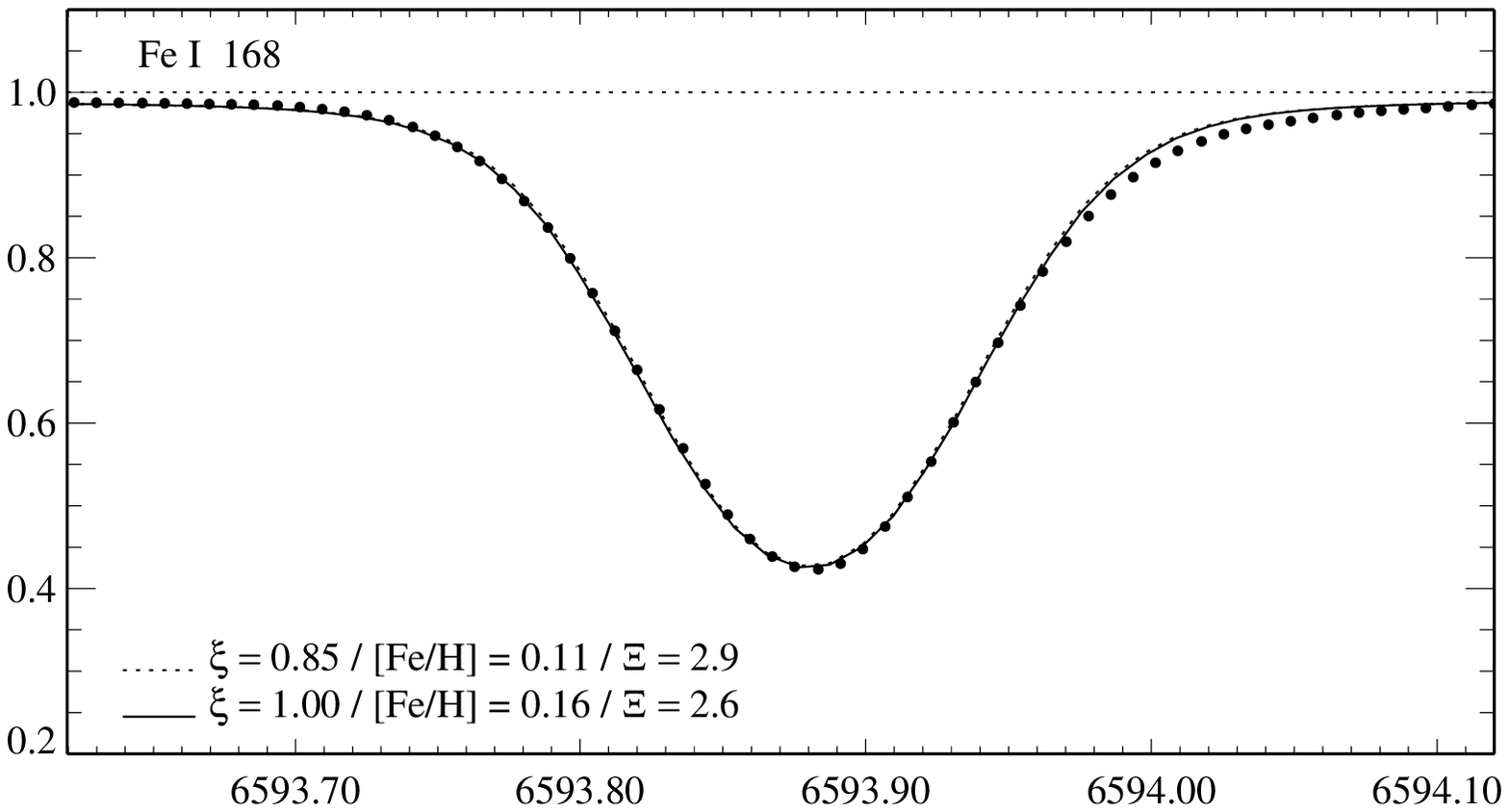}}}
\vspace{0mm} 
\hbox{\resizebox{\columnwidth}{!}{\includegraphics{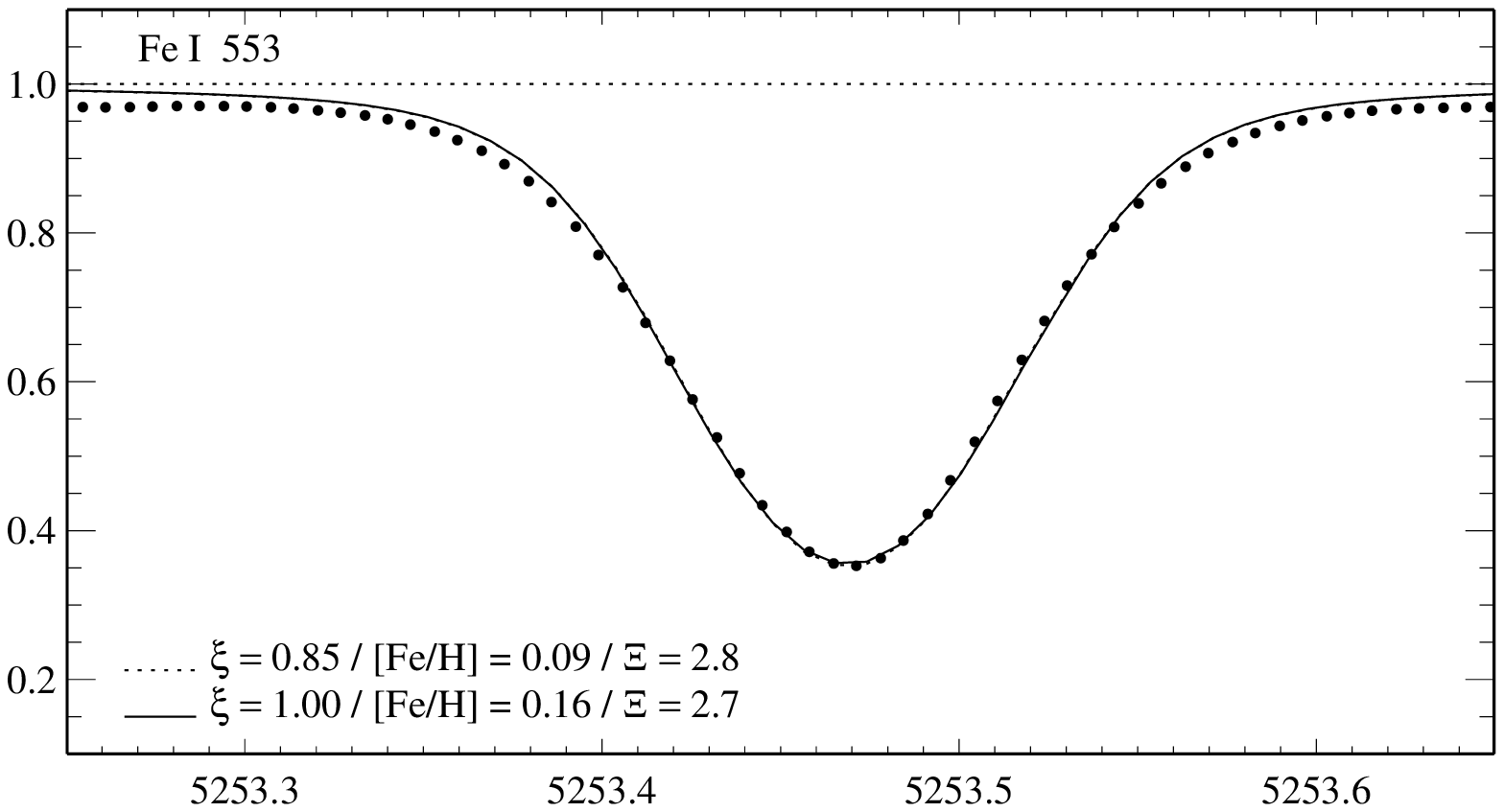}}\hfill 
  \resizebox{\columnwidth}{!}{\includegraphics{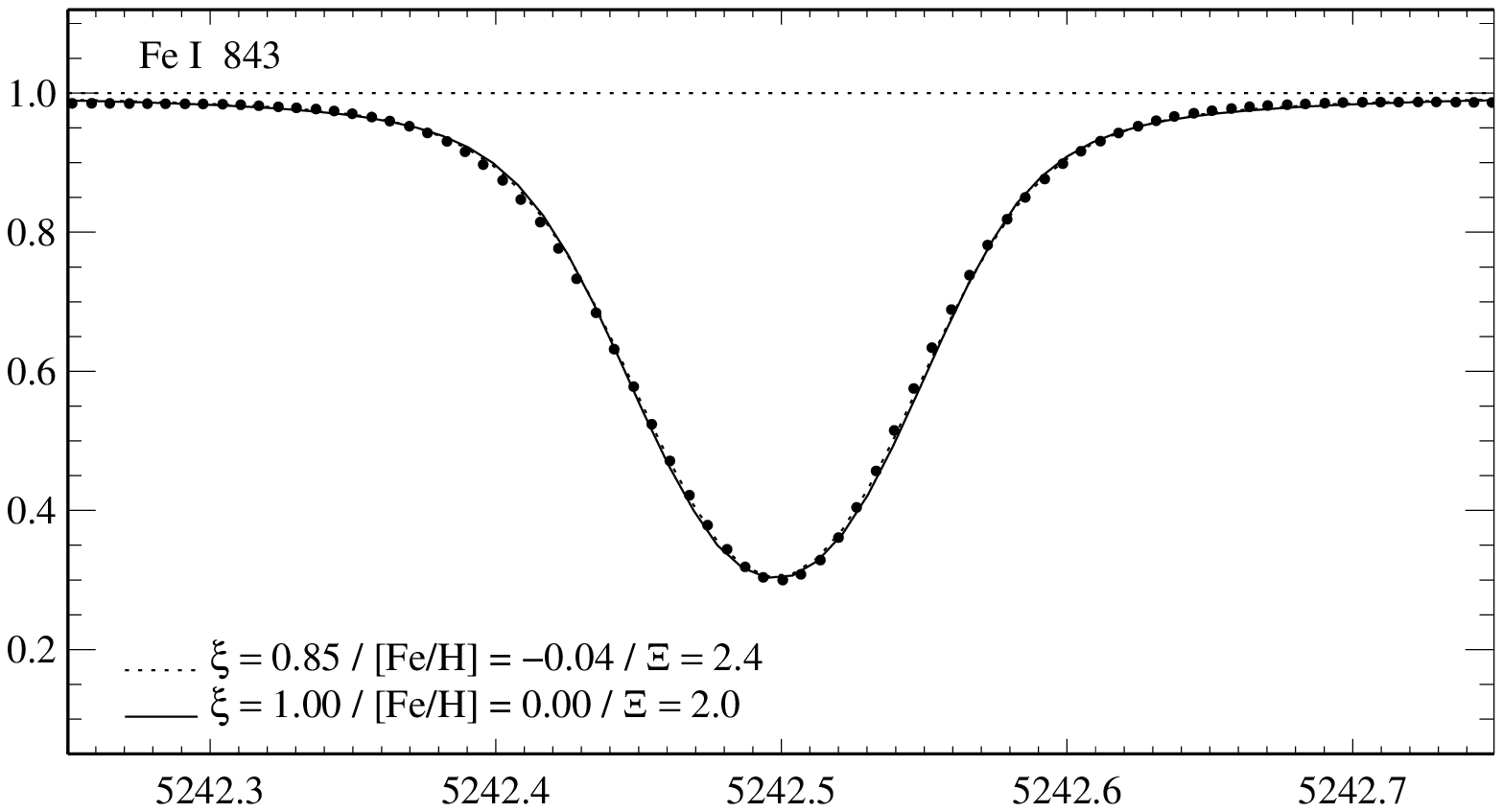}}}
\vspace{0mm} 
\hbox{\resizebox{\columnwidth}{!}{\includegraphics{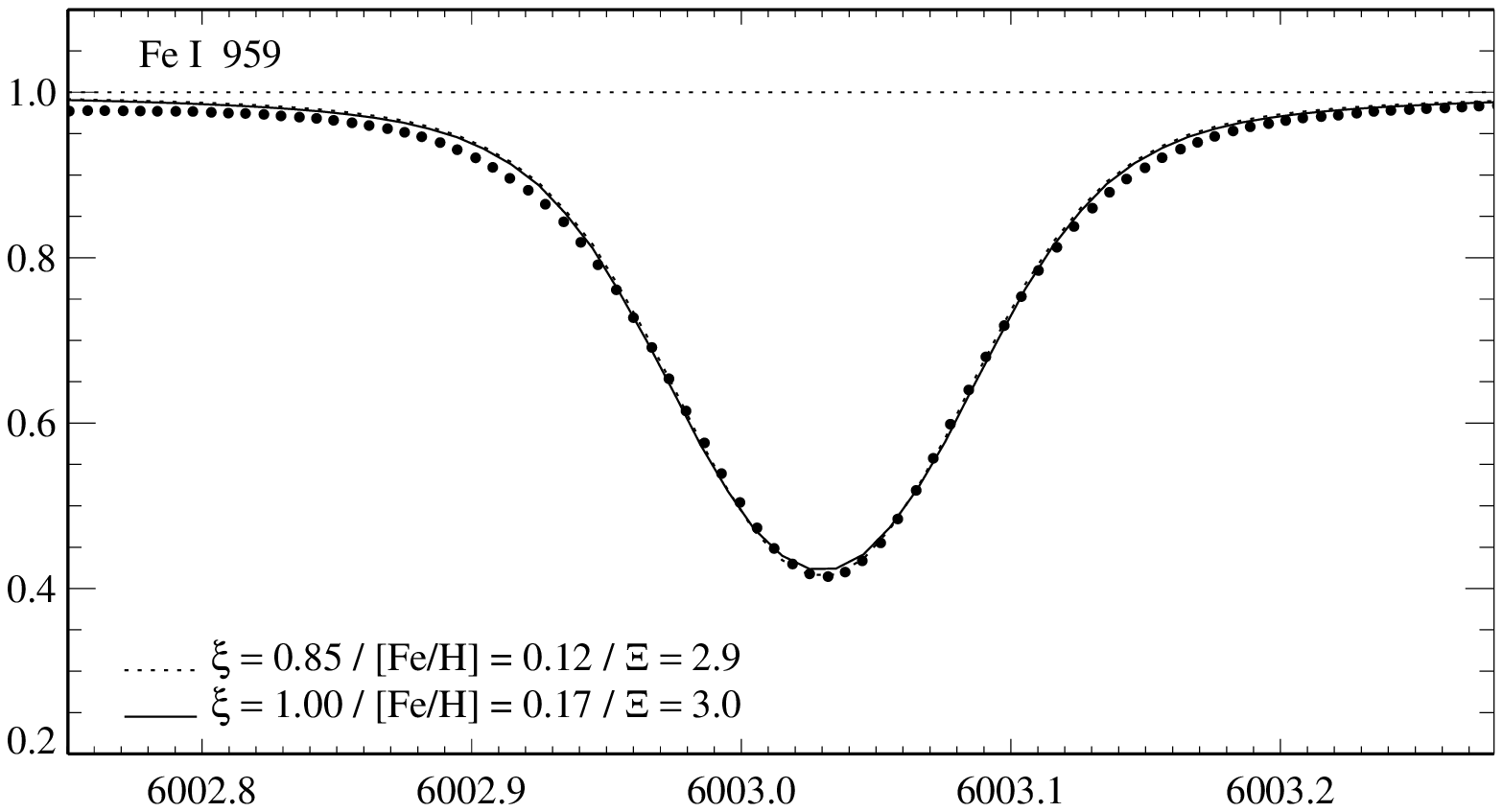}}\hfill 
  \resizebox{\columnwidth}{!}{\includegraphics{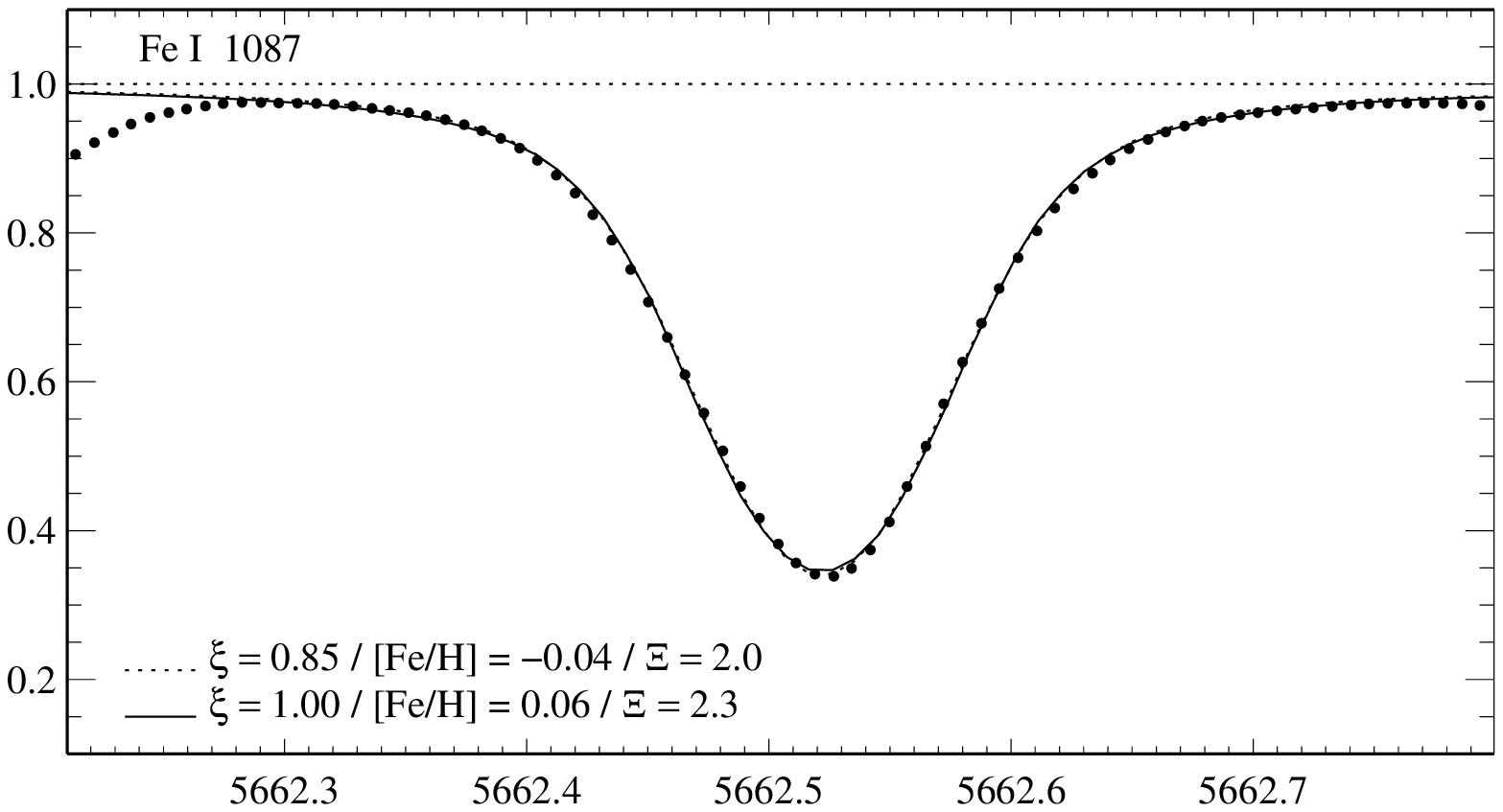}}}
\vspace{0mm} 
\hbox{\resizebox{\columnwidth}{!}{\includegraphics{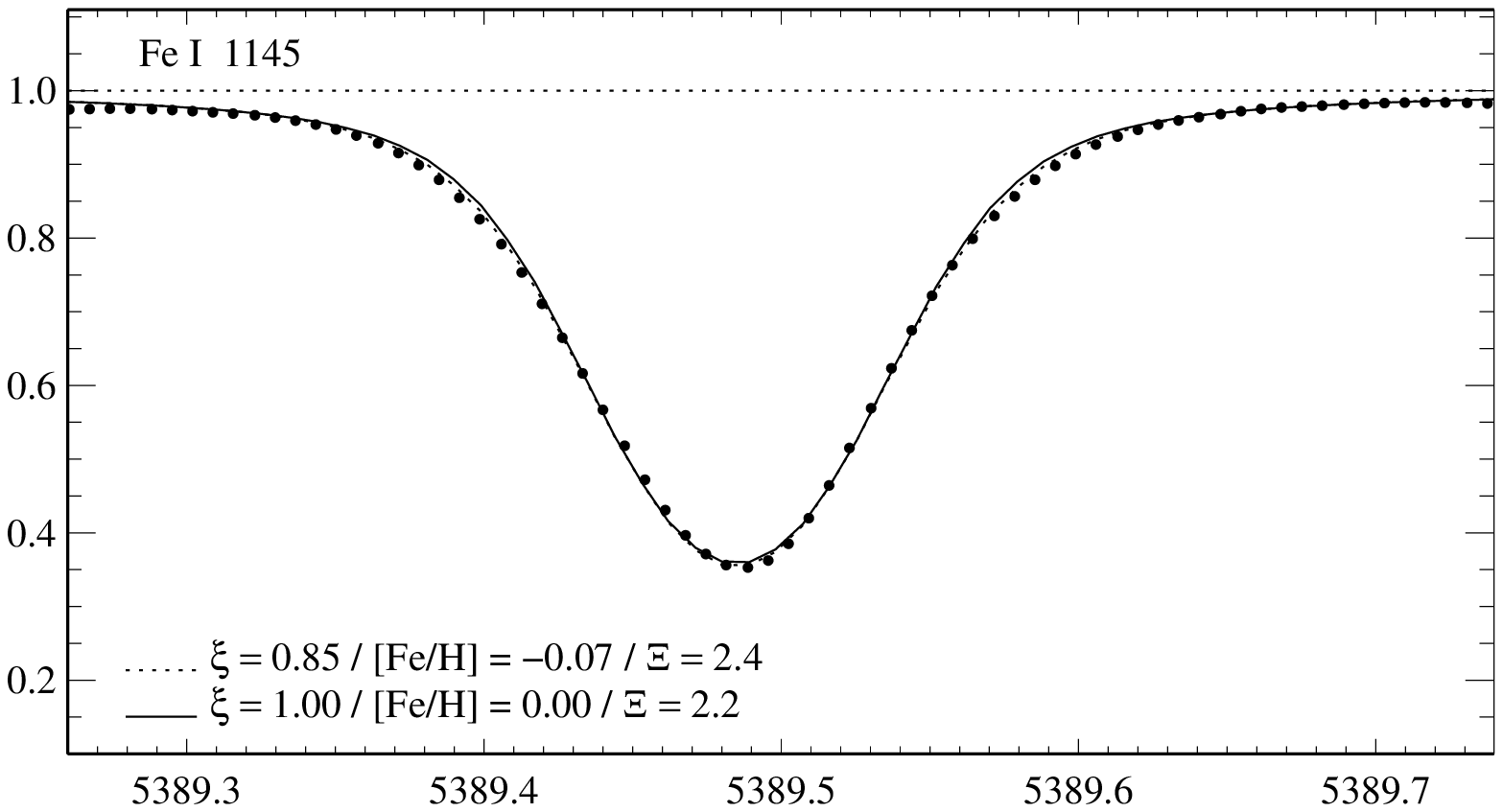}}\hfill 
  \resizebox{\columnwidth}{!}{\includegraphics{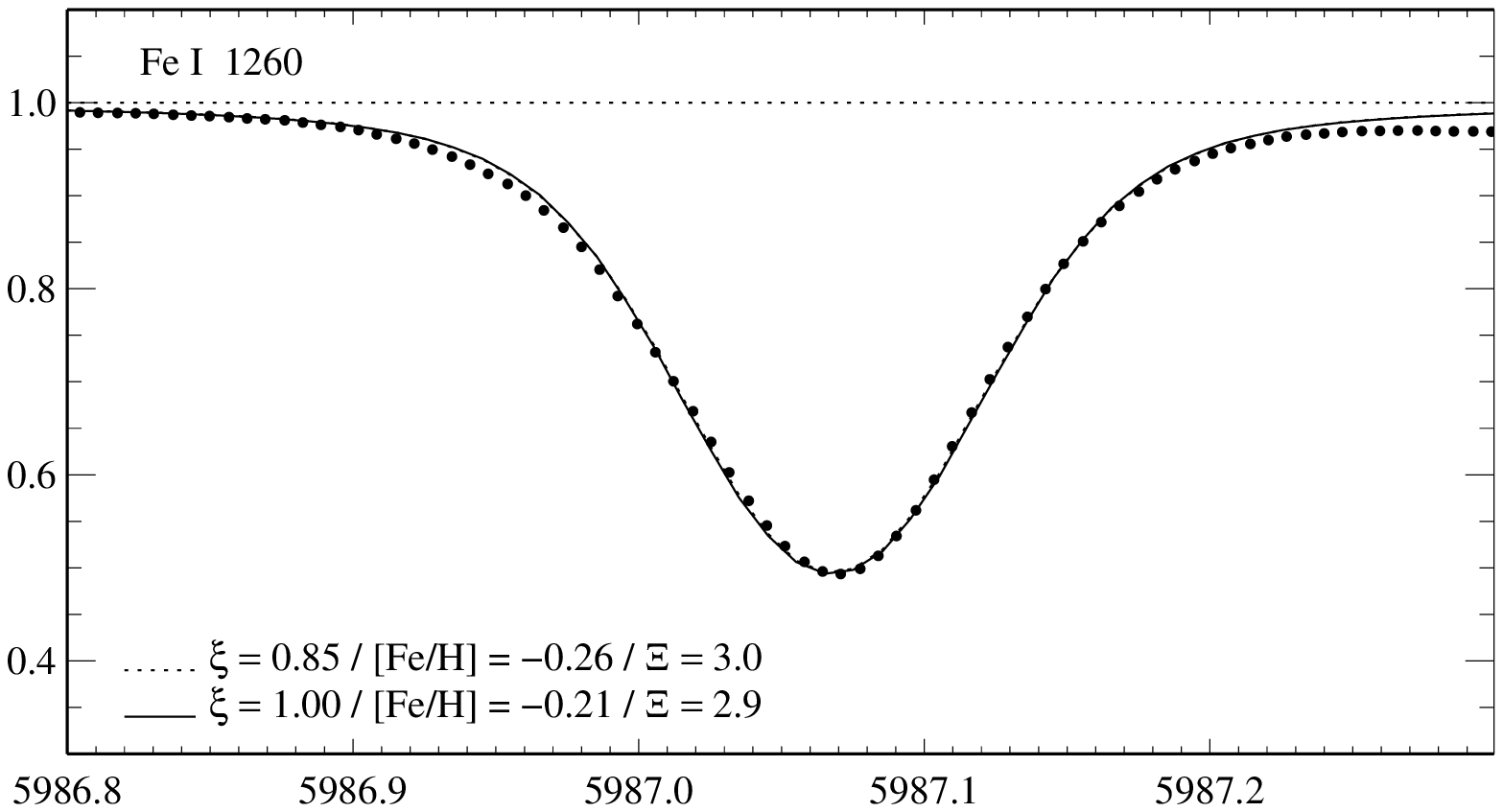}}}
\vspace{0mm} \caption[]{Profiles of turbulence lines ($60 < W_\lambda < 110$ 
m\AA) of \FeI\ in the solar flux spectrum (filled circles). Synthetic profile 
fits are again for LTE and HM (~\makebox[0.7cm]{\hrulefill}~) or TH (~
\makebox[0.7cm]{\dotfill}~) atmospheres. Fit parameters are indicated} 
\label{turblines} 
\end{center}
\end{figure*}
\noindent {\bf Turbulence lines} ($60 < W_\lambda < 110$ m\AA):\\[1mm] Roughly 
20\% of our sample are strong enough for core saturation and are therefore 
shaped by the value of the microturbulence parameter. Naturally, a static model 
atmosphere reproduces such lines only in an approximative way. This is seen in 
Fig. \ref{turblines} where a number of such lines and their synthetic fits are 
presented. Most of these fits require substantially smaller values of the 
macroturbulence velocity $\Xi$, but even then the synthetic core profiles are 
often too broad \emph{and} too shallow. 

In contrast to weaker lines for which the fit with synthetic profiles can be 
made nearly as accurate as desired, the fit of turbulence lines with a 
plane-parallel atmospheric model has its natural limitations which are explained 
by the velocity differences necessary to fit the innermost core and the wings 
simultaneously. Thus, in principle the saturated core seems to require 
relatively small velocity fields, whereas the opposite is required for the 
wings, a modulation that roughly represents the hydrodynamic equation of 
continuity. The microturbulence values used in the LTE models of Fig. 
\ref{turblines} have in fact been chosen so as to fit the line core 
\emph{width}. Using even larger values as would be indicated by comparison with 
weak and strong lines does \emph{not} improve the profile fits although it may 
help to minimize the overall abundance scatter. Fig. \ref{macro} emphasizes the 
difference in core saturation between the two model atmosphere types (HM and 
TH). Due to the temperature differences between the atmospheric models profiles 
synthesized from the HM model always require a \emph{smaller} macroturbulence to 
fit the very line core than do the LTE or NLTE profiles based on the TH model. 

We note that turbulence velocity \emph{gradients} introduced within the scope of 
static plane-parallel models do not improve the profile fits either. The 
\emph{kinematic} fine-tuning of the turbulence lines thus will stay the 
exclusive domain of granular hydrodynamics. 

Again, as with the weaker lines, LTE and NLTE models both tend to produce 
similar profile fits for the turbulence lines provided that the abundances are 
correspondingly adjusted. This is a direct consequence of the source function 
thermalization inherent to our NLTE modelling. As can be seen in Table 
\ref{lintab}, lines with equivalent widths around 100 m\AA\ display an abundance 
spread of $\sim 0.2$ dex among different LTE and/or NLTE models. 

The profiles of the stronger \FeI\ lines ($W_\lambda > 110$ m\AA) have been 
discussed in Paper I. It is therefore sufficient to repeat here, that 
simultaneous fits of line cores and damping wings are only obtained outside the 
range of the inner wings ($\pm 0.1 \ldots 0.4$ \AA). 

\subsection{Abundances} 

Our investigation of NLTE excitation and ionization in the solar photosphere 
would not be complete without mentioning the solar \FeI\ abundance problem. 
Since there exists quite a number of publications on the ''true'' solar \FeI\ 
abundance (e.g. Bi\'emont et al. \cite{BBKAP91}, Blackwell et al. 
\cite{BLS95a},\cite{BSL95b}, Holweger et al. \cite{HKB95}, Kostik et al. 
\cite{KSR96}, Grevesse \& Sauval \cite{GS99}), we will not enter into details 
but simply give our judgement according to the large number of lines of all 
strengths examined with reference to complete profile information (but ignoring 
their center-to-limb variation) and an exhaustive range of NLTE models. 

Current analyses tend to put their results into perspective by denoting the 
differences between \emph{photospheric} and \emph{meteoritic} \FeI\ abundances. 
The latter has been known for many years now (Anders \& Grevesse 
\cite{AG89}), $\eps{\FeI,\odot} = 7.51$. Photospheric abundance determinations, 
however, range from $\eps{\FeI,\odot} = 7.42$ (Schnabel et al. \cite{SKH99}, 
\FeII) to 7.67 (Blackwell et al. \cite{BLS95a}, \FeI). As was pointed out by 
Kostik et al. (\cite{KSR96}) and later iterated by Grevesse \& Sauval 
(\cite{GS99}), the discrepancy between different groups of researchers depends 
on a number of different methods and data sets the influences of which are not 
always easily disentangled. 

\begin{figure}
\resizebox{\hsize}{!}{\includegraphics{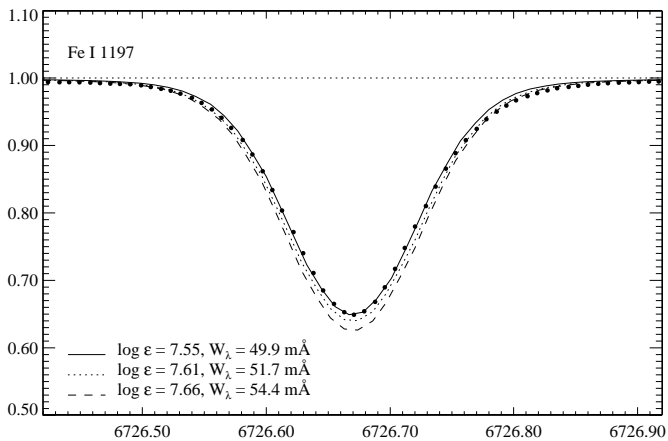}} \caption[]{LTE profiles of 
\FeI\ 1197, 6726.670 \AA, computed with the HM model atmosphere displaying the 
sensitivity of turbulence lines with respect to abundance changes} 
\label{meylan} 
\end{figure}
A few problems have already been discussed above, in particular the important 
influence of selecting a local spectral continuum. Other problems arise when 
determining abundances based on measurements of \emph{equivalent widths}. Thus, 
Meylan et al. (\cite{MFWK93}) have used Voigt profile fits to reproduce their 
observed \FeI\ lines. Their results differ systematically from those of other 
methods produced either by planimeter measurements or -- as in our case - from 
full line profile synthesis. This is an important source of systematic errors 
because anything but fitting \emph{synthesized} profiles requires an 
\emph{estimate} of the line wing extension that is often -- and always 
systematically -- neglecting a weak line haze. One of the more moderate examples 
is reproduced in Fig. \ref{meylan}. For this line Meylan et al. (\cite{MFWK93}) 
list an equivalent width of 53.6 m\AA, obtained from their Voigt profile fit. 
Our synthesis reproduces the observed solar flux spectrum with no continuum 
adjustment applying an \FeI\ abundance of $\eps{\FeI,\odot} = 7.55$, whereas 
their equivalent width requires an abundance $\sim 0.1$ dex higher than ours. 
More importantly, their equivalent width does \emph{not} fit the observed 
profile. Other turbulence lines listed by Meylan et al. show even larger 
discrepancies up to 0.3 dex! Therefore it is not surprising that -- using the 
$f$-values published in that paper -- we derive a mean solar abundance of 
$\eps{\FeI,\odot} = 7.25$. Altogether, at this stage of analyzing the solar 
\FeI\ abundance we ignore solar $f$-values because they would not add to 
abundance information, since their determination requires the input of a mean 
abundance value. 
 
\subsubsection{Sources of oscillator strengths} 

Except for the results of Meylan et al. (\cite{MFWK93}) and Gurtovenko \& Kostik 
(\cite{GK81}) Table \ref{lintab} contains only references to \emph{laboratory} 
$f$-values that cover more than 80\% of the lines. Among them we find 
essentially four different sets of data,
\begin{itemize}
  \item The laser-induced fluorescence measurements of O'Brian et al. 
  (\cite{OWLWB91}),
  \item $f$-values obtained from stabilized arc-emission by May et al. 
  (\cite{MRW74}),
  \item Observations of stabilized furnace absorption by the Oxford group
  of Blackwell et al. (\cite{BIPW76},\cite{BIPS79},\cite{BPS79},\cite{BPSS80},
  \cite{BPSS82a},\cite{BPSS82b})
  \item Hollow-cathode and laser-induced fluorescence measurements 
  performed by the Hannover group of Bard et al. (\cite{BKK91},\cite{BK94})
\end{itemize}
The rest of the sources is not very important for our investigation. The results 
listed in Table \ref{lintab} refer to a broad selection of methods which have 
been repeatedly discussed (see Holweger et al. \cite{HKB95}, Kostik et al. 
\cite{KSR96} or Grevesse \& Sauval \cite{GS99}). We start with a plain 
characterization of the abundance results obtained with the different sets of 
$f$-values. 

\begin{figure}[b]
\resizebox{\columnwidth}{!}{\includegraphics{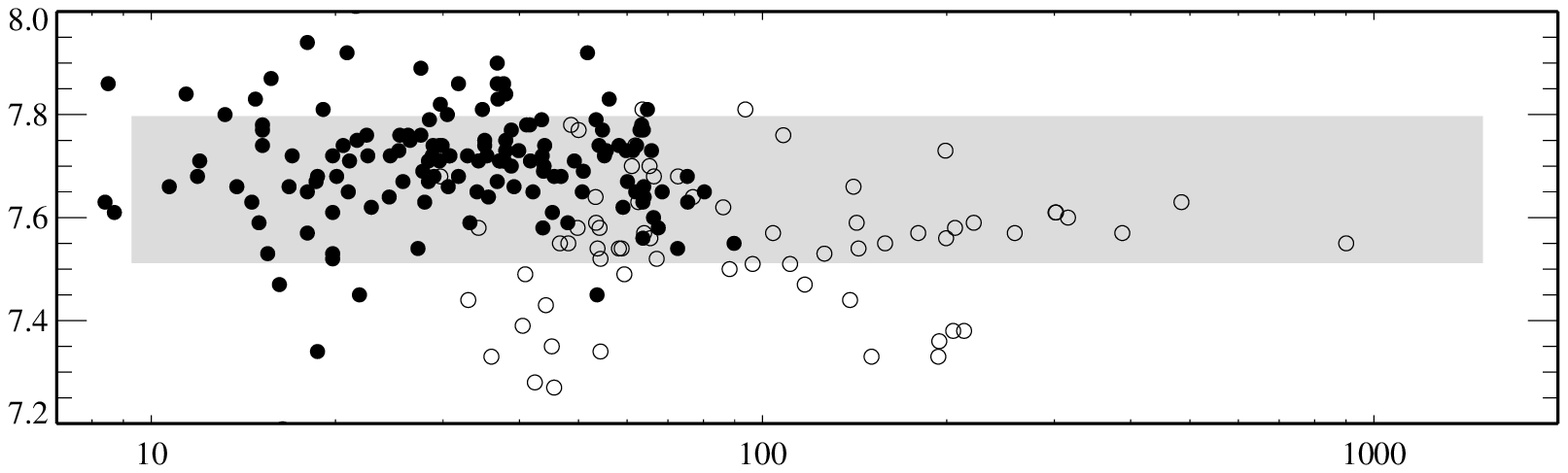}}\\ 
\resizebox{\columnwidth}{!}{\includegraphics{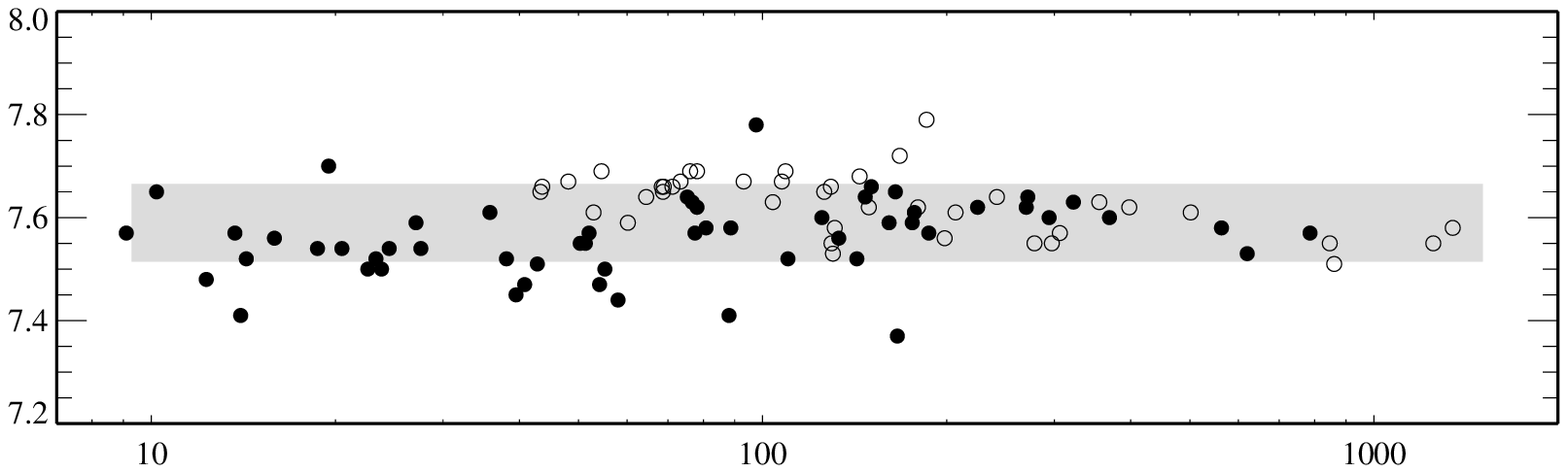}} 
\vspace{0mm} \caption[]{Logarithmic solar abundances as a function of equivalent 
width in m\AA\ determined with the HM solar model in LTE and $\xi = 1.0$ \kms. 
\emph{Top}: Oscillator strengths from May et al. (\cite{MRW74}, filled circles) 
and from  O'Brian et al. (\cite{OWLWB91}, open circles). \emph{Bottom}: 
$f$-values from the Hannover group (sources p,q of Table 2, filled circles) and 
from the Oxford group (sources e,f,g,h,n,o of Table 2, open circles). The range 
of $\pm 1\sigma$ r.m.s. scatter is indicated by the shading} \label{gfabund} 
\end{figure}
The top frame of Fig. \ref{gfabund} shows LTE abundance results obtained from 
the HM empirical model atmosphere using the data of O'Brian et al. 
(\cite{OWLWB91}) and May et al. (\cite{MRW74}), whereas the bottom frame of Fig. 
\ref{gfabund} displays the results for the oscillator strengths determined by 
the Oxford and Hannover groups. While the proper choice of models and parameters 
is discussed in the following subsection, it is already evident here that the 
two frames harbour sources of different quality. Thus, the $f$-values of O'Brian 
et al. or May et al. lead to approximately twice the r.m.s. scatter of the solar 
abundances as compared with the results derived from the $f$-values of the 
Oxford and Hannover groups. The May et al. abundances are also systematically 
higher than the mean. 

The $f$-values of O'Brian et al. and those of Bard and Kock (\cite{BK94}) are on 
the same absolute scale since both have used very similar measurements and 
normalization procedures. In fact, Fig. 3 in Bard \& Kock shows a negligible 
difference of the corresponding $f$-values for the lines in common, although the 
strong scatter is confirmed. What makes the O'Brian et al. sample so suspicious 
is the occurrence of abundance differences between lines in a \emph{common 
multiplet}. An extreme case is Mult 66, where our results for $\lambda5145.099$ 
and $\lambda5250.646$ lead to $\eps{\FeI} = 7.34$ and $7.76$, respectively. 
There are also other lines such as $\lambda4798.267$ and $\lambda4735.845$ of 
Mult 1042 with $\eps{\FeI} = 7.35$ and $7.81$, respectively. 

There is no simple explanation why the oscillator strengths of May et al. and 
those of Bard and Kock (\cite{BK94}) lead to different abundances. The data used 
in our analysis are those in Fuhr et al. (\cite{FMW88}), which had been 
renormalized to the scale of the Oxford measurements. Most of the corrected May 
et al. $f$-values are therefore 0.1 dex \emph{smaller} than the original data. 
Based on the original paper, the May et al. \emph{abundances} thus would be 0.1 
dex smaller. While this accounts for half of the difference between the two 
groups, there remains another 0.1 dex difference which is not seen in Fig. 2 of 
Bard \& Kock. However, the r.m.s. scatter of both the original and the 
renormalized data set of May et al. is even slightly larger than that of O'Brian 
et al., and differences such as in $\lambda5395.250$ and $\lambda5487.160$ of 
Mult 1143 with $\eps{\FeI} = 8.01$ and $7.71$, respectively, are also found in 
their sample. 

Interestingly enough some of the more recent measurements of the Oxford and 
Hannover groups seem to produce substantially smaller scatter. Whereas 
$\sigma(\eps{}) \sim 0.15$ for the O'Brian et al. and May et al. samples, 
$\sigma(\eps{}) \sim 0.05 \ldots 0.07$ for the Oxford and Hannover lines. Fig. 
\ref{gfabund} shows a marginal difference between the two groups, but that 
depends on a particular choice of our models with $(\Delta\eps{\FeI})_{\rm 
Oxf-Han} = 0.067$ for the HM LTE model and $0.026$ for the TH LTE model. Let us 
mention here that \emph{line-by-line} comparison of $f$-values of the two groups 
leads to a difference of $(\Delta\log gf_\FeI)_{\rm Oxf-Han} = -0.029 \pm 
0.009$. 

In order to evaluate the solar iron abundance we thus decided to disregard all 
but the Oxford and Hannover $f$-values. Unfortunately, this choice reduced our 
line sample from 391 to 97 lines. Fig. \ref{gfabund} demonstrates that all of 
the weak lines in this combined sample are from Hannover sources whereas most of 
the strong lines were measured in Oxford. This correlates nicely with excitation 
energies, such that all low-excitation lines come from Oxford sources and all 
high-excitation lines are due to Hannover measurements. 

\subsubsection{The solar iron abundance} 

Irrespective of the choice of the $f$-values the solar \FeI\ abundances as 
calculated from fitting the solar flux spectrum depend sensitively on the model 
assumptions. Blackwell et al. (\cite{BLS95a}) and Grevesse \& Sauval 
(\cite{GS99}) both have reported that the HM empirical solar model leads to 
\FeI\ abundances systematically higher than those obtained from theoretical 
models or other empirical models with a lower temperature in their upper layers. 
This is to be expected under the assumption of LTE since the source function 
then is always Planckian, and the emerging intensities in theoretical models 
will to first order follow the temperature stratification. It is, however, 
\emph{not} evident for NLTE line formation, since there both the source function 
and the optical depth scale may deviate from their thermal behaviour. 

In Paper I the level populations had been discussed for a number of LTE and 
NLTE population models. It was argued there that in most of the NLTE models -- 
at least those with non-zero hydrogen collisions -- the line source functions 
were very close to thermal, and the differences of line profiles with respect to 
LTE occurred essentially due to parametrization of (a) hydrogen collisions and 
(b) a cutoff energy above which all levels were thermalized with respect to the 
\FeII\ ground state. The latter operation had to be included to simulate the 
missing ionization/recombination channels. The different populations are shown 
in Fig. 6 of Paper I, and as yet we have not been able to choose a best case 
model on the basis of comparison with the strong lines only.

\begin{table}
\centering \caption{Solar \FeI\ abundances based exclusively on the $f$-values 
of the Oxford and Hannover groups, calculated for different models of line 
formation. Note that $\Delta\log C_6$ refers to Anstee \& O'Mara's damping 
constants. It was chosen so that the mean abundances did not depend on 
equivalent width (see left panels in Fig. \ref{modabund}). See text for further 
discussion} 
\begin{tabular}{rlccc}
    & Model   & $\xi$ [\kms] & $\Delta\log C_6$ & $\eps{\FeI,\odot}$\\
\noalign{\smallskip}\hline\noalign{\smallskip}
  0 & TH LTE  & 0.85 & -0.12 & 7.508 $\pm$ 0.080 \\
  1 & NLTE 0+ & 0.85 & -0.23 & 7.605 $\pm$ 0.087 \\
  2 & NLTE 5+ & 0.85 & -0.10 & 7.521 $\pm$ 0.089 \\
  3 & NLTE 5- & 0.85 & -0.15 & 7.629 $\pm$ 0.094 \\
  4 & HM LTE  & 1.00 & ~0.09 & 7.574 $\pm$ 0.074 \\
  5 & TH LTE  & 1.00 & -0.14 & 7.477 $\pm$ 0.070 \\
  6 & NLTE 5+ & 1.00 & -0.12 & 7.488 $\pm$ 0.075 \\
  7 & NLTE 1+ & 1.00 & -0.13 & 7.503 $\pm$ 0.077 \\
  8 & NLTE 1+ & 1.00 & -0.16 & 7.499 $\pm$ 0.075 \\
  9 & NLTE 1/2+ & 1.00 & -0.17 & 7.509 $\pm$ 0.077 \\ 
\noalign{\smallskip}\hline
\end{tabular}
\label{modmod} 
\end{table}

\begin{figure*}
\begin{center}
\hbox{\resizebox{85mm}{!}{\includegraphics{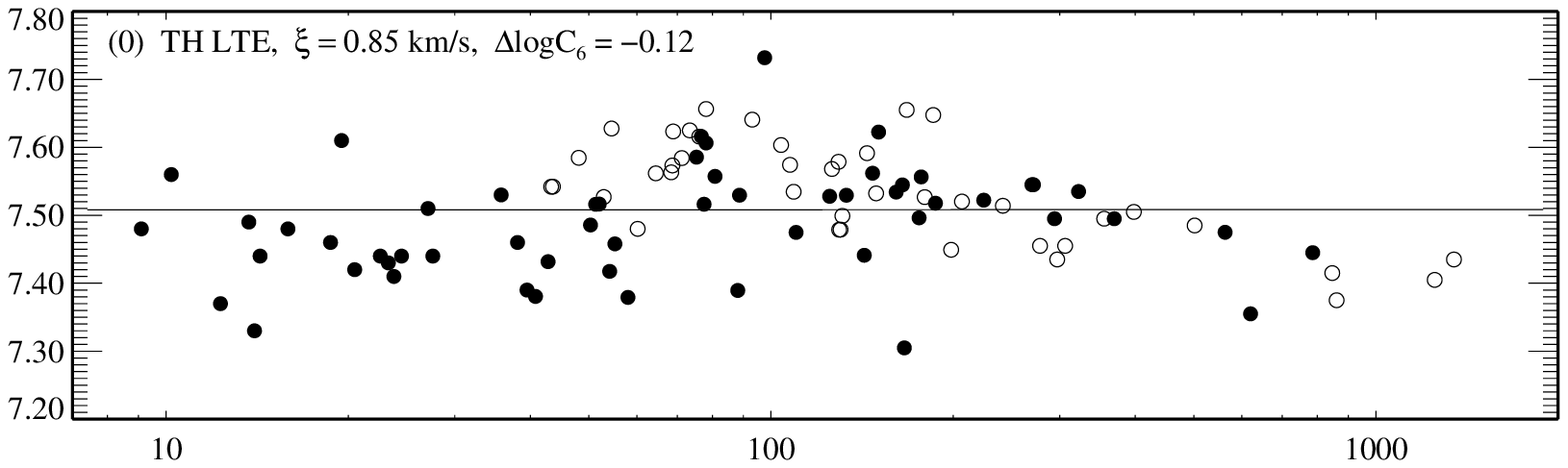}}\hspace{5mm} 
  \resizebox{85mm}{!}{\includegraphics{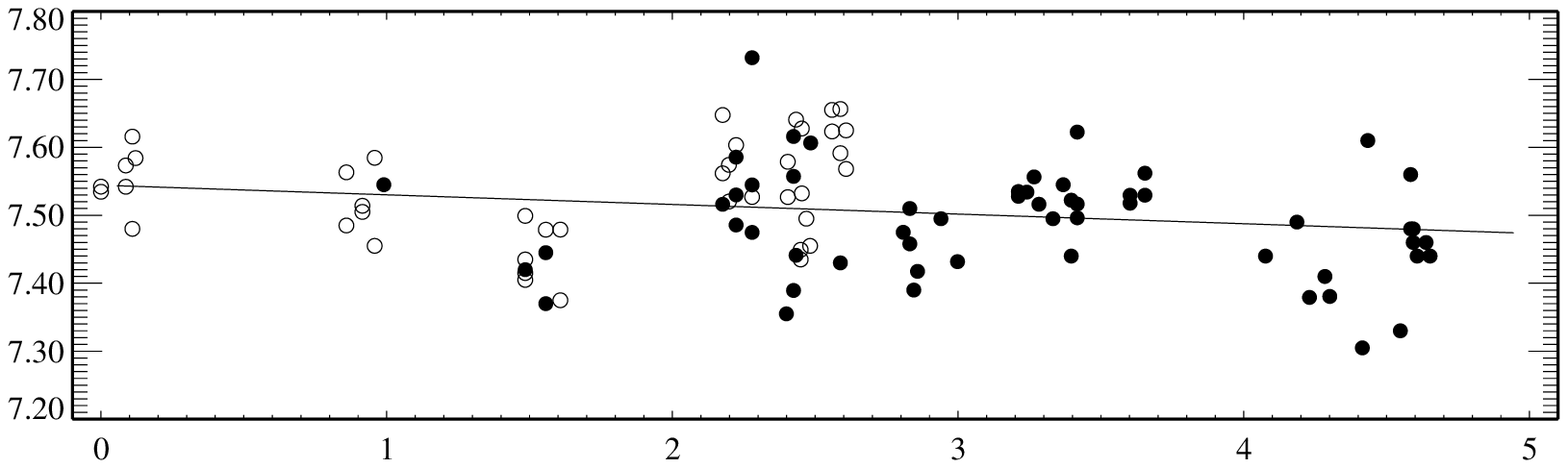}}}
\vspace{1mm} 
\hbox{\resizebox{85mm}{!}{\includegraphics{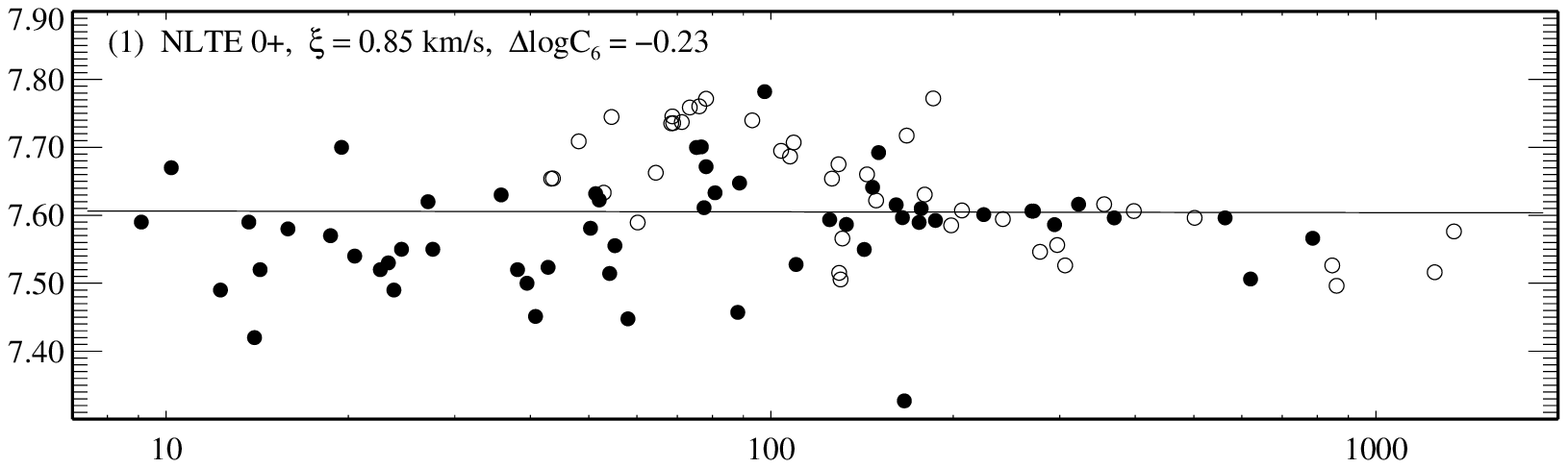}}\hspace{5mm} 
  \resizebox{85mm}{!}{\includegraphics{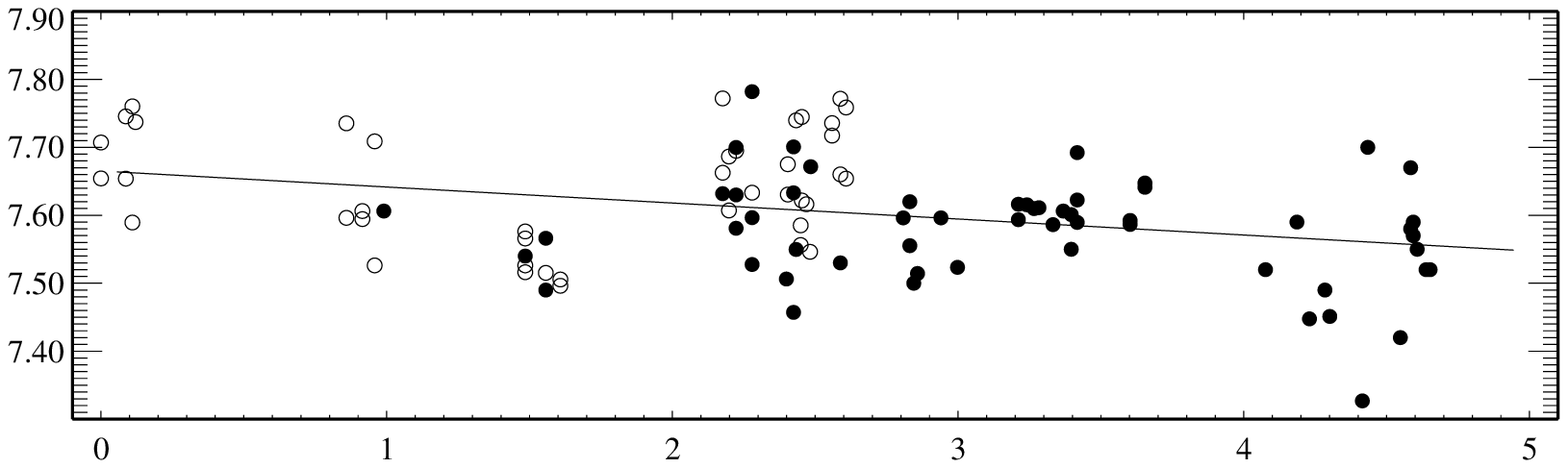}}}
\vspace{1mm} 
\hbox{\resizebox{85mm}{!}{\includegraphics{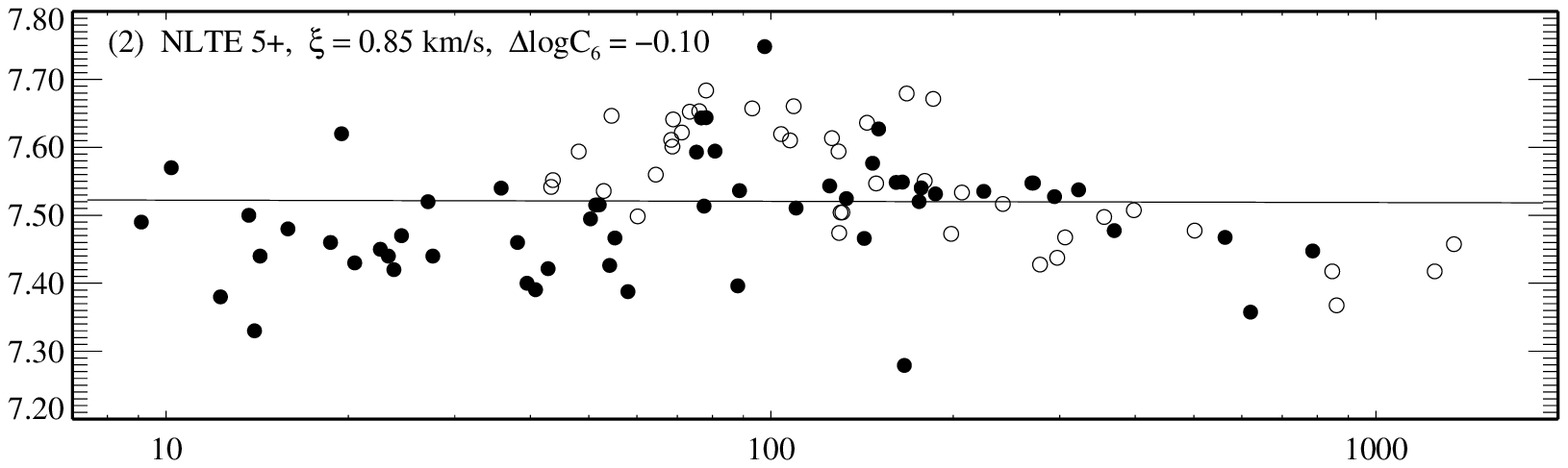}}\hspace{5mm} 
  \resizebox{85mm}{!}{\includegraphics{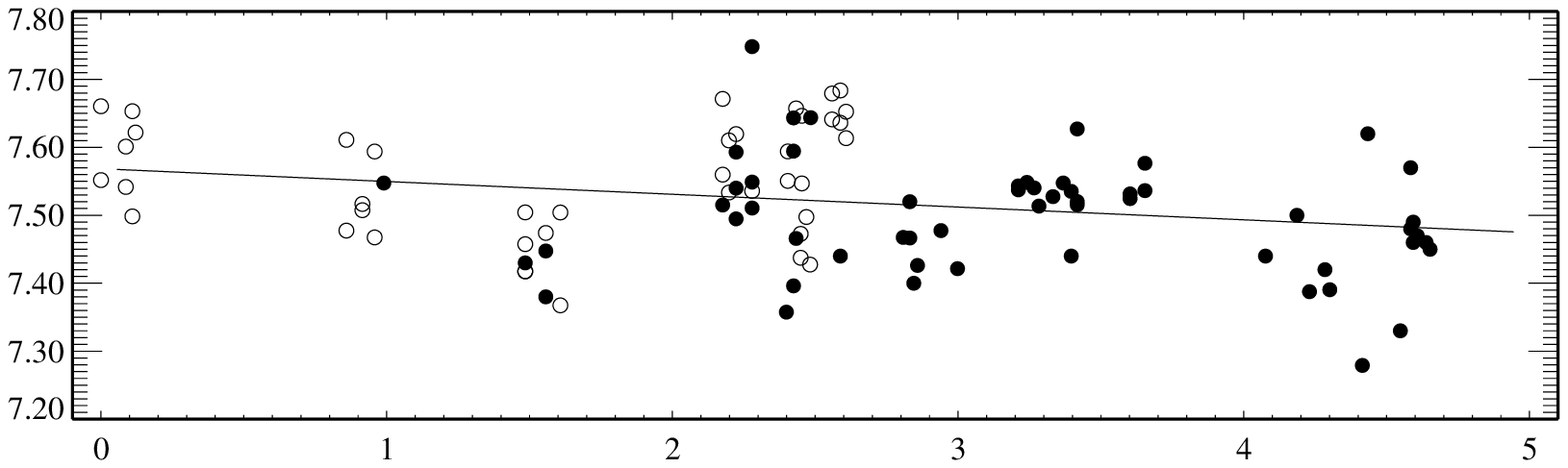}}}
\vspace{1mm} 
\hbox{\resizebox{85mm}{!}{\includegraphics{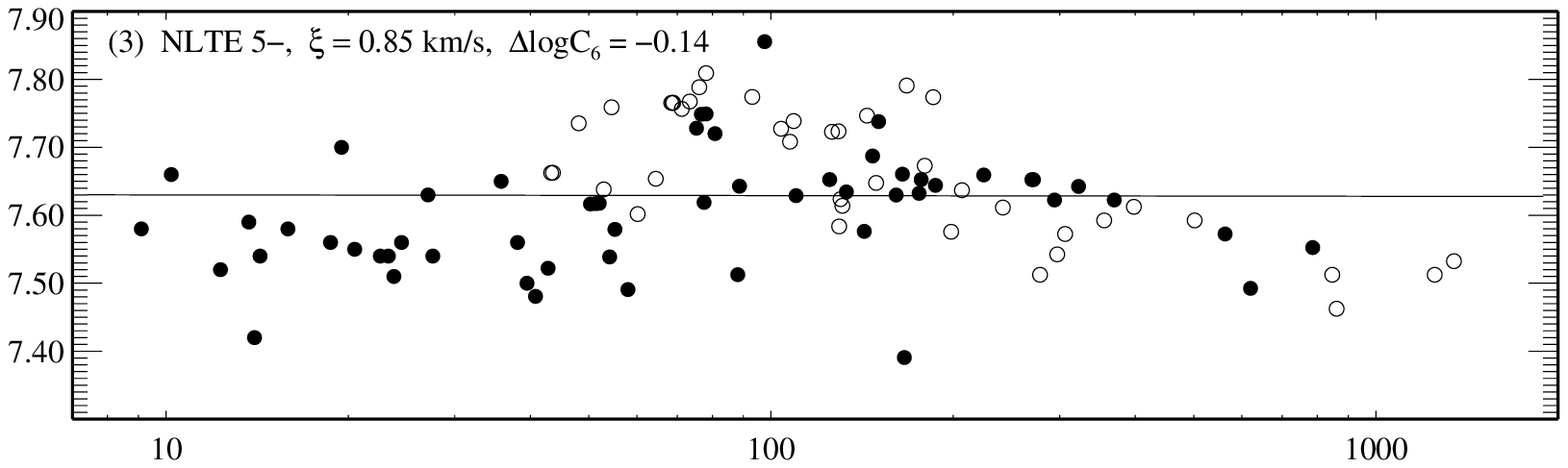}}\hspace{5mm} 
  \resizebox{85mm}{!}{\includegraphics{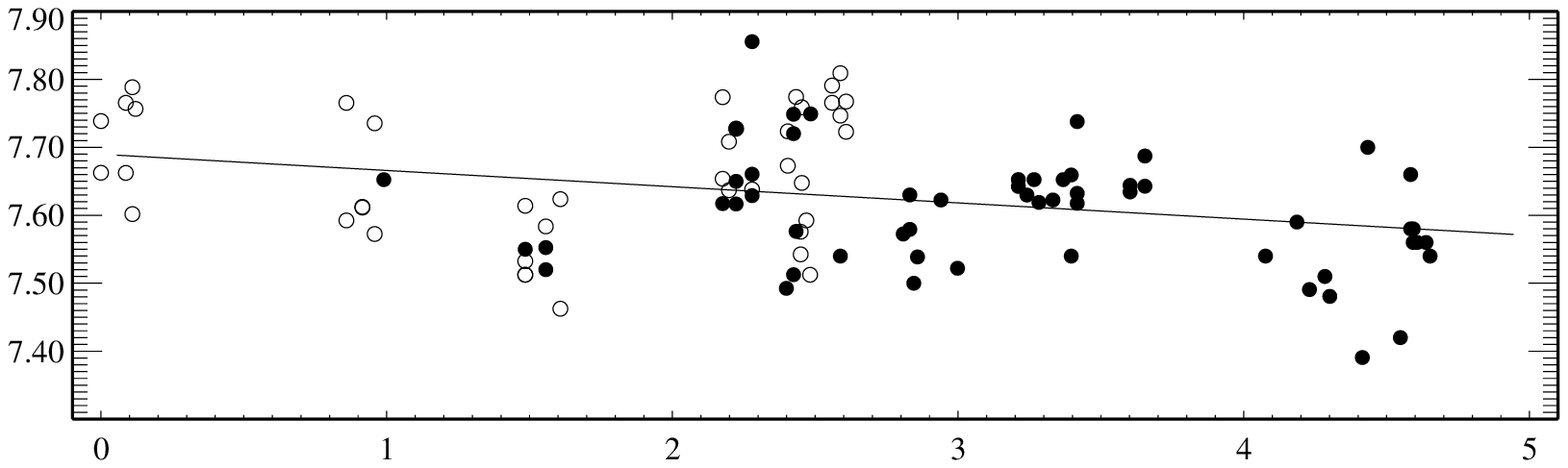}}}
\vspace{1mm} 
\hbox{\resizebox{85mm}{!}{\includegraphics{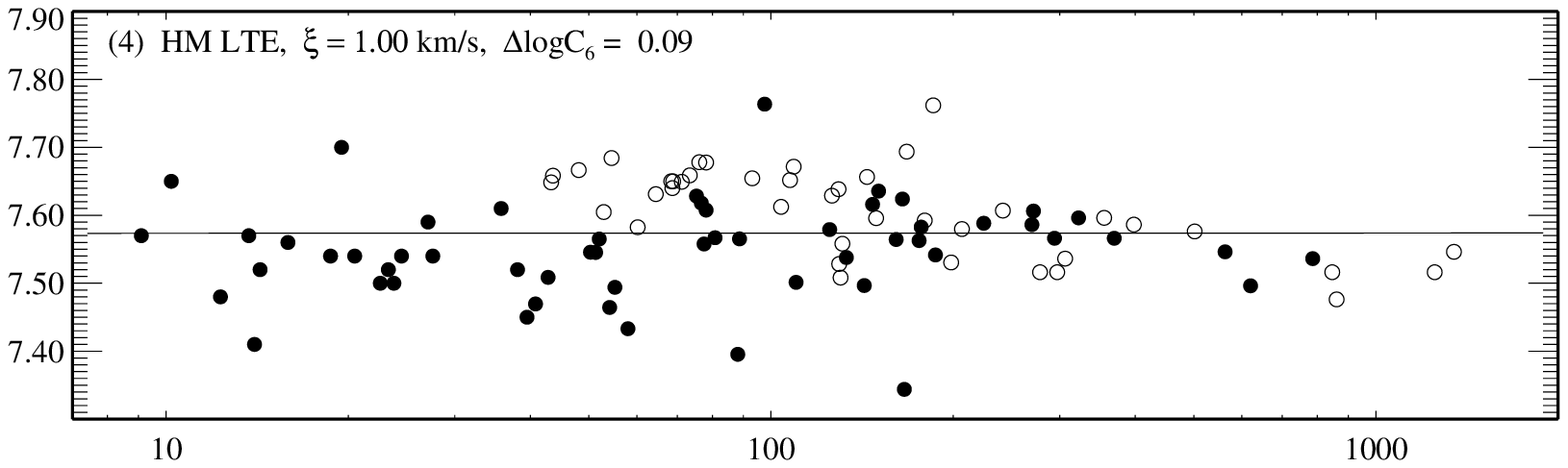}}\hspace{5mm} 
  \resizebox{85mm}{!}{\includegraphics{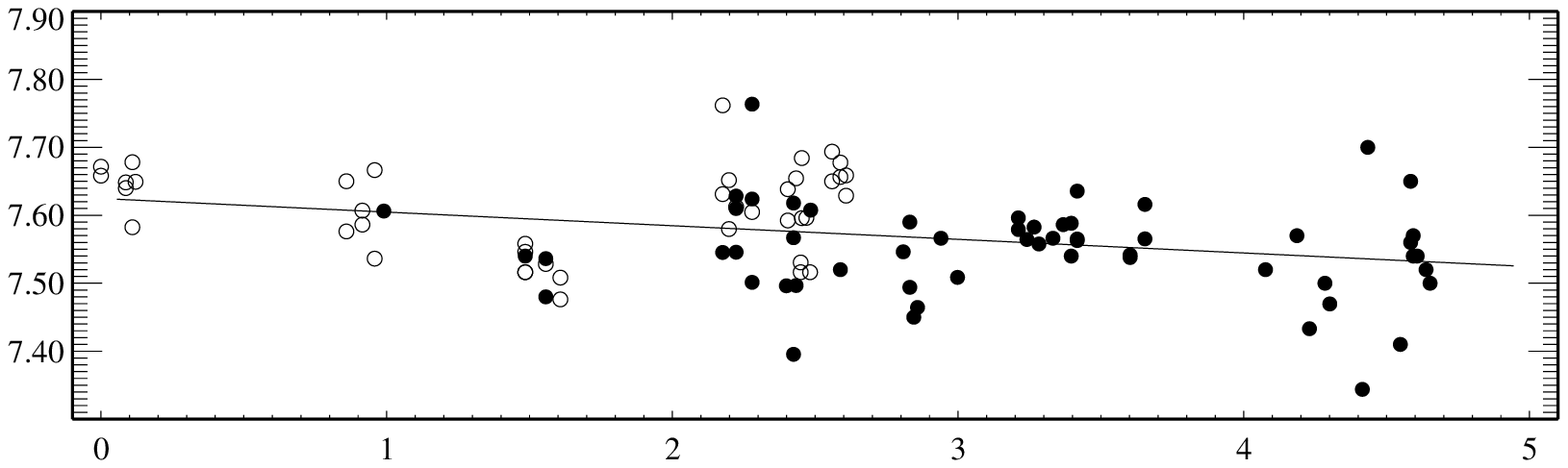}}}
\vspace{1mm} 
\hbox{\resizebox{85mm}{!}{\includegraphics{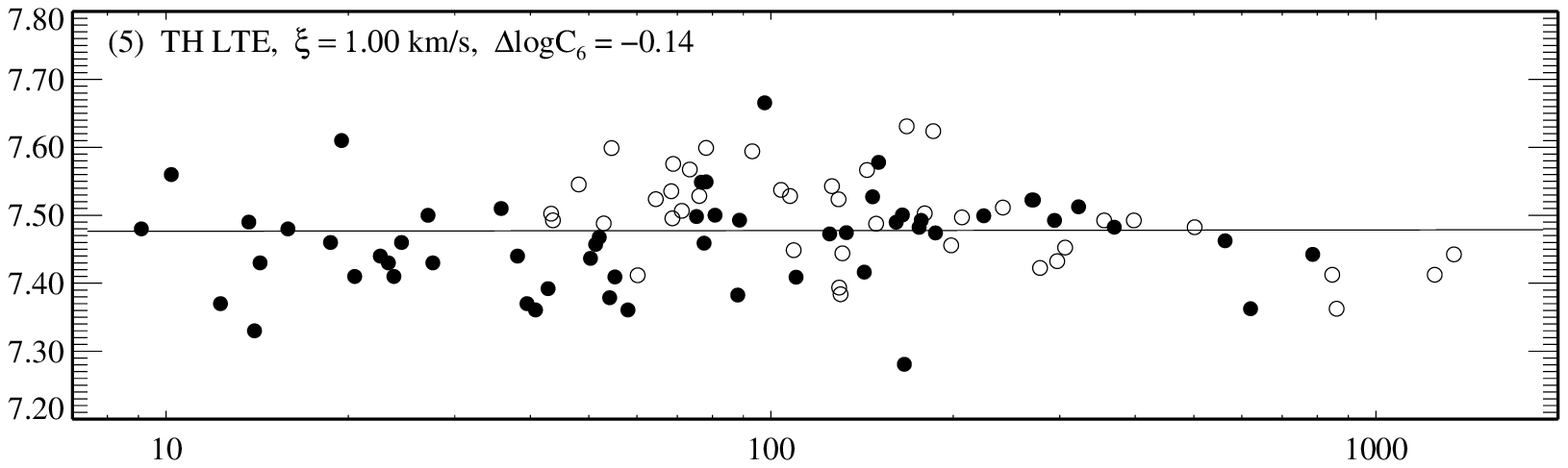}}\hspace{5mm} 
  \resizebox{85mm}{!}{\includegraphics{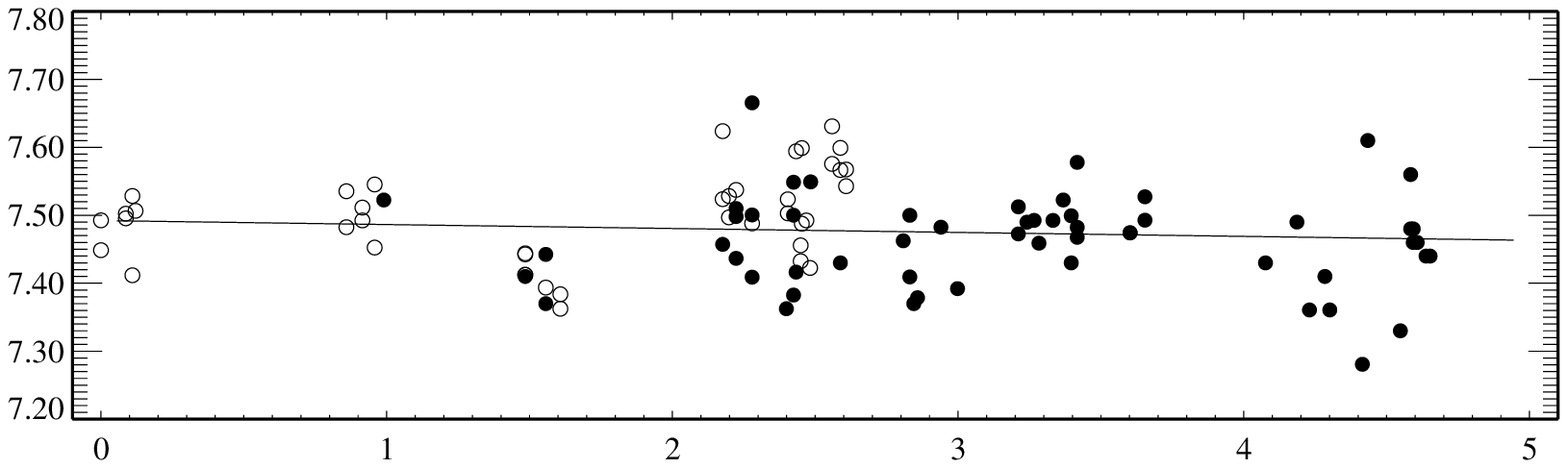}}}
\vspace{1mm} 
\hbox{\resizebox{85mm}{!}{\includegraphics{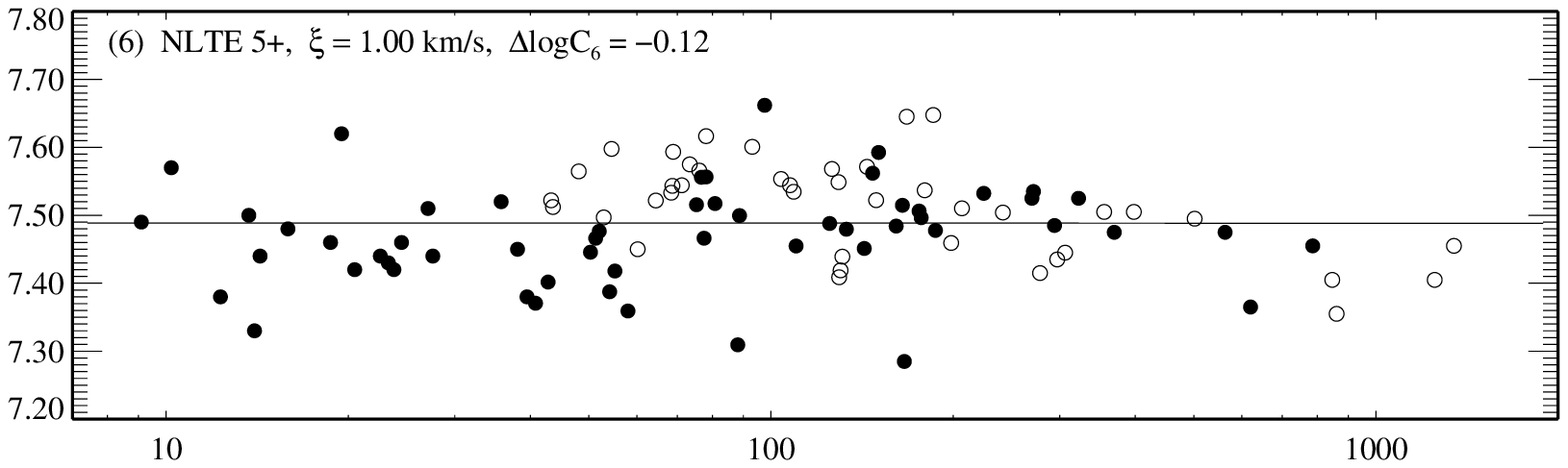}}\hspace{5mm} 
  \resizebox{85mm}{!}{\includegraphics{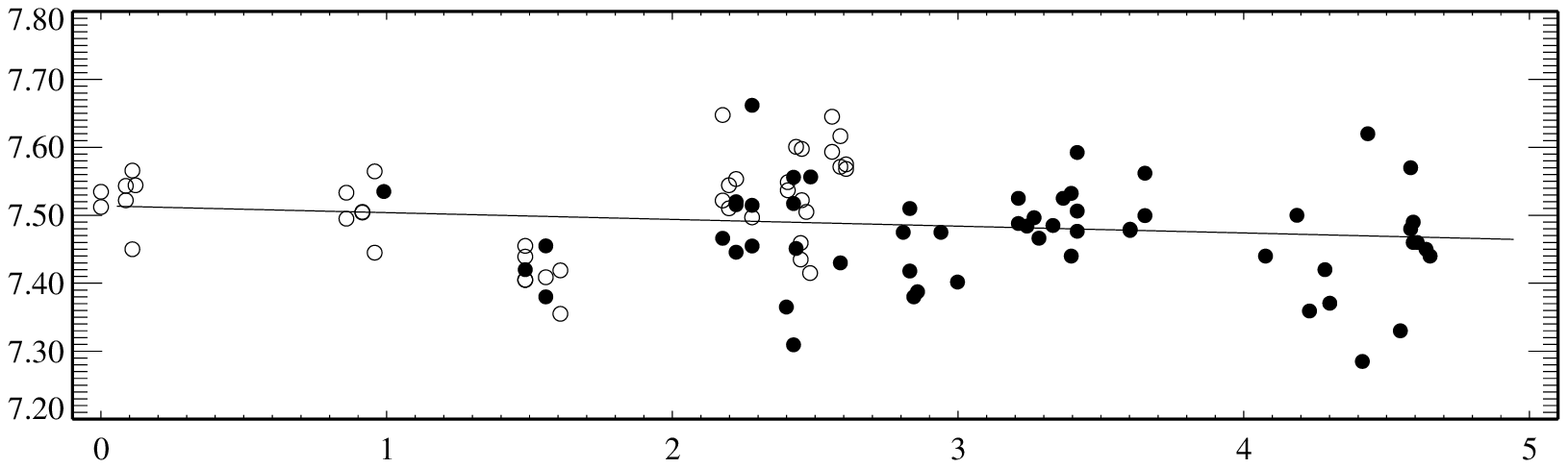}}}
\vspace{1mm} 
\hbox{\resizebox{85mm}{!}{\includegraphics{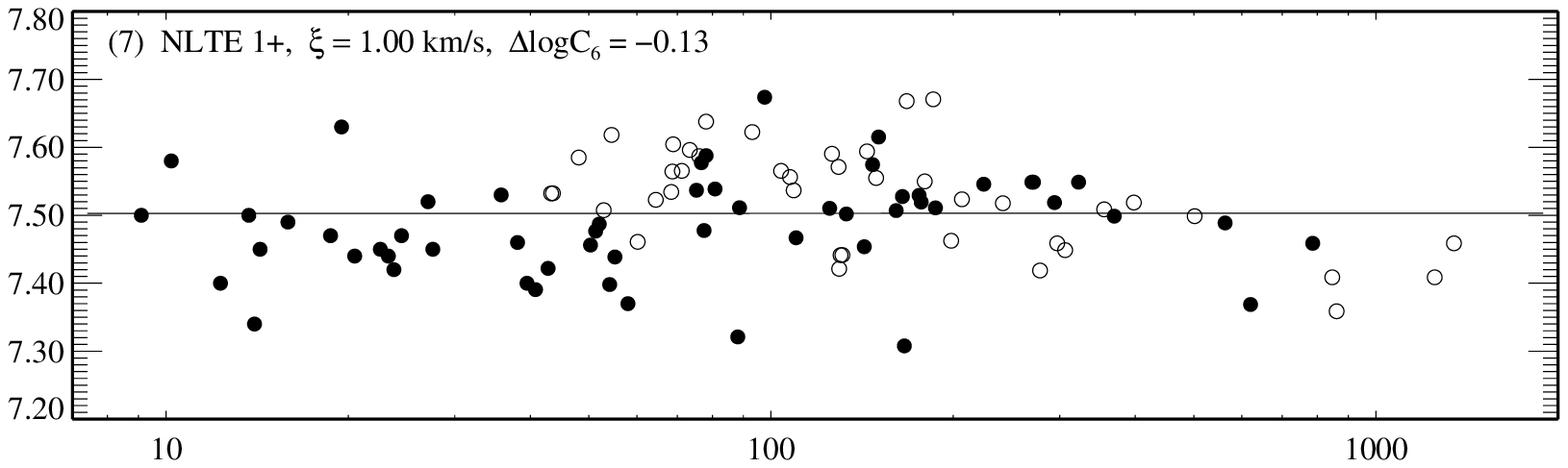}}\hspace{5mm} 
  \resizebox{85mm}{!}{\includegraphics{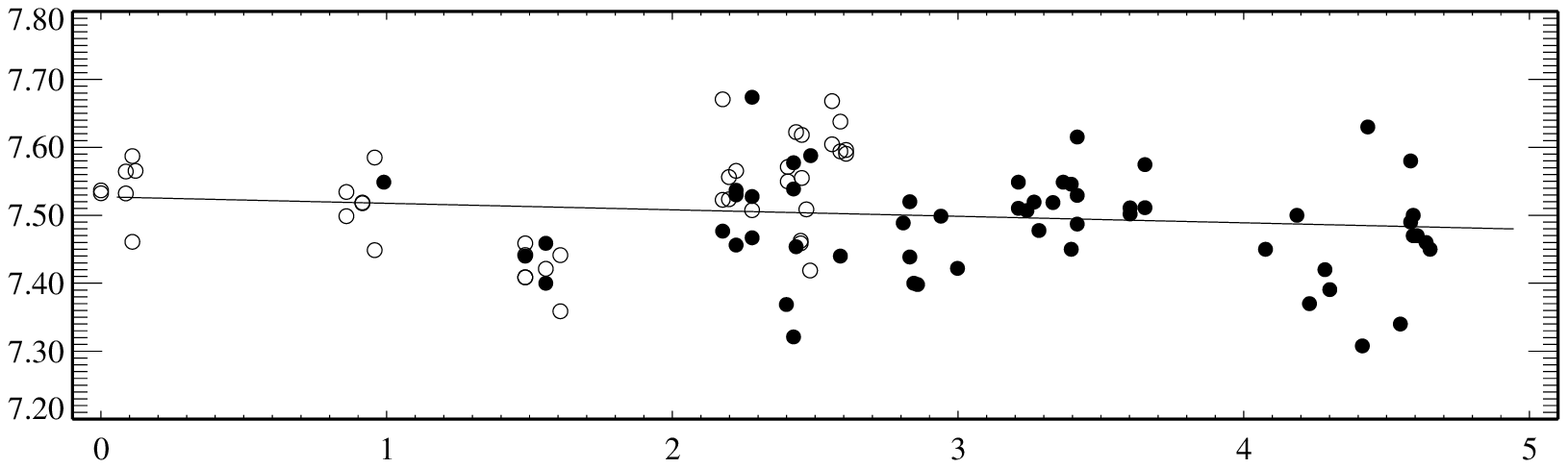}}}
\vspace{1mm} 
\hbox{\resizebox{85mm}{!}{\includegraphics{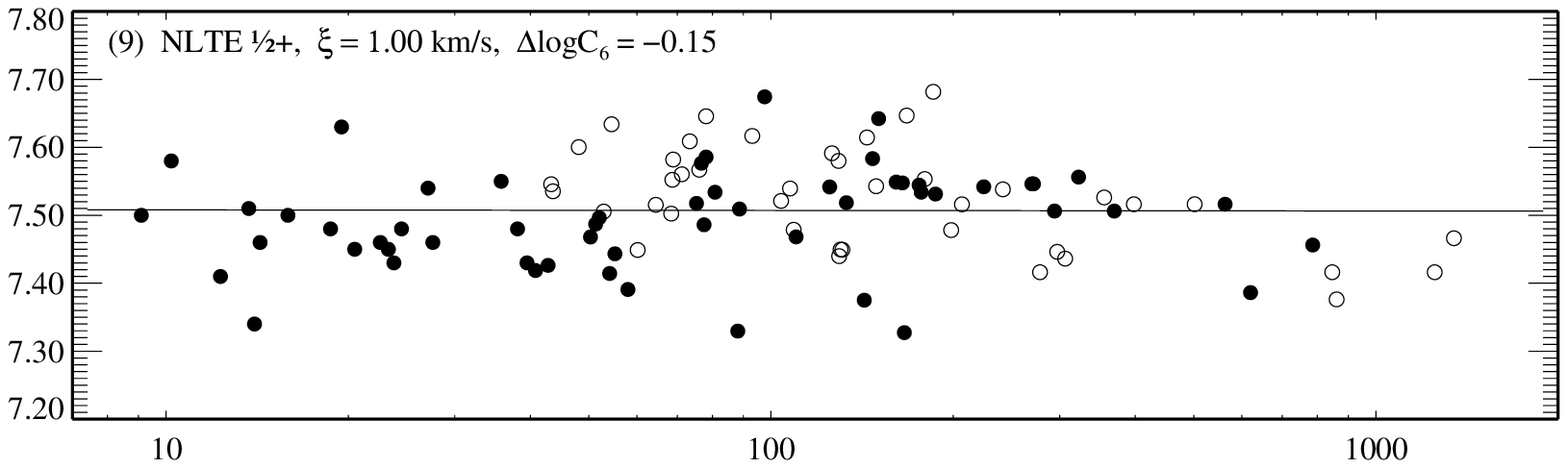}}\hspace{5mm} 
  \resizebox{85mm}{!}{\includegraphics{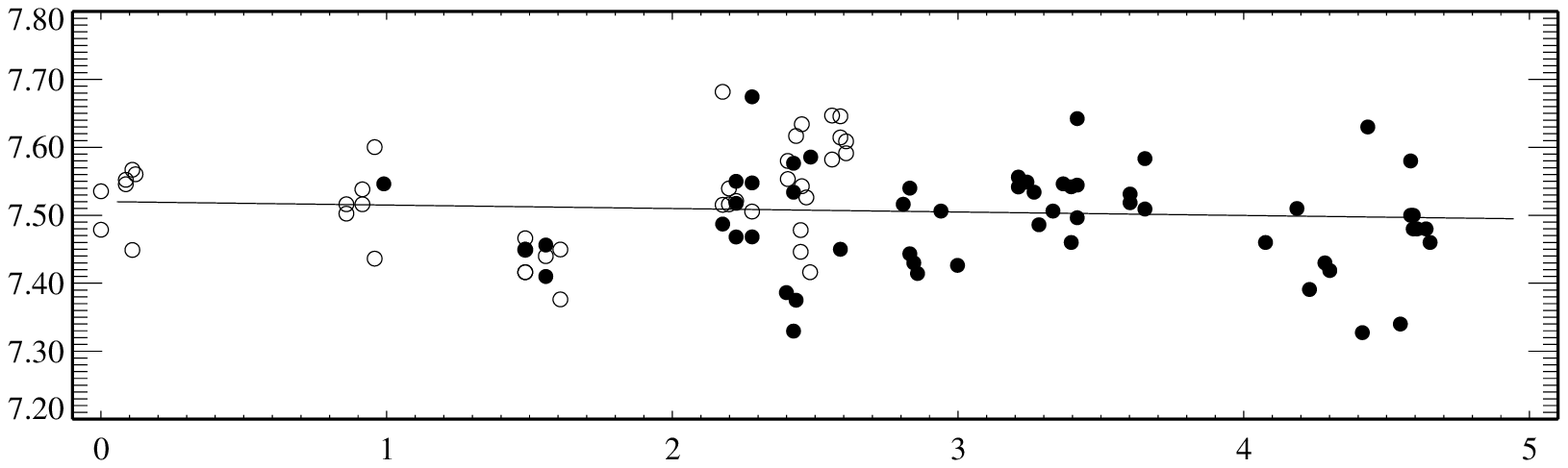}}}
\vspace{-1mm} \caption[]{Solar \FeI\ abundances as a function of $W_\lambda$ 
(left) and lower-level excitation energy $E_{\rm low}$ (right), calculated for 
different models of line formation. Oscillator strengths are from the Oxford 
(open circles) and Hannover groups (filled circles). See text for further 
explanations} \label{modabund} 
\end{center}
\end{figure*}
Fig. \ref{modabund} therefore gives an impression of how the solar \FeI\ 
abundances obtained from line profile fits based on different LTE and NLTE 
models with different line-broadening parameters depend on the model 
assumptions. As mentioned above, only the Oxford and Hannover group $f$-values 
have been considered. With respect to Table \ref{models} the models in Fig. 
\ref{modabund} are modified using the original models 7 and 8 of Table 
\ref{models} to interpolate corrections of the damping constant \emph{so that 
the resulting mean abundances are independent of line strength}. As documented 
in Table \ref{modmod} these additional corrections are always small. Comparing 
models 7 and 8 in Table \ref{modmod} it is evident that the two interpolated 
results do not differ significantly. 

\begin{figure}
\resizebox{\columnwidth}{!}{\includegraphics{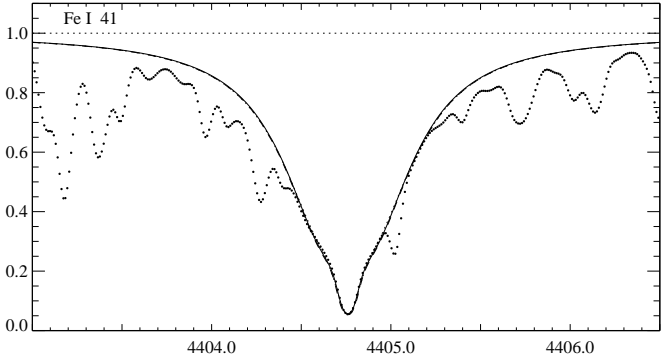}} 
\caption[]{Solar flux spectrum of \FeI\ 41, $\lambda 4404.750$ \AA, together 
with three nearly identical synthetic profile fits using the HM LTE, the TH LTE 
and the TH NLTE 1+ models. See text for a discussion of strong lines} 
\label{strongline} 
\end{figure}
Four characteristic features are displayed in Fig. \ref{modabund}, 
\begin{enumerate}
\item There is still a systematic difference between abundances 
(oscillator strengths) of the Oxford and Hannover groups which is seen best in 
the domain of the turbulence lines around 80 to 100 m\AA. It is also found as a 
difference between lines of low and high excitation. This would be even easier 
to detect if the adjustment of the damping constants were applied to the 
individual sets of lines calculated from a common base of $f$-values. For our 
model (4) in Fig. \ref{modabund} the Oxford data alone then would require a 
damping correction of $\Delta\log C_6 = -0.28$, and they would lead to the value 
of $\eps{\FeI,\odot} = 7.693 \pm 0.052$, reasonably close to that of Blackwell 
et al. (\cite{BLS95a}). Vice versa, LTE in the HM model (4) applied only to the 
Hannover $f$-values would require $\Delta\log C_6 = +0.13$, and result in 
$\eps{\FeI,\odot} = 7.543 \pm 0.070$. Our compromise to fit the combined data 
set thus does not at all \emph{resolve} the long-standing discrepancy. It is 
important to recognize that this problem does \emph{not} seem to depend on the 
particular LTE or NLTE model chosen. The difference between the Oxford and 
Hannover line abundances is only slightly smaller (0.10 dex) for the TH LTE 
model. It is removed here only by adjustment of the damping constants for the 
individual models, the shortcomings of which are hidden in a slightly increased 
scatter. 
\item The \emph{turbulence lines} deviate from both weak and strong lines in 
models (0) to (3), and perhaps in the HM LTE model (4) because the mean 
microturbulence is relatively low. This choice was made in Paper I mostly to 
model a number of the stronger turbulence lines with equivalent widths around 
100 m\AA. After having examined a series of tests with different values we 
concluded that a value of $\xi = 1.0$ \kms\ produced profile fits of 
approximately the same quality. As is evident from comparing models (0) and (5) 
in Fig. \ref{modabund} the higher value of $\xi$ tends to improve the uniformity 
of the abundances. A similar increase would also improve the results of the HM 
model. 
\item Even after having adjusted the \emph{strong lines} to fit to a common mean 
abundance with the weaker lines it is surprising how they lead to systematically 
\emph{lower} abundances than the sample mean. Part of this difference may be 
attributed to a relatively bad fit of the line wings. Fig. \ref{strongline} 
shows the discrepancy between inner and outer wing synthesis. Both parts of the 
profile are of photospheric origin. We note that a slightly better fit of the 
outer wings can be achieved with an increase of the iron abundance by $\simeq 
0.03$ dex which, however, would not remove the trend. Moreover, it would destroy 
the fit of the inner wing to an unacceptable degree. 
\item The run of abundances with excitation energies displays a 
decrease with $E_{\rm low}$ for most of the models. As was emphasized by 
Blackwell et al. (\cite{BLS95a}) and Grevesse \& Sauval (\cite{GS99}) this 
tendency is relaxed or even removed by introducing atmospheric models with lower 
temperatures in the upper photosphere. This trend is confirmed when comparing 
the HM and TH LTE models in our analysis. However, care must be taken not to 
confuse it with a similar one produced by the dependence upon microturbulence. 
The current sample of \FeI\ lines includes quite a number of low-excitation 
lines in the turbulence regime (Mults 1, 2, 3 and 13), which dominate the 
least-squares approach in Fig. \ref{modabund}. This becomes particularly evident 
by comparison of the LTE model (0) and (5), and by the NLTE models (2) and (6), 
where the increase of $\xi$ from 0.85 to 1.00 \kms\ removed most of the energy 
dependence. 
\end{enumerate}

The solar iron abundance determined by even the most careful spectral analysis  
thus depends on the proper choice of both the atmospheric model \emph{and} the 
oscillator strengths. While Grevesse \& Sauval (\cite{GS99}) claim to have 
solved the discrepancies of the long-standing debate on the solar iron abundance 
by introducing their special semi-empirical adjustment to the HM atmospheric 
model, it is only fair to notice that even their final data produce an abundance 
difference with mean values of $(\eps{\FeI,\odot})_{\rm Han} = 7.476 \pm 0.053$, 
and $(\eps{\FeI,\odot})_{\rm Oxf} = 7.514 \pm 0.036$. What makes this result 
less useful is the neglect of all strong lines. As was shown above it is the 
\emph{strong} lines in the Oxford sample that -- having been adjusted to the 
weaker lines by a corresponding decrease of the damping constants -- confirm the 
high solar \FeI\ abundance claimed by Blackwell et al. (\cite{BLS95a}). 
Different from the Kiel-Hannover group the Oxford group does not cover the full 
range of line strengths and excitation energies encountered in the solar 
spectrum. In particular the weak lines are missing, for which an analysis would 
allow a direct comparison of the $f$-value sources without reference to the 
uncertainties of line broadening processes. 

There is no use ignoring the fact that either the oscillator strengths currently 
available are discrepant at a level that cannot be explained by laboratory 
measurement errors alone, or that the solar spectral line identifications are 
erroneous at an equally unacceptable level, or that atmospheric inhomogeneities 
are much more important for individual lines than expected. Let us discuss all 
three possibilities. 

Much of the different \emph{absolute} scales of $f$-values is due to the 
necessary normalization which can be improved; however, an \emph{individual} 
scatter of lines in a common multiplet is obtained even for experimental methods 
thought to be very accurate. As an example let us consider the abundance scatter 
of lines in Mult 114. All lines have been measured by the Hannover group, and 
the abundances spread from $7.41$ at $\lambda 5141.739$ to $7.65$ at $\lambda 
5049.819$ to a value as high as $7.78$ for $\lambda 4924.769$ if the HM LTE 
model is applied. These are not faint lines for which high measurement errors 
could be accepted; the experi\-mental error estimates range from 0.04 to 0.07 
dex for these lines, which transforms to the fact that our abundances lead to 
results that are discrepant on much more than a $3\sigma$ level. Of course, the 
results may tell us that the hollow-cathode measurements of $\lambda 5141.739$ 
are not of the same quality as the other two lines which were measured by 
laser-induced fluorescence, but that would invalidate the experimental error 
estimates. 

Comparison of such multiplet abundance scatter based on common source $f$-values 
with that already discussed above indicates that this does not depend very much 
on the experimental methods either, although there may exist still a number of 
problems that are connected with the control of experimental environment 
parameters as discussed by  Holweger et al. (\cite{HKB95}). Thus we conclude 
that agreement of \emph{mean} abundance values between different sources of 
oscillator strengths (often claimed for the O'Brian et al. data) is not a 
significant measure of methodical accuracies. Taken at face value the r.m.s. 
scatter of abundances obtained from a single set of oscillator strengths such as 
that of O'Brian et al. is a measure of the accuracy of the mean solar \FeI\ 
abundance that can be reached with these data. In fact the accuracy is then even 
less due to blends and other problems referring to the profile fits, and to the 
ambiguities of atmospheric modelling. 

There exists a number of lines in the iron spectrum that could be misidentified 
in that the spectral features could be blends that are not only unresolved but 
also fall within a few m\AA\ of the same center wavelength. As with other 
undetected blends such profiles will be fitted with too large abundances. This 
should produce abundance distributions that are systematically shifted to the 
high-abundance side, something that is not detected in the results. To reduce 
the dominating intrinsic abundance scatter to reasonable amounts it would mean 
that more than half of the lines would have to be corrected for such blend or 
identification problems, a situation that seems highly unlikely. We note that 
many blend problems of the kind producing too large fit abundances are avoided 
by our profile fitting method which allows an exchange of certain fit parameters 
such as abundance, microturbulence or damping parameters only within a narrow 
region. In such cases the profile fit procedure always tends to produce 
\emph{higher} abundances.  

Our discussion of line broadening in subsection 3.1.2 and Fig. \ref{asplund} has 
shown that the true abundance differences resulting from line formation in 
plane-parallel and in hydrodynamic atmospheres are quite small. They are even 
negligible taking into account the large abundance differences that appear 
between sets of different $f$-values. The mere change of \emph{atmospheric 
models} affects the mean abundance but not the r.m.s. scatter as can be found in 
Table \ref{modmod}, and it is obvious that changing the microturbulence has a 
greater influence on such results. Thus it is doubtful if any other atmospheric 
model could significantly reduce the abundance scatter.

Our results then indicate that it is the atomic data, in particular the 
oscillator strengths, that presently do not allow the determination of the solar 
\FeI\ abundance with an accuracy better than $\sim 0.1$ dex. Based on the most 
reliable sets of $f$-values (Oxford and Hannover data) and on the model 
producing the smallest overall dependence on excitation energy (TH NLTE 1/2+) we 
find a value of $\eps{\FeI,\odot} = 7.509 \pm 0.075$ with no dependence on line 
strength but a small residual gradient with energy, $\Delta\eps{}/\Delta E_{\rm 
low,\eV} = -0.005$. In view of the differences between the Oxford and Hannover 
$f$-values it is important to notice that this value is only 0.02 dex above that 
obtained from the Hannover data alone, while it is 0.09 dex below the pure 
Oxford value.  This apparent contradiction is resolved by inspection of the 
corresponding energy dependence of the respective sources. Whereas the Hannover 
results show no energy gradient, the Oxford data -- after having adjusted the 
damping constants to remove a line strength trend -- keep a strong gradient with 
excitation energy for which $\Delta\eps{}/\Delta E_{\rm low,\eV} = 0.034$. The 
last three models in Fig. \ref{modabund} show only a small residual energy 
dependence of the \FeI\ abundances ranging from  $\Delta\eps{}/\Delta E_{\rm 
low,\eV} = -0.0094$ for the TH NLTE5+ model to $\Delta\eps{}/\Delta E_{\rm 
low,\eV} = -0.0054$ for the TH NLTE1/2+ model. 

The above results are to be understood as a clear report of our failure to solve 
the photospheric solar \FeI\ abundance problem if more than the Hannover data 
set were involved. Using this data set \emph{alone} with the HM LTE model, a 
microturbulence of $1.05$ \kms\ together with damping corrections $\Delta\log 
C_6 = 0.11$ (above the Anstee \& O'Mara damping constants) yields 
$\eps{\FeI,\odot} = 7.535 \pm 0.070$. The energy gradient for that result is 
$\Delta\eps{}/\Delta E_{\rm low,\eV} = -0.008$. The overall best NLTE model (TH 
NLTE 1/2+) applied to the Hannover data alone leads to $\eps{\FeI,\odot} = 7.480 
\pm 0.072$ with no dependence on energy. 

\section{Conclusions}

The choice of a particular model to determine the solar \FeI\ line formation 
with a valid parametrization of the atomic collisions is not possible even when 
including the weak solar lines. Arguments referring only to the solar abundance 
problem with or without inclusion of the \FeII\ lines are not conclusive since 
both sets of $f$-values (\FeI\ and \FeII) are far from producing homogeneous 
results. One marginal result is that the models of Paper I with their low 
microturbulence are no longer competitive because they all display a relatively 
strong gradient with excitation energy (see Fig. \ref{modabund}). This does no 
longer appear when increasing the microturbulence from $\xi = 0.85$ \kms\ to 
$1.00$ \kms\ as in our present models 5 to 9. All the TH models are roughly 
compatible with meteoritic abundance. Small corrections for dynamic line 
formation such as suggested by comparison with hydrodynamic results of Asplund 
et al. (\cite{ANTS00}) in section 3.1.2 are of the order of $-0.03$, which would 
bring the solar abundance to a value slightly below that of the carbonaceous 
chondrites. 

The quality of individual line fits are significantly different for the HM and 
TH model atmospheres only for the cores of strong lines. In Paper I this was 
demonstrated for a number of lines of various excitation energies. The line 
center flux reflects essentially the different temperatures in the upper 
photosphere with a 150 \ldots 200 K difference predicting $\Delta F \sim 4$\% as 
observed. However, these differences vanish when a compromise is accepted for a 
profile fit of the inner wings (see Fig. \ref{strongline}) allowing the 
synthetic profile to fall \emph{below} the observed flux by a small amount. The 
evaluation of profile fits thus has changed marginally as compared with Paper I. 
For the weaker lines Figs. \ref{weaklines} and \ref{turblines} document the 
independence of fit quality from the model atmosphere if abundances and 
macroturbulence velocities are adjusted accordingly. 

The selection of a particular atmospheric/atomic model on the grounds of profile 
synthesis of the solar \FeI\ flux spectrum is therefore still somewhat 
ambiguous. This would be different if the abundance determinations were of 
higher quality. For \emph{differential} analyses of stellar spectra it is 
obvious that our atmospheric model can be only one of the TH models because only 
they allow a physically consistent change of parameters such as $\Teff$, $\log 
g$ or [Fe/H]. Since strong lines in the solar spectrum reduce to weak or 
turbulence lines in stars of low metal abundance, it is most important to 
install a unique recipe for the determination of the damping parameter. This can 
be done with reference to Table \ref{modmod} where a good mean value for the 
correction would be $\Delta\log C_6 = -0.15$. We should, however, bear in mind 
that this deviation from the Anstee \& O'Mara results is essentially necessary 
to correct the \emph{strong} lines with $f$-values from the Oxford group. The 
error introduced to differential abundance determinations in metal-poor stars 
thus will have to include a systematic uncertainty of $\sim 0.04$ dex due to 
inconsistencies in the interpretation of the \emph{solar} lines. 

Current investigations of a small number of \emph{reference} stars with 
different iron abundances will have to show how to select a common NLTE model 
that fits the \FeII/\FeI\ ionization equilibria of all stars.   

\begin{acknowledgements}
Part of this work was funded by the Deutsche Forschungsgemeinschaft under grant 
Ge 490/12-2. AJK benefitted from a stipend of the Studienstiftung des Deutschen 
Volkes. JS is grateful for support from the National Natural Science Foundation 
of China 
\end{acknowledgements}

\end{document}